%% file: main.tex
\documentclass[useAMS,usenatbib]{mn2e}

\input{macros}

\begin{document}

\label{firstpage}

\title[Scan strategies for CMB satellite experiments]{Optimal scan strategies for future CMB satellite experiments}
\author[Wallis et al.]{\parbox[t]{\textwidth}{Christopher G. R. Wallis$^{1,2}$\thanks{E-mail:
chris.wallis@ucl.ac.uk}
, Michael L. Brown$^1$, Richard A. Battye$^1$, Jacques Delabrouille$^3$}\vspace*{6pt}\\
$^1$Jodrell Bank Centre for Astrophysics, School of Physics and Astronomy, The University of Manchester, Oxford Road, Manchester, M13 9PL, U.K.\\
$^2$Mullard Space Science Laboratory, University College London, Surrey RH5 6NT, U.K.\\
$^3$APC, 10, rue Alice Domon et Leonie Duquet, 75205 Paris Cedex 13, France
}

\date{Accepted 2015 XXXXX XX. Received 2015 XXXXX XX; in original form 2015 XXXXX XX}

\pagerange{\pageref{firstpage}--\pageref{lastpage}} \pubyear{2015}

\maketitle

\begin{abstract}
The $B$-mode polarisation power spectrum in the Cosmic Microwave
Background (CMB) is about four orders of magnitude fainter than the
CMB temperature power spectrum. Any instrumental imperfections that
couple temperature fluctuations to $B$-mode polarisation must therefore
be carefully controlled and/or removed. We investigate the role
that a scan strategy can have in mitigating certain common systematics
by averaging systematic errors down with many crossing angles. We present
approximate analytic forms for the error on the recovered $B$-mode
power spectrum that would result from differential gain, differential pointing and
differential ellipticity for the case where two detector pairs are used in a polarisation
experiment. We use these analytic predictions to search the 
parameter space of common satellite scan strategies in order to identify
those features of a scan strategy that have most impact in mitigating
systematic effects. As an example we go on to identify a scan strategy suitable {\color{black}for the
CMB satellite proposed for the ESA M5 call.}
considering the practical considerations of fuel requirement, data rate and
the relative orientation of the telescope to the earth. Having chosen a scan strategy we
then go on to investigate the suitability of the scan strategy.

%We find that, as long as the satellite spin
%period is much less than the precession period, and that both are much less
%than 1 year, the exact values of the timescales are unimportant from
%the point of view of mitigating systematics. We conclude that the main
%parameters of interest are the precession angle and the boresight
%angle. By reducing the boresight angle and increasing the precession
%angle, a scan strategy will scan the sky in small circles. These
%small circles are beneficial for creating a wide range of orientation
%angles, and are therefore effective in mitigating systematic effects. 
\end{abstract}


\input{sections/intro}

\input{sections/analytic}

\input{sections/full_sims}

\input{sections/strat_sims}

\input{sections/scan_investigation}

\input{sections/discussion}

\section*{Acknowledgments}
{\color{black}The authors thank Clive Dickinson and Patrick Leahy 
for useful discussions and comments on the manuscript.} 
{\color{black}We also acknowledge useful discussions with Pascal Rideau and 
Joel Michaud concerning practical constraints related to data transfer and to attitude and orbit control.}
CGRW acknowledges the award of a STFC quota studentship. MLB
acknowledges the European Research Council for support through the
award of an ERC Starting Independent Researcher Grant (EC FP7 grant
number 280127). MLB also thanks the STFC for the award of Advanced and
Halliday fellowships (grant number ST/I005129/1). Some of the results in this paper have been derived using
the {\sevensize HEALPix}~\citep{2005ApJ...622..759G} package.

\bibliography{references}{}
\bibliographystyle{mn2e_arXiv}
%\bibliographystyle{mn2e}

\input{sections/appendix}

\label{lastpage}

\end{document}

%% file: macros.tex
\usepackage{times}
\usepackage{amsmath}    % need for subequations
\usepackage{graphicx}   % need for figures
\usepackage{verbatim}   % useful for program listings
\usepackage{color}      % use if color is used in text
\usepackage{subfigure}  % use for side-by-side figures
\usepackage{amssymb}
\usepackage{cite}
\usepackage{setspace}
\usepackage{gensymb}
\newcommand{\threej}[6]{\left(\begin{array}{ccc} #1 & #2 & #3 \\
                                                 #4 & #5 & #6
                        \end{array}\right)}
\newcommand{\be}{\begin{equation}}
\newcommand{\ee}{\end{equation}}
\newcommand{\ba}{\begin{eqnarray}}
\newcommand{\ea}{\end{eqnarray}}
\newcommand{\bnum}{\begin{enumerate}}
\newcommand{\enum}{\end{enumerate}}

\newcommand{\ft}[1]{\tilde{#1}}

\newcommand{\bom}{\bomega}

%% file: sections/intro.tex
% !TEX TS-program = compile

\section{Introduction}

The cosmic microwave background (CMB) $B$-mode polarisation represents
a powerful cosmological probe. In particular, certain early Universe
models predict large scale $B$-mode polarisation due to gravitational
waves created during inflation. On smaller scales gravitational
lensing of the stronger $E$-mode polarisation creates a lensing
$B$-mode polarisation signal. Observationally, a great deal of effort
has been devoted in recent times to measuring the $B$-mode
polarisation signal. The {\it Keck}/BICEP2 series of experiments
\citep{2015arXiv151009217A} has produced the deepest polarization maps
at 95 and 150 GHz, thus providing tight constraints on the amplitude of the
large scale ($\ell \lesssim 100$) $B$-mode polarisation power spectrum.
This constraint was only made possible due to the wide frequency range of
polarisation maps provided by the WMAP and {\it Planck} experiments 
\citep{2013ApJS..208...20B, 2015arXiv150201582P} --- the low frequency
maps constrain the foreground synchrotron emission and the
high frequency maps constrain the level of foreground polarized dust
emission. The smaller scale ($\ell \approx 500-2000$)
lensing $B$-mode signal has been detected by a number of groups thus
providing constraints on large scale structure
\citep{2014ApJ...794..171T,2013PhRvL.111n1301H}.

The $B$-mode polarisation power spectrum is approximately four orders
of magnitude smaller than the temperature power spectrum. Therefore,
any coupling between the two must be carefully controlled and/or
removed. {\color{black}Previous analytic work on this topic has focused 
on predicting the levels of leakage 
\citep{2003PhRvD..67d3004H,2007MNRAS.376.1767O,2008PhRvD..77h3003S,2014MNRAS.442.1963W}
while experimental teams have included a careful assessment of its
impact. For example, the POLARBEAR collaboration addressed the issue of temperature to polarisation 
leakage with detailed simulations \citep{2014ApJ...794..171T}.} Using
these, they showed that the leakage was significantly below the $B$-mode
polarisation power spectrum signal that they had detected. The {\it Planck}
Low Frequency Instrument examined temperature to polarisation leakage
extensively pre-launch \citep{2010A&A...520A...8L}. The BICEP2
team found levels of temperature leakage that were a factor of a few
above the $B$-mode polarisation power spectrum signal. They therefore
developed and applied techniques to remove the leakage by fitting for it in the
polarisation timestream \citep{2015arXiv150200608B}. In addition to
removing the temperature leakage, this approach removes
genuine polarisation signal that must be accounted for in the
subsequent estimation of polarisation power spectra. It would
therefore be beneficial to either use techniques that do not require
such fitting (e.g.~\citealt{2015arXiv150303285W}), or preferably, to
design an experiment sufficiently well such that the leakage is 
insignificant.

Here, we examine how satellite scan strategies can be designed to mitigate
various systematic effects that couple temperature to polarisation for
satellite-based experiments. In our study, we focus on differencing
experiments, consisting of pairs of detectors. Within each pair, one
detector is sensitive to the CMB temperature and polarisation
signal in a particular direction on the sky, convolved with some
detector response function. The other detector is in
principle sensitive to the same temperature signal but has a
polarisation sensitivity that is rotated by 90 degrees. By
differencing the timestreams of these two detectors, the temperature
response is removed. However, any differences between the two
temperature response functions of the detectors will couple
temperature fluctuations to the polarisation map.

In this study, we consider three types of mismatch between the two
temperature response functions, all of which were found to be present
in the BICEP2 experiment \citep{2015arXiv150200608B}. Firstly, we
consider a difference in the gain calibration of the two detectors
which we term ``differential gain''. We note that a difference in the 
spectral windows of the two detectors, as has been shown to exist 
for instance in Planck bolometer pairs \citep{2015arXiv150201586P},
 is equivalent to a ``differential" 
gain effect that depends on the emission law, and thus is different 
for different astrophysical components. We also consider a difference in
the pointing direction of the two detector response functions, which
we call ``differential pointing''. Finally we investigate the impact
of a ``differential ellipticity'' arising from a difference in the
beam ellipticities of the two detectors.

{\color{black}The systematics that we consider are typical for the type of
experimental set up that we have chosen. By using bolometers, which
do not conserve the phase of the incoming radiation, either detector differencing 
or polarisation angle rotation must be used to disentangle the
polarisation signal from the temperature response of the
detector. Modern bolometers are close to photon-noise limited and are
therefore prefered by state-of-the-art CMB polarisation experiments. 
The South Pole Telescope \citep{2008ApOpt..47.4418P},
BICEP2 \citep{2014arXiv1403.4302B} and the POLARBEAR collaboration 
\citep{2014ApJ...794..171T} all use bolometers. Using a half wave plate
(HWP) can allow one to make maps of temperature and polarisation without
differencing. If the HWP is continually rotating then certain ``lock-in''
techniques can be used to isolate the polarisation signal from the
systematic errors \citep{2007ApJ...665...55W}. However, maintaining
continuous rotation of the HWP can cause its own wealth of
systematic errors. A stepped HWP can also be used to increase the
polarization angle coverage but it is less effective than a rotating HWP
in terms of mitigating systematic effects \citep{2009MNRAS.397..634B}. 
However, systematic effects associated with the HWP itself are much
easier to control when the HWP is stepped.  

Previous work has been undertaken to identify optimal scan
strategies for CMB satellite experiments. \citet{2000ApL&C..37..259D} 
identified the requirements for the scan strategy for the {\it Planck}
mission. One of these requirements was the need to have multiple
crossing angles in order to mitigate systematic errors that depend on the orientation, 
or parallactic angle, of the telescope -- a problem that we also consider in
this paper. \citet{1998MNRAS.298..445D} looked at the increase of noise due to 
instrumental drift, or $1/f$ noise, and the benefit that a well chosen 
scan strategy can have on the final power spectrum analysis. The benefits 
of different scan strategies for {\it Planck} were also investigated in 
\citet{2005A&A...430..363D}, where the authors attempt to
maximise the uniformity of the integration time over the sky.}

The aim of our study is to examine the degree to which a scan strategy
can mitigate systematic errors by averaging their effects through
multiple observations of the same sky pixel with different instrument
orientations. To do this we first derive a set of simple equations
that predict the error on the recovered $B$-mode polarisation power
spectrum given a few characteristics of the scan strategy and the
amplitude of the systematic effect. We then go on to use these simple
equations to predict the error on the $B$-mode polarisation power
spectrum for different satellite scan strategies. This allows us to
clearly identify those features of a scan strategy that
have the most impact in controlling the level of instrumental $B$-mode
polarisation.

{\color{black} At the time of writing there are a number of major proposed CMB
polarization satellite missions for which our work is relevant.  An
improved Cosmic Origins Explorer (COrE+) proposal is currently being
prepared for submission to the European Space Agency's anticipated
``M5'' call for proposals for a medium sized mission. The primary
science goal of the improved COrE+ will be to contrain the inflationary $B$-mode
polarisation signal to a precision of $\sigma_r \sim 10^{-3}$ (where $r$
is the tensor-to-scalar ratio). However, the COrE++ concept will also
facilitate many other science goals including precision CMB lensing
measurements and Sunyaev-Zel'dovich cluster counts. The Japan
Aerospace Exploration Agency has a well developed proposal named
``Lite (Light) satellite for the studies of $B$-mode polarisation and
Inflation from cosmic background Radiation Detection (LiteBIRD,
\citealt{2014JLTP..176..733M}). The LiteBIRD team have built in many
features into the design of the experiment to mitigate systematic
effects, including a HWP providing the experiment with additional
polarisation angle modulation.  Another proposed mission is the
Primordial Inflation Expolore (PIXIE, \citealt{2011JCAP...07..025K}),
which also focusses on inflationary $B$-modes as its primary science
goal. The PIXIE concept makes use of a polarizing Fourier Transform
Spectrometer (FTS) to measure both the linear polarisation and the
spectral dependence of the microwave sky over a large range in
frequency, from 30 GHz to 6 THz. In this paper, we have particularly focussed
on the scan-strategy options for the improved COrE+ mission concept. However,
our results are general enough that they should also be useful for the 
design of scan strategies for these other satellite missions. }

The paper is organised as follows. In Section~\ref{sec:ana} we derive
the equations that predict the temperature-to-polarisation leakage due
to three main systematic effects that are of concern for CMB
polarization measurements. In Section~\ref{sec:sims} we use time
order data simulations to demonstrate the validity of the equations
derived in Section~\ref{sec:ana}. In Section~\ref{sec:scan_params} we
search the main parameter space of satellite scan strategies to
identify those key features that have the largest impact in terms of
mitigating systematic effects. In Section \ref{sec:scan_investigation}
we demonstrate how the tools developed in the previous sections can be
used to identify an ``optimal'' scan strategy for COrE++. Finally in
Section~\ref{sec:discussion} we summarise our results.

%% file: sections/analytic.tex
% !TEX TS-program = compile

\section{Impact of systematics on the $B$-mode power spectrum}\label{sec:ana}

To assess the impact of the systematic effects on the recovered
$B$-mode power spectrum we begin by considering the detected signal
from a single pair of detectors. The detected signal from a single
detector pointing at sky position $\Omega$ is
\ba
d_{i}^X = \int d\Omega [B^T(\Omega)T(\Omega) +B^Q(\Omega)Q(\Omega)
+B^U(\Omega)U(\Omega)],
\label{eq:beam_int}
\ea where $i$ denotes the pair that the detector belongs to and
$X=\{A,B\}$ distinguishes between the two detectors within a
pair. $B^Y(\Omega)$ is the beam response function of the detector to
the Stokes parameters $Y=\{T,Q,U\}$. In this work we focus on the most
problematic systematic effects for CMB $B$-mode experiments -- those
which couple the temperature signal to polarisation maps. We consider
a differencing experiment, i.e.\ one where each instrument ``pixel'' is
composed of two detectors which are sensitive to orthogonal
polarisation directions. The timestream from the two detectors in a
pair can be summed to obtain the temperature of the sky and
differenced to obtain a measurement of the polarisation. Therefore any
mismatch in the response of the two detectors to the temperature sky
will result in this type of leakage. Explicitly the differenced signal
is,
\ba
d_i = \frac{1}{2}(d_i^A - d_i^B).
\ea

We describe the pointing of the telescope using Euler angles, $\bom =
(\theta,\phi,\psi)$. The Euler angles represent a sequence of three
active rotations (starting from some fiducial initial
orientation). They are active in the sense that the beam moves with
respect to the coordinate system. The following series of steps
describe how to rotate the beam from the fiducial orientation to the
orientation described by $\bom$, all rotations being performed
anticlockwise when looking down the axis by which they are defined,
that is they are performed in a right handed sense.

\bnum
\item The beam is rotated around the $z$ axis by $\psi$.
\item The beam is rotated by $\theta$ around the $y$ axis.
\item The beam is rotated around the $z$ axis again by $\phi$.
\enum

One important characteristic of a systematic error when considering
the impact on the recovered polarisation maps and power spectra is the
spin of the systematic error. The spin is defined by how the
temperature leakage rotates with the orientation of the telescope for
a particular sky pixel. Here we have defined the orientation angle by
$\psi$, which is the angle between the orientation of the focal plane
and the direction to the North pole. This is often referred to as the 
parallactic angle of the telescope. If the systematic error is of a
different spin to the spin-2 polarisation signal, then if a sky pixel
is observed at many instrument orientations the resulting bias on
the measured polarisation is reduced. The primary goal of this study
is to examine the effectiveness of different scan strategies to
mitigate systematic errors in this way. To faciliate this we derive 
a set of simple analytic equations to model the leaked $B$-mode
polarisation power spectrum. The first stage of the derivation is to
calculate the leakage of the temperature signal into the polarisation
timestream for each systematic error.

\subsection{Leakage in the differenced timestream}

We start by examining the effect of differential gain on the
differenced signal from a detector pair. This is simply a
mis-calibration between the two detectors of $\delta g_i$. The
temperature leakage, $\delta d^{\rm g}_i$, due to differential gain
($\delta g_i$) in detector pair $i$ is,
\ba
\delta d^{\rm g}_i &=& \frac{1}{2}(T^{\rm B}(\Omega) - (1-\delta g_i)T^{\rm B}(\Omega))\\
&=& \frac{1}{2}\delta g_i T^{\rm B}(\Omega),\label{eq:dg_time}\\
&=& G^{i} \label{eq:dg_G}
\ea
where $T^{\rm B}(\Omega)$ denotes the CMB temperature sky convolved
with the axisymmetric part of the temperature beam and this defines the 
level of the systematic gain, $G^{i}$. This systematic
effect is independent of the orientation of the telescope with respect
to the sky coordinates. It depends only on the size of the temperature
signal in the direction in which the telescope is pointing at any
given time. Differential gain is, therefore, a spin-0 systematic effect.

The second systematic effect we consider is differential
pointing. {\color{black}This is a misalignment of the two detector beams by some
angle $\rho_i$ in a direction $\chi_i$ with respect to the orientation
of the telescope from North $(\psi)$, see Fig.~\ref{fig:dif_pointing_schem}.}
As the differential pointing
will be a small angle, we make the flat sky approximation. We consider a
Cartesian coordinate system where the $y$-axis is aligned with
North. The error in the differenced timestream can then be modelled as,
\ba
\delta d^{\rm p}_i &=& \frac{1}{2}[T^{\rm B}(x,y)\\ \nonumber 
&&- T^{\rm B}(x-\rho_i \sin(\psi+\chi_i),y-\rho_i\cos(\psi + \chi_i))].
\ea
%
\input{sections/fig/dif_pointing_schematic}
If we Taylor expand around $(x,y)$ to first order in $\rho_i$, we find
\ba
\delta d^{\rm p}_i &=&  \frac{1}{2}\left[ \frac{\partial T^{\rm B}}{\partial x}\rho_i \sin(\psi+\chi_i)  + \frac{\partial T^{\rm B}}{\partial y}\rho_i\cos(\psi + \chi_i)\right],\\
&=& \frac{1}{4}\left[\left(\frac{\partial T^{\rm B}}{\partial y}-i\frac{\partial T^{\rm B}}{\partial x}\right)\rho_ie^{i(\psi + \chi_i)}+ {\rm c.c.}\right]\label{eq:dp_time},\\
&=& \frac{1}{2}(M^{i}e^{i\psi} + {\rm c.c.})\label{eq:dp_M}
\ea
where ${\rm c.c.}$ denotes the complex conjugate of the first term
inside the square bracket and this defines the systematic due to 
differential pointing $M^{i}$. The $e^{i\psi}$ term signifies the known
result that differential pointing is a spin-1 systematic effect. 

The final systematic effect we consider is a differential ellipticity
between the detector pairs. To treat this, it is convenient to write
the integration in equation \eqref{eq:beam_int} in spherical harmonic
space. For simplicity we use a coordinate system with the North pole
coincident with the pointing centre. We denote the beam decomposed into
spherical harmonics as $b^X_{\ell k}$ and the temperature sky is
denoted by $a_{\ell m}^T$. The error on the differenced signal
between the two detectors within a pair is then
\ba
\delta d^{\rm e}_i &=& \frac{1}{2}\sum_{\ell m} \sqrt{\frac{4\pi}{2\ell+1}}\left(b^{A}_{\ell m} a^{T*}_{\ell m} - b^{B}_{\ell m} a^{T*}_{\ell m}\right),\\
&\approx& \frac{1}{2} \sum_{\ell} \sqrt{\frac{4\pi}{2\ell+1}}\left(\delta b^i_{\ell, 2}a_{\ell, 2}^{T*}+\delta b^i_{\ell, -2}a_{\ell, -2}^{T*}\right)\label{eq:de_time},
\ea
where we have assumed that the axisymmetric components of the two
detector beams cancel and that the dominant remaining term is
the second azimuthal mode. This is demonstrated to be true for an
elliptical Gaussian beam in figure 2 of
\citet{2014MNRAS.442.1963W}. When the telescope orientation is
changed, but the same patch of sky is observed, this differential
ellipticity will rotate as, $\delta b_{\ell, 2}' = e^{i2\psi}\delta
b_{\ell, 2}$. This rotation makes the differential ellipticity effect a
spin-2 effect. We can, therefore, rewrite the temperature leakage due
to differential ellipticity as a function of the orientation of the telescope as
\ba
\delta d^{\rm e}_i &\approx& \!\!\!\!\frac{1}{2} \sum_{\ell} \sqrt{\frac{4\pi}{2\ell+1}}(\delta b^i_{\ell, 2}a_{\ell, 2}^{T*}e^{i2\psi}{+}\delta b^{i}_{\ell, -2}a_{\ell, -2}^{T*}e^{-i2\psi}),\\
&=& \!\!\!\!\frac{1}{2}(E^ie^{i2\psi} + {\rm c.c.}),\label{eq:de_E}
\ea
which defines $E^i$.

\subsection{Temperature leakage to the polarisation map}

We can combine the above understanding of the temperature leakage
effects due to the various systematics to create a model for the
differenced signal as a function of the orientation of the telescope.   
The differenced signal from a detector pair $i$ for a single sky pixel
is given by
\ba
S^i(\psi) = G^i + \frac{1}{2}[Pe^{i2\psi} + M^ie^{i\psi} +
  E^ie^{i2\psi} + {\rm c.c.}]\label{eq:s_of_psi},
\ea
where $P=Q+iU$ is the complex representation of the polarisation
signal in the sky pixel. $G^i$, $M^i$ and $E^i$ are the contributions
to the differenced signal from the three systematic effects we are
considering --- differential gain, pointing and ellipticity
respectively. The exact form of $G^i$, $M^i$ and $E^i$ are defined in
equations (\ref{eq:dg_G},\ref{eq:dp_M},\ref{eq:de_E}). To understand the effect that
these systematics have on the recovered power spectrum we first
examine the effect that they have on the map.

In general, each pixel of a map will be observed at a variety of
orientation angles $\psi$. We define the detected signal as $S^d$
which is the information that we have about the pixel as a function of
the orientation,
\ba
S^d(\psi) &=& h(\psi)S(\psi), \text{where,}\\
h(\psi) &=& \frac{2\pi}{N_{\rm hits}}\sum_j \delta(\psi - \psi_j) \label{eq:h_def}.
\ea
Here, $\psi_j$ is the orientation of the $j$th observation (or
``hit'') of the pixel and the sum is over all such observations,
$N_{\rm hits}$. 

Let us now consider the signal seen by two detector pairs. To do this
imagine an ``instrument-$Q$'' detector pair whose orientation with respect to
the sky coordinates for the $j$th observation is labelled $\psi_j$
and an ``instrument-$U$'' pair of detectors whose
polarization sensitivity directions are rotated by $\pi/4$ with respect
to those of the ``instrument-$Q$" detectors. For every hit, there will
then be two orientations, one at $\psi_j$ and the other at
$\psi_j+\pi/4$, each with different systematic effects. We can
therefore write
\ba
S^d_{\rm tot}(\psi) = h(\psi)S^1(\psi) + h(\psi-\pi/4)S^2(\psi),
\ea
which we can write in Fourier space as a convolution:
\footnote{The Fourier transform and inverse we use are,
\ba
\tilde{f}_k &=& \frac{1}{2\pi}\int_0^{2\pi} d \psi e^{ik\psi}f(\psi),\nonumber\\
f(\psi) &=& \sum_{k=-\infty}^{\infty} \tilde{f}_k e^{-ik\psi}. \nonumber
\ea }
\ba
\tilde{S}^d_k = \sum_{k'=-\infty}^{\infty}  \tilde{h}_{k-k'}\tilde{S}^1_{k'} + \tilde{h}_{k-k'}e^{i\pi(k'-k)/4}\tilde{S}^2_{k'}. \label{eq:conv_sum}
\ea

From equation~\eqref{eq:s_of_psi} we can see that $S^i(\psi)$ is
made up of only a few Fourier terms. Therefore equation~\eqref{eq:conv_sum}
can also be limited to just a few terms, explicitly the $k'{=}{-}2$ to
$k'{=}2$ terms. We can use these simplifications to write out the
$k=2$ row of $\tilde{S}^d_k$:
\ba
\tilde{S}^d_2 &=& \tilde{h}_0(\ft{S}^1_2 + \ft{S}^2_2) + \ft{h}_1(\ft{S}^1_1 + e^{i\pi/4}\ft{S}^2_1)+ \ft{h}_2(\ft{S}^1_0 + i\ft{S}^2_0) \nonumber \\
&& + \ft{h}_3(\ft{S}^1_{-1} + e^{i3\pi/4}\ft{S}^2_{-1}) + \ft{h}_4(\ft{S}^1_{-2} - \ft{S}^2_{-2}). \label{eq:row_2}
\ea

We can see from the definition of $h(\psi)$ in equation
\eqref{eq:h_def} that $\ft{h}_0 {=}1$; also from equation
\eqref{eq:s_of_psi} we can see that $S^i_2{=}P/2$ and
$S^i_{-2}{=}P^*/2$. Therefore, in the absence of any systematic
effects or instrumental noise, $\ft{S}_2^d {=}P$. In a simple binning
map-making scheme this leads to an estimate of the polarisation in a
pixel. We can now deduce the spurious polarisation signal by examining
equation~\eqref{eq:row_2}. Any additional terms, beyond the expected
$P$, will be spurious. Performing this analysis for the systematic
effects we have considered, and defining the differential operator 
$\nabla = (\partial/\partial y - i\partial/\partial x)$, we find the
following systematic terms in the polarization maps:
\ba
\Delta P^{\rm g} &=& \frac{1}{2} \ft{h}_2 (\delta g_1 + i\delta g_2) T^{\rm B}, \label{eq:deltap_dg}\\
\Delta P^{\rm p} &=& \frac{1}{4} \ft{h}_1 \nabla T^{\rm B}(\rho_1e^{i\chi_1} + \rho_2e^{i(\chi_2 + \pi/4)}) \label{eq:deltap_dp}\\
&& + \frac{1}{4} \ft{h}_3 \nabla^* T^{\rm B}(\rho_1e^{-i\chi_1} + \rho_2e^{-i(\chi_2 - 3\pi/4)}),\nonumber\\
\Delta P^{\rm e} &=& \frac{1}{2} \sqrt{\frac{4\pi}{2\ell+1}}(\delta b^1_{\ell, 2} + \delta b^2_{\ell, 2})a_{\ell, 2}^{T*} \label{eq:deltap_de}\\
&&+ \frac{1}{2}\ft{h}_4\sqrt{\frac{4\pi}{2\ell+1}}(\delta b^1_{\ell, -2} - \delta b^2_{\ell, -2})a_{\ell, -2}^{T*}.\nonumber 
\ea
\subsection{Temperature leakage to the $B$-mode power spectrum}

We now wish to calculate the error on the $B$-mode polarisation power
spectrum. As the window function, $h(\psi)$, will be different for
each pixel, an exact calculation is a difficult computational
task. However, we can simplify the problem with a few
approximations. Firstly, we assume that the only effect the $\ft{h}_k$
term of equations (\ref{eq:deltap_dg})--(\ref{eq:deltap_de}) have on
the power spectrum is to damp the resulting bias by a factor of
$\langle|\ft{h}_k|^2\rangle$, where the average is over all pixels of
the sky. The phase of $\ft{h}_k$ across the sky, and its coupling to
the temperature sky, will dictate whether or not the spurious
polarisation is of a $E$-mode or $B$-mode form and therefore we expect 
this aproximation to give an indication of the amplitude of the effect. As we do not expect
the scan strategy to correlate with the temperature
sky, we assume that half the resulting bias power will be of $E$-mode
and half of $B$-mode form. The temperature terms in the spurious
polarisation will result in an error on the $B$-mode power spectrum
whose size is proportional to the temperature power spectrum. In
equation \eqref{eq:deltap_dg} the temperature field is convolved with
the axisymmetric component of the beam. Therefore, the resulting bias
will be proportional to $B_\ell^2C_{\ell}^T$, where
$B_\ell{=}\sqrt{\frac{4\pi}{2\ell+1}}b_{\ell 0}$ is the smoothing
function due to the beam. The differential pointing is dependent on
the gradient of the convolved temperature sky and therefore the bias
will be proportional to $\ell^2B_\ell^2C_{\ell}^T$. The differential
ellipticity is dependent on the temperature field directly and the
resulting bias will be proportional to $C_\ell^T$. For the systematic
error terms, we simply take the modulus squared. We provide
justification for, and examine the impact of, these approximations in
Appendix~\ref{sec:pseudo_cl_coupling}. In Section~\ref{sec:sims}, we
demonstrate with full time-ordered data (TOD) simulations the accuracy
of the approximations for a selection of representative scan
strategies.

%Note that the first term of equation~\eqref{eq:deltap_de} has no
%$\ft{h}_k$ contribution. This is a consequence of the fact that
%differential ellipticity and the polarisation signal are both spin-2
%quantities. It has been shown previously that this leakage will only
%couple to $B$-modes if the polarisation sensitivity is offset from the
%direction of the ellipticity. This will manifests itself as $\delta
%b_{\ell 2}$ having an imaginary component and only this part couples
%temperature to polarisation \citep{2014MNRAS.442.1963W}.

With the approximations described above, we find the following
expressions for the bias on the $B$-mode power spectrum resulting from
the systematic effects we have considered:
\ba
\Delta \ft{C}_\ell^{BB{\rm g}} &=& \frac{1}{8}\langle |\tilde{h}_2|^2\rangle\left|\delta g_1 + i\delta g_2\right|^2 B_\ell^2C_{\ell}^{TT},\label{eq:psdg}\\
\Delta \ft{C}_\ell^{BB{\rm p}} &=& \frac{1}{32}\langle |\tilde{h}_1|^2\rangle\left|\rho_1e^{i \chi_1} {+} \rho_2e^{i(\chi_2+\pi/4)}\right|^2\ell^2B_\ell^2C_{\ell}^{TT}, \label{eq:psdp}\\ \nonumber
&+&  \frac{1}{32}\langle |\tilde{h}_3|^2\rangle\left|\rho_1e^{-i \chi_1} {+} \rho_2e^{-i (\chi_2-3\pi/4)}\right|^2\ell^2B_\ell^2C_{\ell}^{TT}\\
\Delta \ft{C}_\ell^{BB{\rm e}} &=&  \frac{4\pi}{4(2\ell+1)}\left|\Im\left[\delta b^1_{\ell 2} + \delta b^2_{\ell 2}\right]\right|^2C^{TT}_\ell\label{eq:psde}\\ \nonumber
&&+ \frac{4\pi}{8(2\ell+1)}\langle |\tilde{h}_4|^2\rangle\left|\delta b^1_{\ell 2} - \delta b^2_{\ell 2}\right|^2 C_{\ell}^{TT}.
\ea

Finally we must also consider the beam deconvolution, as the
reconstructed polarisation map will be smoothed with the beam. When
analysing a real experiment the recovered $B$-mode power spectrum will
be deconvolved for the beam. To take this into account we divide
equations~(\ref{eq:psdg})--(\ref{eq:psde}) by $B_\ell^2$ giving us,
\ba
\Delta C_\ell^{BB{\rm g}} &=& \frac{1}{8}\langle |\tilde{h}_2|^2\rangle\left|\delta g_1 + i\delta g_2\right|^2 C_{\ell}^{TT},\label{eq:dg}\\
\Delta C_\ell^{BB{\rm p}} &=& \frac{1}{32}\langle |\tilde{h}_1|^2\rangle\left|\rho_1e^{i \chi_1} + \rho_2e^{i(\chi_2+\pi/4)}\right|^2\ell^2C_{\ell}^{TT}, \label{eq:dp}\\ \nonumber
&+& \frac{1}{32}\langle |\tilde{h}_3|^2\rangle\left|\rho_1e^{-i \chi_1} + \rho_2e^{-i (\chi_2-3\pi/4)}\right|^2\ell^2C_{\ell}^{TT}\\
\Delta C_\ell^{BB{\rm e}} &=&  \frac{1}{4}\left|\Im\left[\frac{\delta b^1_{\ell 2} + \delta b^2_{\ell 2}}{b_{\ell 0}}\right]\right|^2C^{TT}_\ell\label{eq:de}\\ \nonumber
&&+ \frac{1}{8}\langle |\tilde{h}_4|^2\rangle\left|\frac{\delta b^1_{\ell 2} - \delta b^2_{\ell 2}}{b_{\ell 0}}\right|^2 C_{\ell}^{TT},
\ea
where the various terms are described in Table~\ref{tab:terms}. We can
immediately see the effect that a good scan strategy can have on
mitigating systematic effects. By providing us with a range of
instrument orientations, $|\tilde{h}_n|$, where $n\ne 0$, will be
lowered for each pixel. This in turn reduces the impact of the systematics on the
recovered power spectrum.

We note that a differential ellipticity of the beams within a
detector pair can couple temperature to polarisation with a spin-2
systematic effect (see equation~\ref{eq:s_of_psi}). This means that no
scan strategy can mitigate the effects of this systematic error. It
has been shown
previously~\citep{2007MNRAS.376.1767O,2008PhRvD..77h3003S,2014MNRAS.442.1963W}
that if the orientation of the ellipticity is in the same direction or
perpendicular to the polarisation sensitivity then this systematic
effect will only couple temperature fluctuations to $E$-mode
polarisation. Any misalignment however will couple to $B$-modes. This
can be seen in the first term of equation~\eqref{eq:de} as the
imaginary part of $\delta b_{\ell 2}$ coupling temperature power to
$B$-mode polarisation. The second term of equation~\eqref{eq:de} does
allow coupling between temperature fluctuations and $B$-mode
polarisation regardless of the orientation of the ellipticity as long
as the two detector pairs have different differential
ellipticity. However, this effect can be mitigated by the scan
strategy, as a result of the $\langle|\tilde{h}_4|^2\rangle$ factor.

\input{sections/tables/parameters}

%\ba
%G_i(\Omega) &=& \frac{1}{2}\delta g_i T(\Omega)\\
%M_i &=& \frac{1}{2}p_i e^{i\chi_i}\nabla T(\Omega)\\
%E_i &=& \sqrt{\frac{4\pi}{2\ell +1}}b_{\ell 2} \sum_{\ell m} Y_{\ell m}(\Omega) a_{\ell m}^T
%\ea

%% file: sections/fig/dif_pointing_schematic.tex
% !TEX TS-program = compile

\begin{figure}
\begin{center}
\begin{tabular}{c}
~\\
~\\
\includegraphics[width=0.8\linewidth, trim=14cm 9cm 9cm 8cm, clip=true]{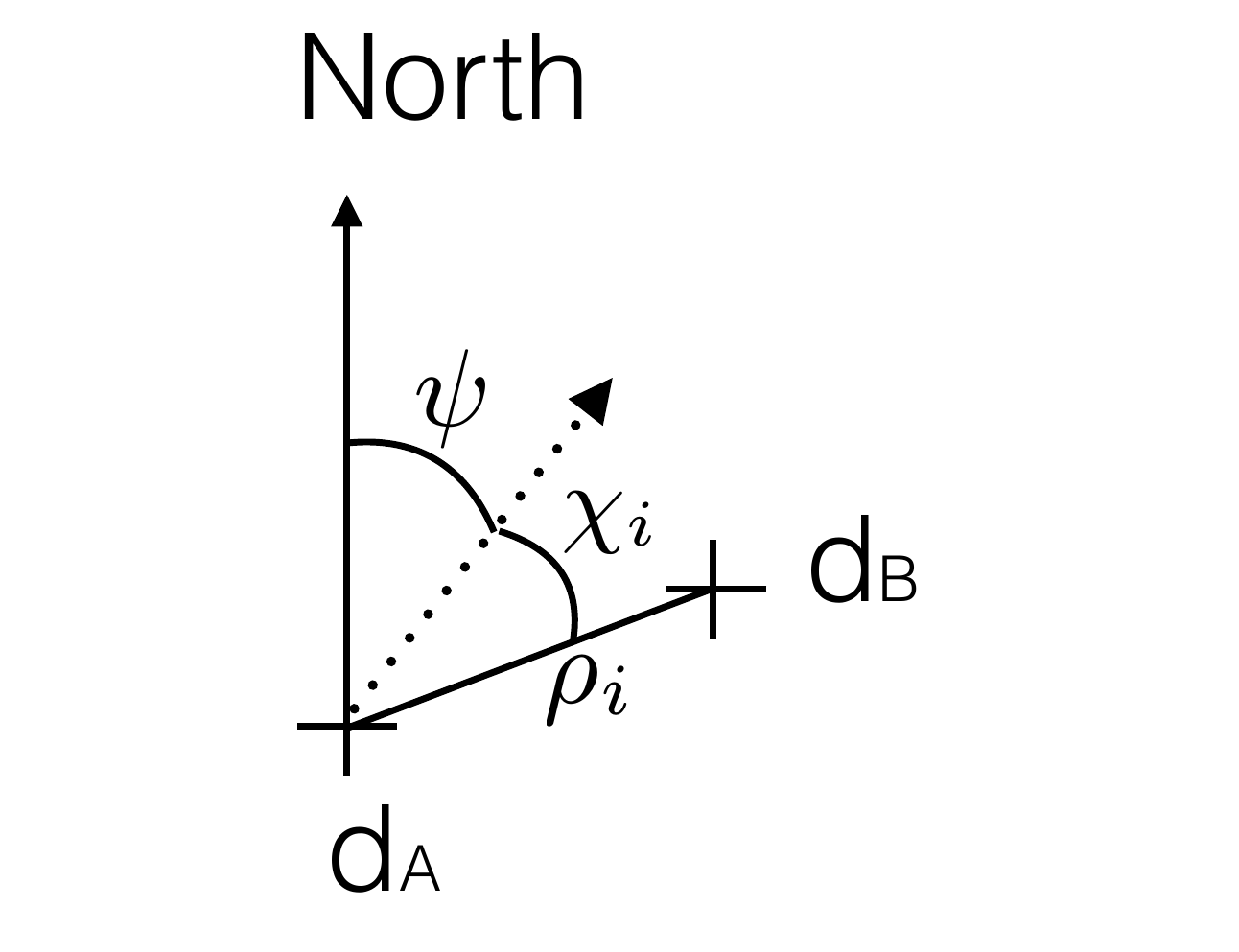}\\
\end{tabular}
\caption{{\color{black}Schematic diagram showing the parameters used to
    define the differential pointing of a detector pair. The detector
    beams are centred at an angle $\rho_i$ with respect to each other, in a direction which is an angle $\chi_i$ with respect to the scan direction, which in turn is an angle $\psi$ from North.}}
\label{fig:dif_pointing_schem}
\end{center}
\end{figure}

%% file: sections/tables/parameters.tex
% !TEX TS-program = compile

\begin{table*}
  \centering
\begin{tabular}{|c|p{6cm}|p{6cm}|}%\label{table:terms}
  \hline
{\bf  Symbol} &{\bf Description} &{\bf Value set to in relevant simulation}\\\hline
$\psi$ & The orientation of the scan direction with respect to North &
Varies with scan strategy, position and time\\ \hline
$\tilde{h}_n$ &The average of the complex exponential of the orientations for a pixel, $\langle e^{in\psi} \rangle_{\rm hits}$ & Varies with scan and pixel\\  \hline
FWHM & the full width at half the maximum of the beam & 7 arcmin for all the simulations \\ \hline
$\delta g_i$ & The differential gain between the two detectors in pair $i$ & 0.01 for both detector pairs\\ \hline
$\rho_i$ & The angle between the two beam centres in pair $i$ & 0.1 arcmin for both (1.5\% of the FWHM)\\ \hline
$\chi_i$ & The orientation of the second beam from the first in a detector pair $i$ relative to the direction of the scan  & 0 and $\pi/4$\\ \hline
$b_{\ell m}$ & The spherical harmonic decomposition of the temperature
beam& That of an elliptical Gaussian --- see equation \eqref{eq:gaus_beam} \\ \hline
$\delta b^i_{\ell m}$ & The spherical harmonic decomposition of the
difference of the temperature beams of pair $i$ & That of an
elliptical Gaussian --- see equation \eqref{eq:gaus_beam} \\ \hline
$q$ & Ellipticity parameter for the elliptical Gaussian beam. Note
that $q=1$ is axisymmetric (see equation~\ref{eq:gaus_beam}). $q$ is
also the ratio of the major and minor axes of the ellipse. & 1.05 and 1\\ \hline
\end{tabular}
  \caption{Description of the variables used in the analysis (see
    Sections \ref{sec:ana} \& \ref{sec:sims} in the main text) and the values adopted for the simulations.}
  \label{tab:terms}
\end{table*}

%% file: sections/full_sims.tex
% !TEX TS-program = compile

\section{Temperature leakage Simulations}\label{sec:sims}

Equations~\eqref{eq:dg}--\eqref{eq:de} provide a fast method to
predict the contamination in the recovered $B$-mode power spectrum for
a given set of systematics and a specified scan strategy. However in
deriving these equations a number of approximations were made. In
particular, our derivation assumes that the systematics contribute to
the polarisation leakage only to first order in the size of the
systematic. We also assumed that the effect of the scan strategy in
mitigating the systematic can be modelled as a simple damping of the power
spectrum and therefore does not couple multiple temperature
$\ell$-modes on to a single $B$-mode scale. In this section, we perform
full timeline simulations to demonstrate that
equations~\eqref{eq:dg}--\eqref{eq:de} nevertheless provide an
accurate prediction for the effect of the systematics on
the $B$-mode power spectrum for a selection of scan
strategies.       
 
We create TOD simulations for two detector pairs, one ``instrument-$Q$"
detector and one ``instrument-$U$" detector with different
systematics. We consider the scan strategies adopted for the {\it
  Planck} \citep{2011A&A...536A...1P} and WMAP
\citep{2003ApJS..148....1B} satellites. In addition, we consider the
scan strategy suggested for the proposed EPIC satellite
\citep{2009arXiv0906.1188B}. The parameters used to
model these scan strategies are listed in
Table~\ref{tab:scan_params}. The input signal for our simulations
consists of a fiducial set of CMB power spectra with 
parameters: $\Omega_{\rm b} {=} 0.04612$, $\Omega_{\rm c} = 0.233$,
$\Omega_\Lambda {=} 0.721$, $H_0 {=} 70$ kms$^{-1}$Mpc$^{-1}$,
$\tau = 0.09$ and $n_{\rm s} {=} 0.96$.
An input tensor-to-scalar
ratio of $r=0.1$ was used for all simulations and a lensing $B$-mode
contribution was included. The input CMB power spectra were created
with CAMB \citep{2011ascl.soft02026L} and we used the {\sevensize
  HEALPix}\footnote{See http://healpix.sourceforge.net} package
\citep{2005ApJ...622..759G} to create simulated maps, and to estimate
power spectra. 

In Fig.~\ref{fig:scan_info} we show the hit maps and maps of
$|\tilde{h}_n|$ for $n=\{1,2,3,4\}$ for one year of observations at a
data sampling rate of 500 Hz. These maps have been constructed using
{\sevensize HEALPix} resolution parameter, $N_{\rm side}{=}2048$,
corresponding to a pixel size $\sim2.7$ arcmin. The lower these
$h$-values are, the better the scan strategy will be at mitigating
different systematics. {\color{black}From equations
  \eqref{eq:dg}--\eqref{eq:de} we can see that the average value of
  these maps acts a scaling factor for the leakage from temperature
  power spectrum to $B$-mode polarisation power spectrum.}

%We simulate TODs from two detector pairs that suffer from the
%systematic effects in question. For each scan strategy we simulate 4
%To mimic the data for two pairs of detectors, for each scan strategy,
%we simulate four sets of TOD, each with a different systematic effect. 
%We do this to test the effectiveness of equations
%(\ref{eq:dg},\ref{eq:dp},\ref{eq:de}) to predict the temperature
%leakage into the B-mode power spectrum. 

\input{sections/tables/scan_parameters}

\input{sections/fig/h_numbers}

For each scan strategy, we simulated the effects of each systematic
(differential gain, pointing and ellipticity) individually assuming
the parameters listed in Table~\ref{tab:terms}. For all the
simulations, except for the differential ellipticity simulations, a
Gaussian beam with FWHM of 7 arcmin was used. To assess the
differential ellipticity mitigation an elliptical Gaussian beam was
used,
\ba
B(\theta,\phi) = \frac{1}{2\pi q \sigma^2} e^{-\frac{\theta^2}{2\sigma^2}(\cos^2\phi + q^{-1}\sin^2\phi)}\label{eq:gaus_beam}.
\ea

In order to separately assess the impact of the two contributions to
equation~\eqref{eq:de}, we run two sets of simulations.
The effect described by the first term in equation~\eqref{eq:de} is
simulated by each of the four detectors having a beam described by
equation~\eqref{eq:gaus_beam} with $q=1.05$, but with the beam such
that the ellipticity is rotated by $\pi/4$ with respect to the
polarisation sensitivity direction. A beam of this type has the
property that the second azimuthal mode of the spherical harmonic
decomposition is imaginary. Such a setup will therefore strongly
contribute a systematic of the type corresponding to the first term of
equation~\eqref{eq:de}. As both the ``instrument-$Q$" and ``instrument-$U$"
detectors have the same differential ellipticity in this setup, there
will be no systematic of the type corresponding to the second term in
equation~\eqref{eq:de}.

The second term of equation~\eqref{eq:de} is non-zero when the
differential ellipticities within the two detector pairs are
different. We create this effect by simulating both detectors within
one pair to have a symmetric Gaussian beam of FWHM $=$ 7 arcmin, while
the detectors of the other pair are modelled as having elliptical
beams, aligned with the polarization sensitivity direction, and described by
equation~\eqref{eq:gaus_beam}. Both detectors in this latter pair 
are set up to have $\sigma =3$ and $q = 1.05$ but for one
of these, the beam is rotated by $\pi/2$ in order to create the
required  differential ellipticity.

The results from these simulations are shown in Fig.~\ref{fig:B_rec}
in terms of the $B$-mode power spectrum recovered from $Q$ and $U$
maps which are constructed from the TOD using a simple binned
map-making algorithm. We plot the recovered power spectrum for each of the
systematics considered, and for each of the three scan strategies tested. In this
figure, we have also plotted the theoretical predictions for the
spurious signal from equations~(\ref{eq:dg})--(\ref{eq:de}). We see
that the theoretical predictions are accurate for $\ell \lesssim
1000$ which, for a beam of FHWM = 7 arcmin, is approaching the beam
scale. {\color{black} The purpose of this plot is to show that the analytical predictions
are consistent will the simulations, therefore, justifying their validity.}

\input{sections/fig/b_mode_error}

%% file: sections/tables/scan_parameters.tex
% !TEX TS-program = compile

\begin{table*}
  \centering
\begin{tabular}{|c|c|c|c|c|}%\label{table:terms}
  \hline
{ \bf Scan } &  { \bf Boresight angle} ($\beta$) & {\bf Precession angle }($\alpha$)& {\bf Spin period} $(T_{\rm spin})$&{\bf Precession period} $(T_{\rm prec})$\\ \hline
{\it Planck} & 85$\degree$ & 7.5$\degree$ & 1 min & 6 months\\ \hline
WMAP & 70$\degree$ & 22.5$\degree$ & 129 s & 1 hr \\ \hline
EPIC & 50$\degree$ & 45$\degree$ & 1 min & 3 hrs \\ \hline
\end{tabular}
  \caption{Observational parameters used to generate the scan
    strategies for the simulations described in Section~\ref{sec:sims}.}
  \label{tab:scan_params}
\end{table*}

%% file: sections/fig/h_numbers.tex
% !TEX TS-program = compile

\begin{figure*}
\begin{center}
\begin{tabular}{c c c}
~\\
~\\
EPIC & WMAP & {\it Planck} \\
%&Hit Map&\\
\includegraphics[width=0.33\linewidth, trim=0cm 0cm 0cm 1cm, clip=true]{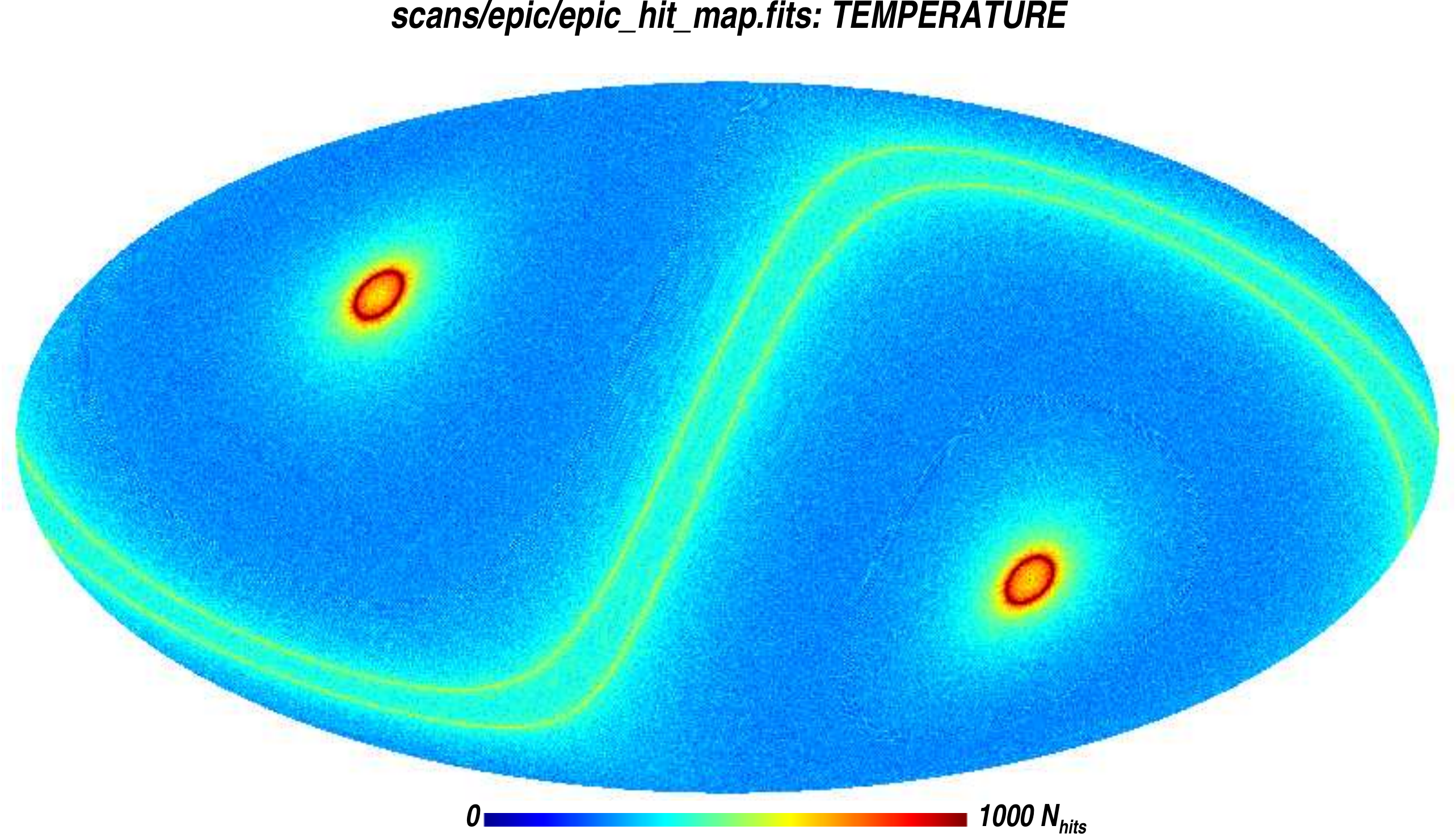}  &
\includegraphics[width=0.33\linewidth, trim=0cm 0cm 0cm 1cm, clip=true]{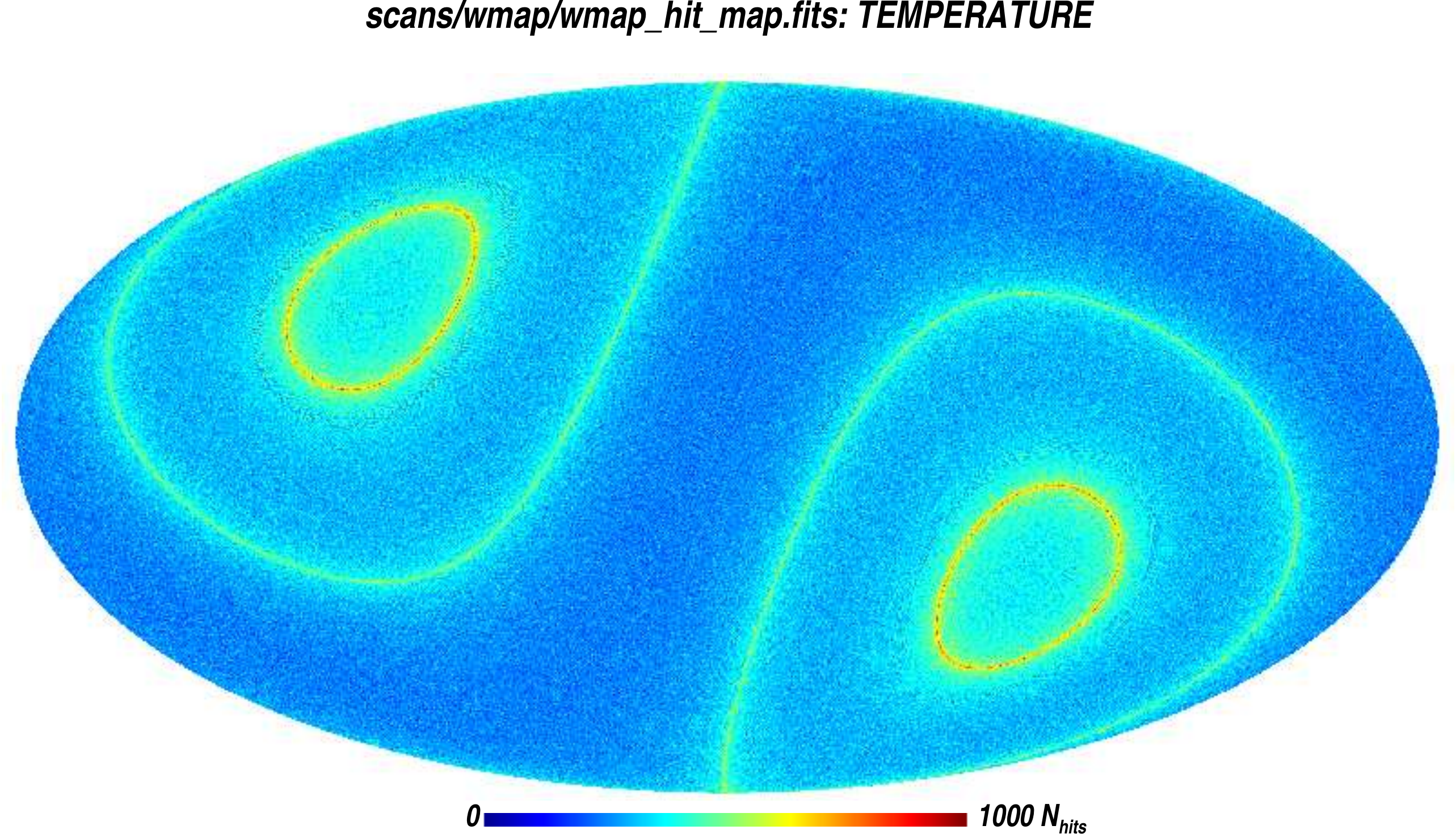}&
\includegraphics[width=0.33\linewidth, trim=0cm 0cm 0cm 1cm, clip=true]{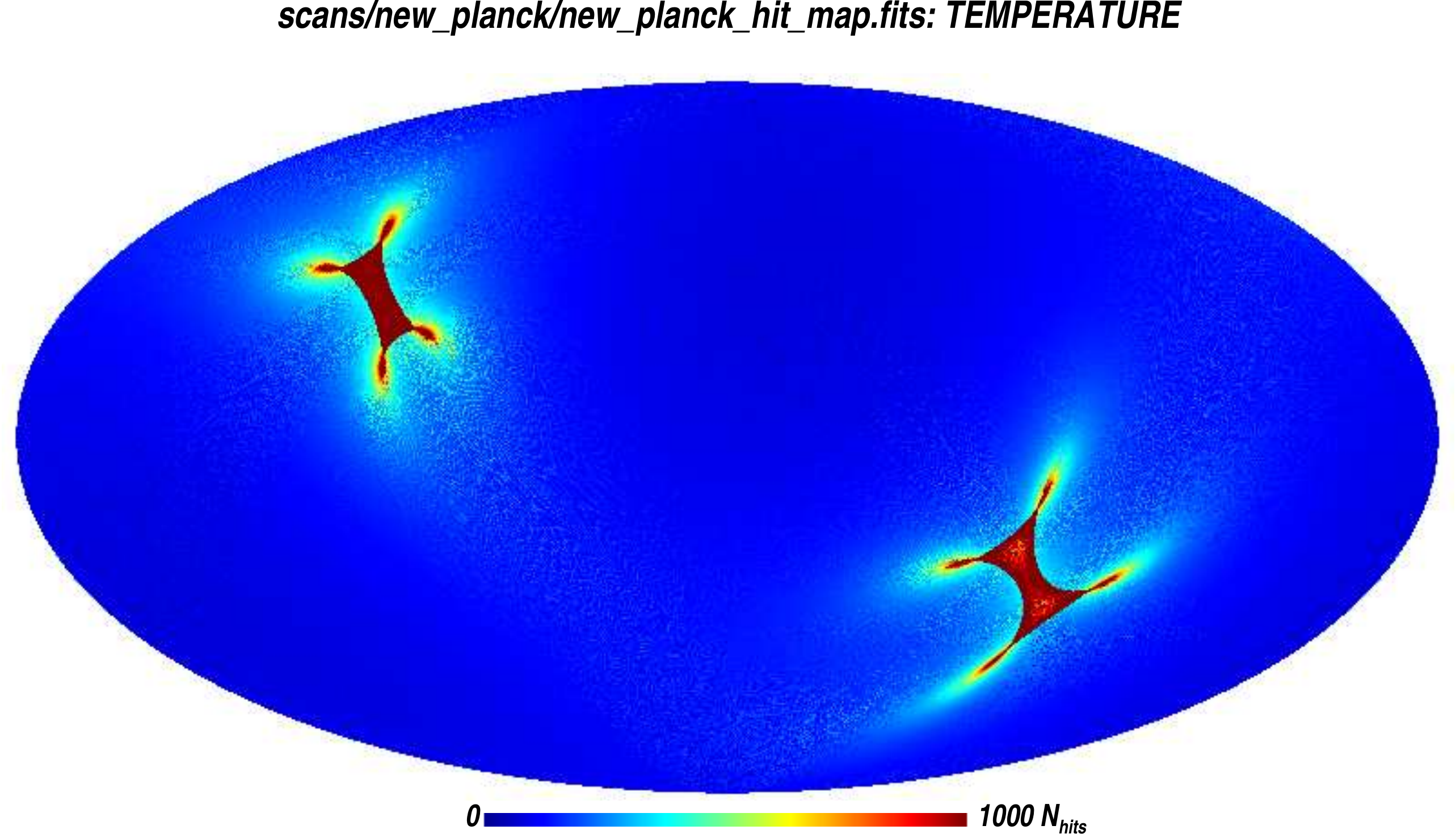}\\
%&$|\tilde{h}_1|^2$&\\
\includegraphics[width=0.33\linewidth, trim=0cm 0cm 0cm 1cm, clip=true]{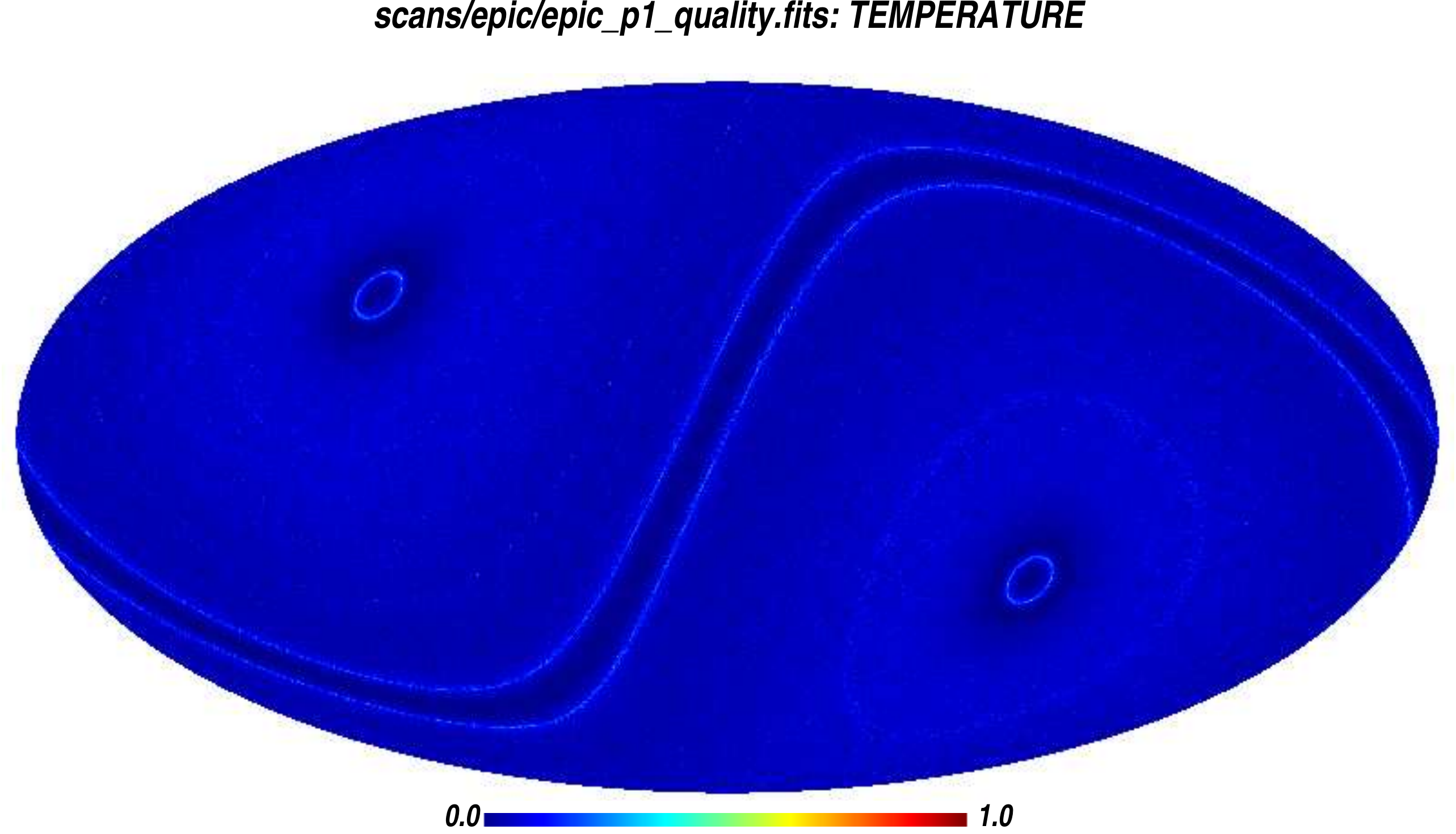}  &
\includegraphics[width=0.33\linewidth, trim=0cm 0cm 0cm 1cm, clip=true]{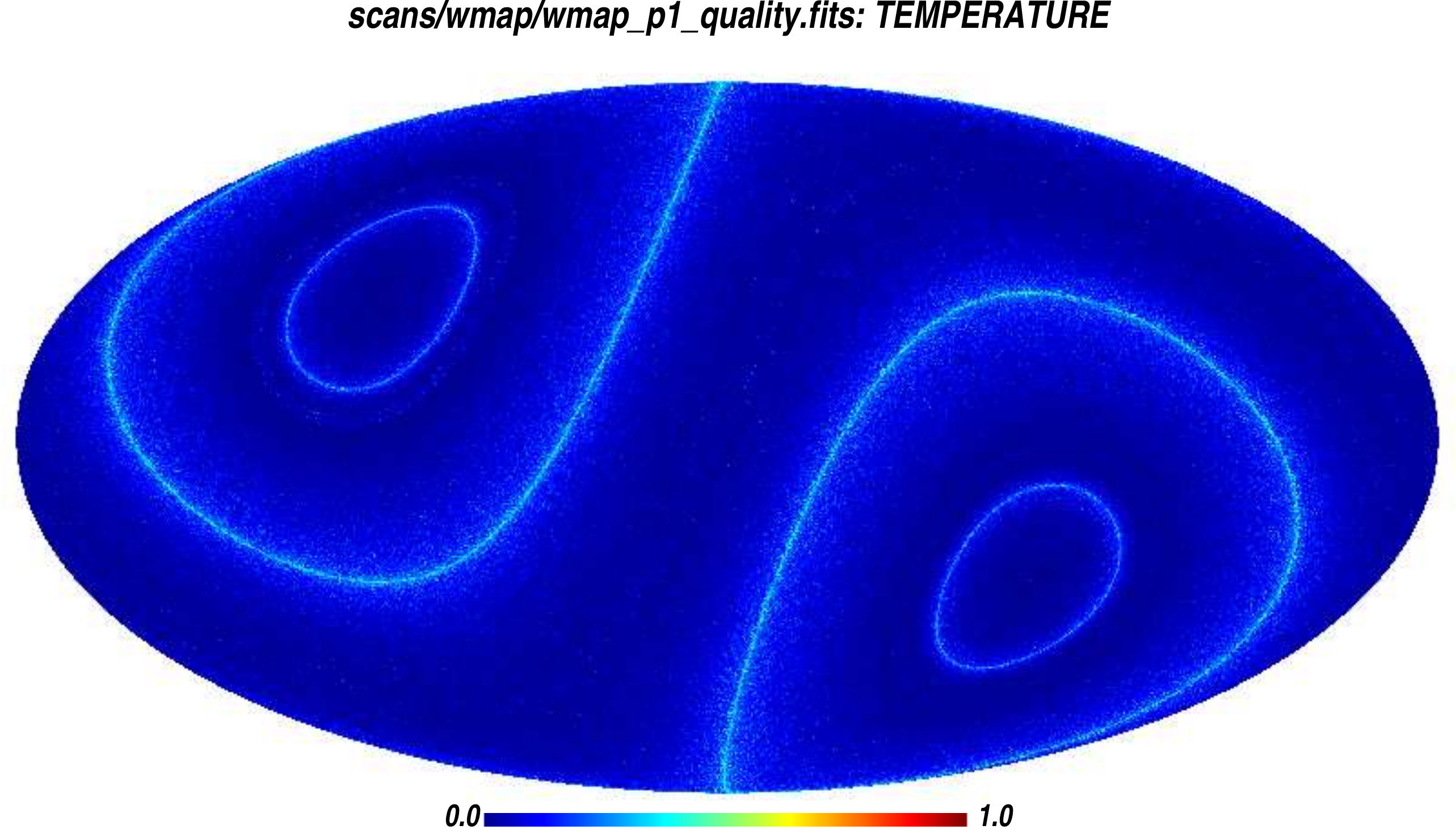}&
\includegraphics[width=0.33\linewidth, trim=0cm 0cm 0cm 1cm, clip=true]{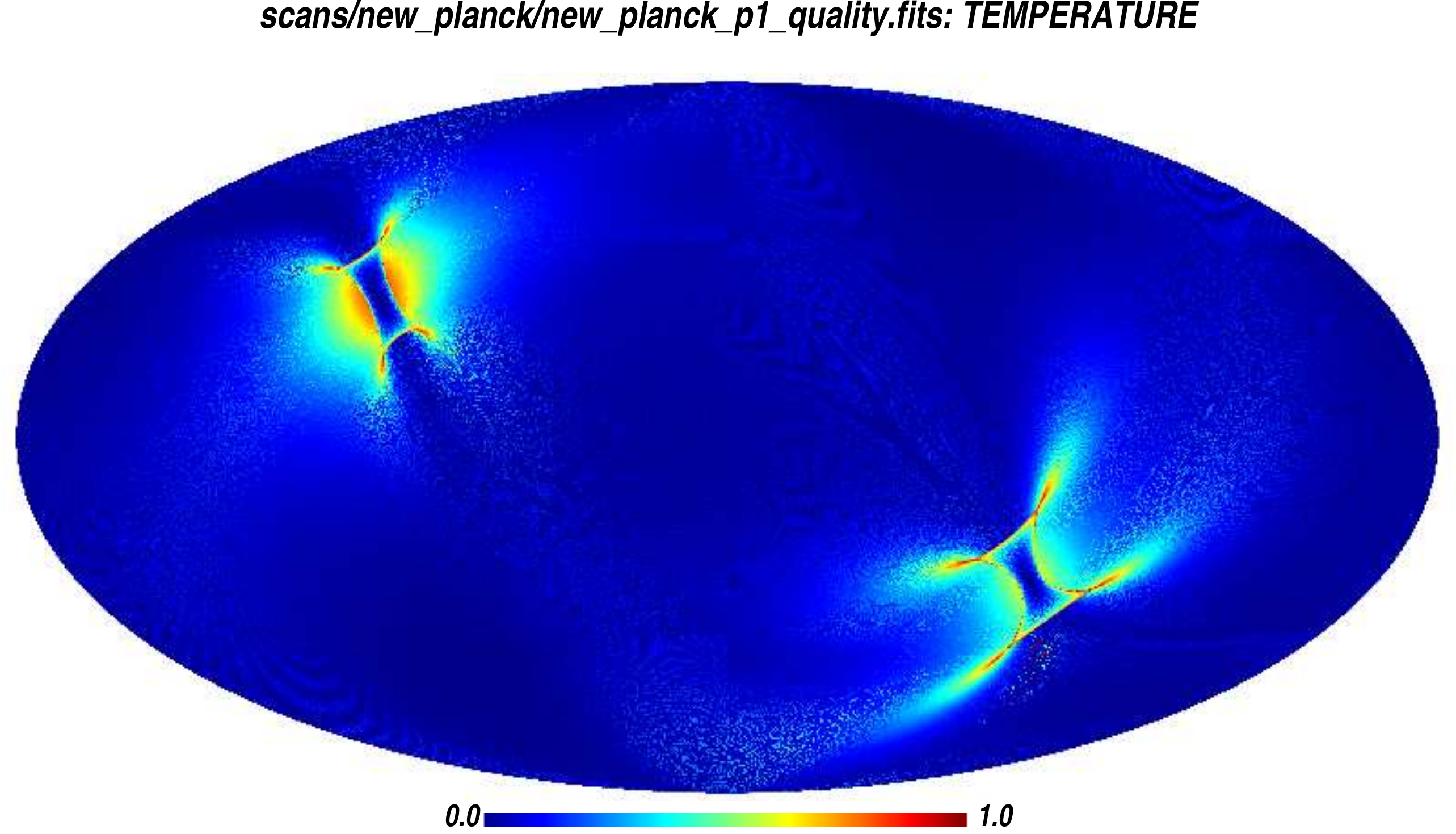}\\
%&$|\tilde{h}_1|^2$&\\
\includegraphics[width=0.33\linewidth, trim=0cm 0cm 0cm 1cm, clip=true]{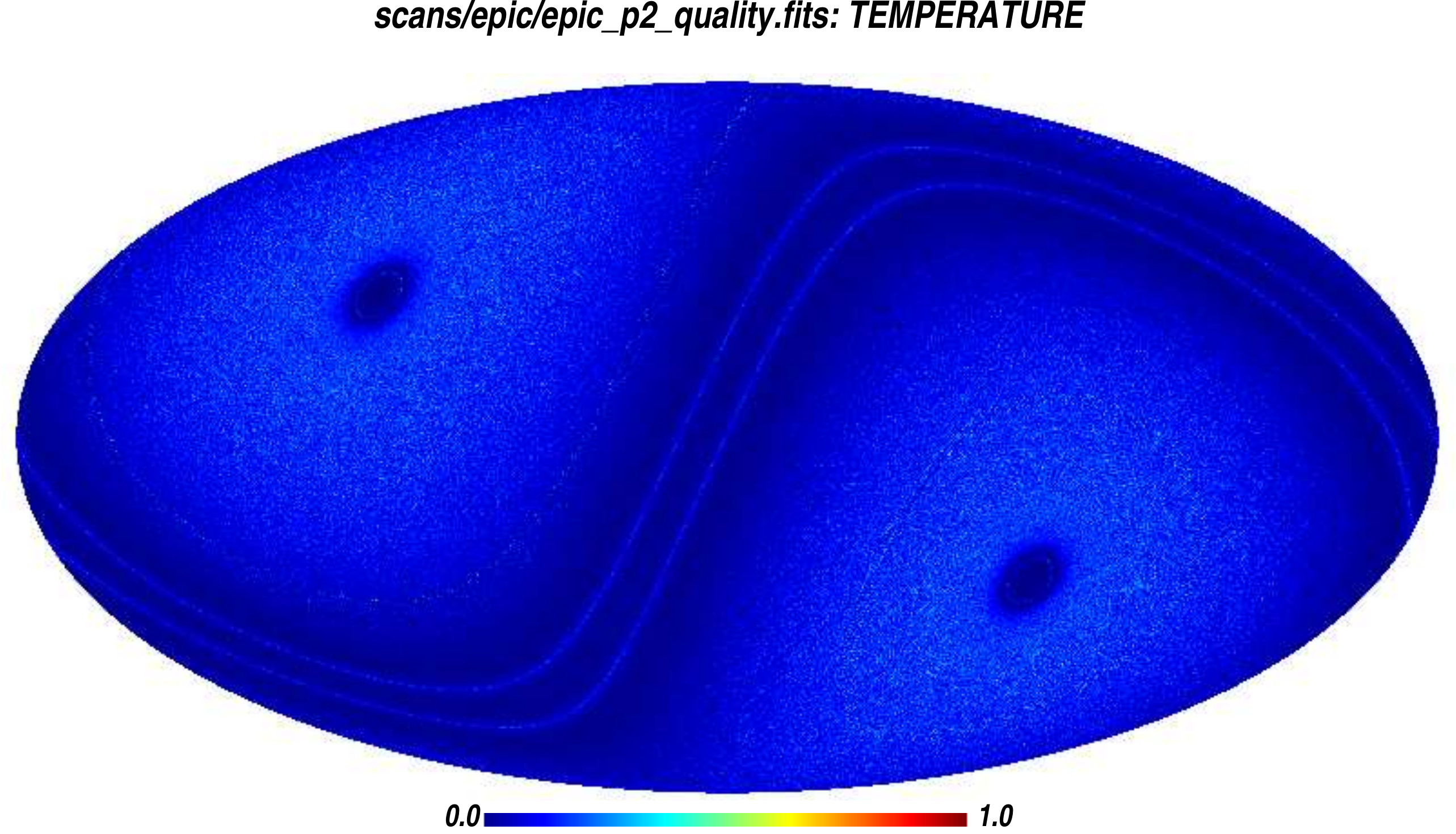}  &
\includegraphics[width=0.33\linewidth, trim=0cm 0cm 0cm 1cm, clip=true]{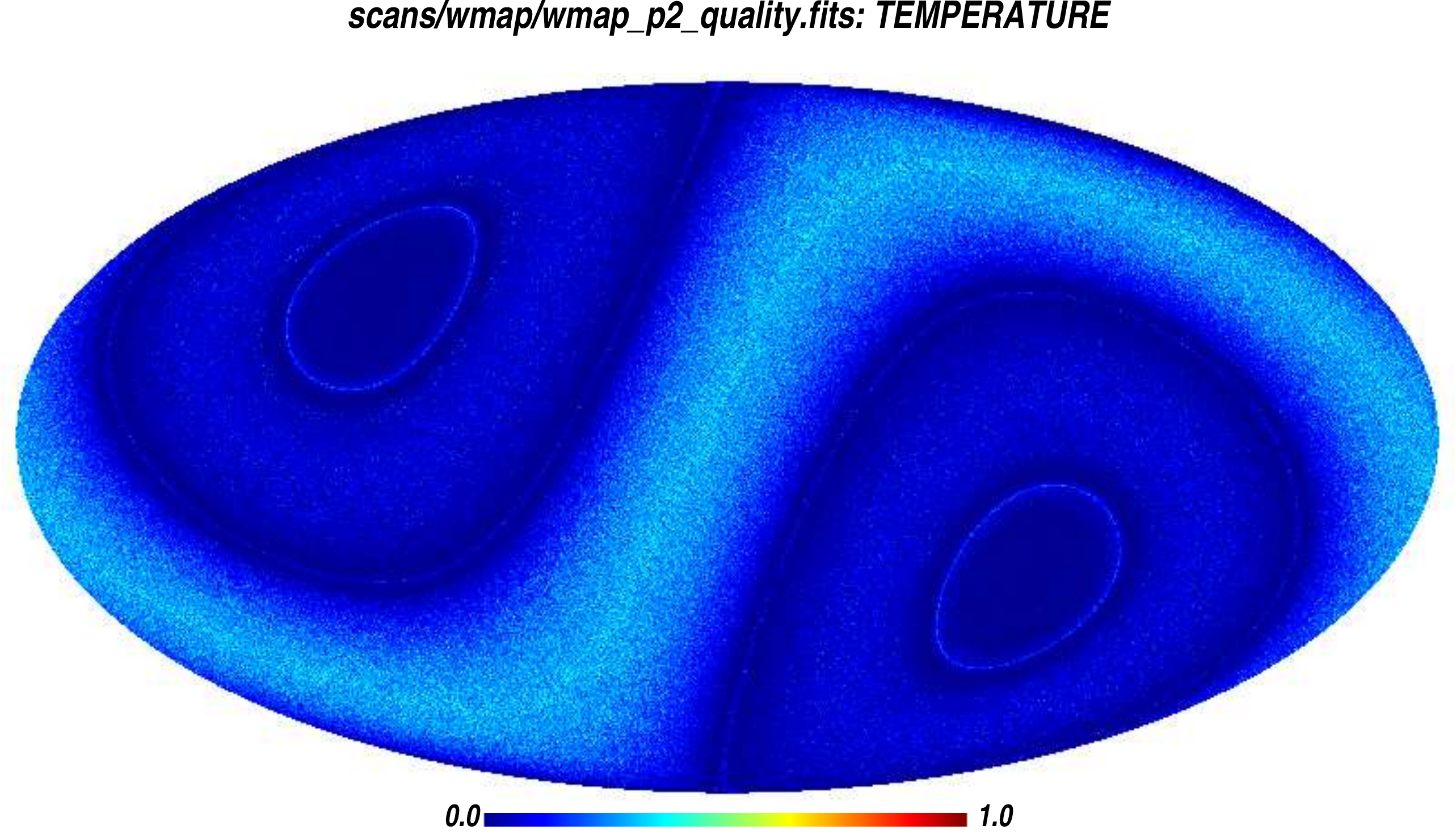}&
\includegraphics[width=0.33\linewidth, trim=0cm 0cm 0cm 1cm, clip=true]{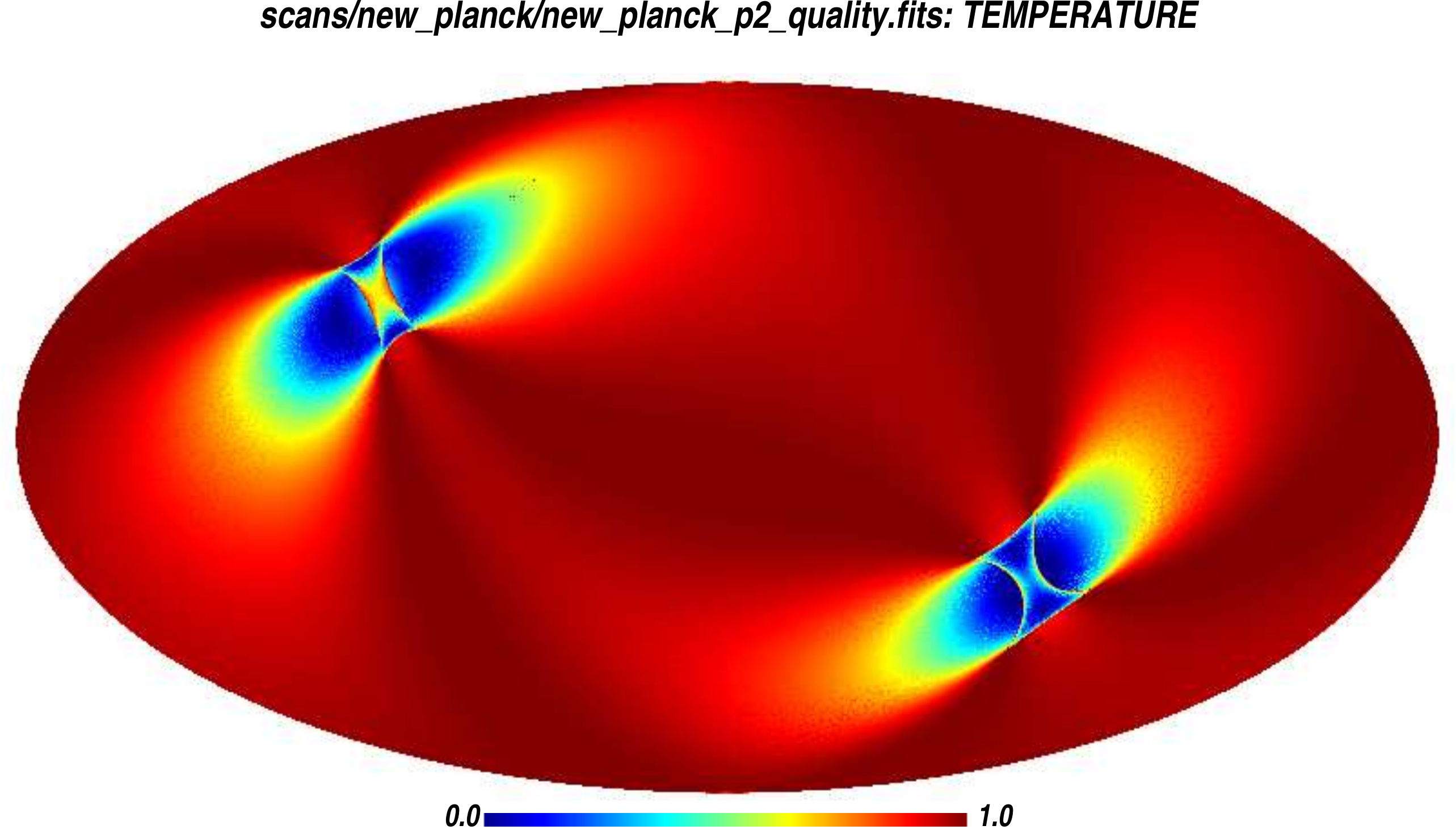}\\
%&$|\tilde{h}_1|^2$&\\
\includegraphics[width=0.33\linewidth, trim=0cm 0cm 0cm 1cm, clip=true]{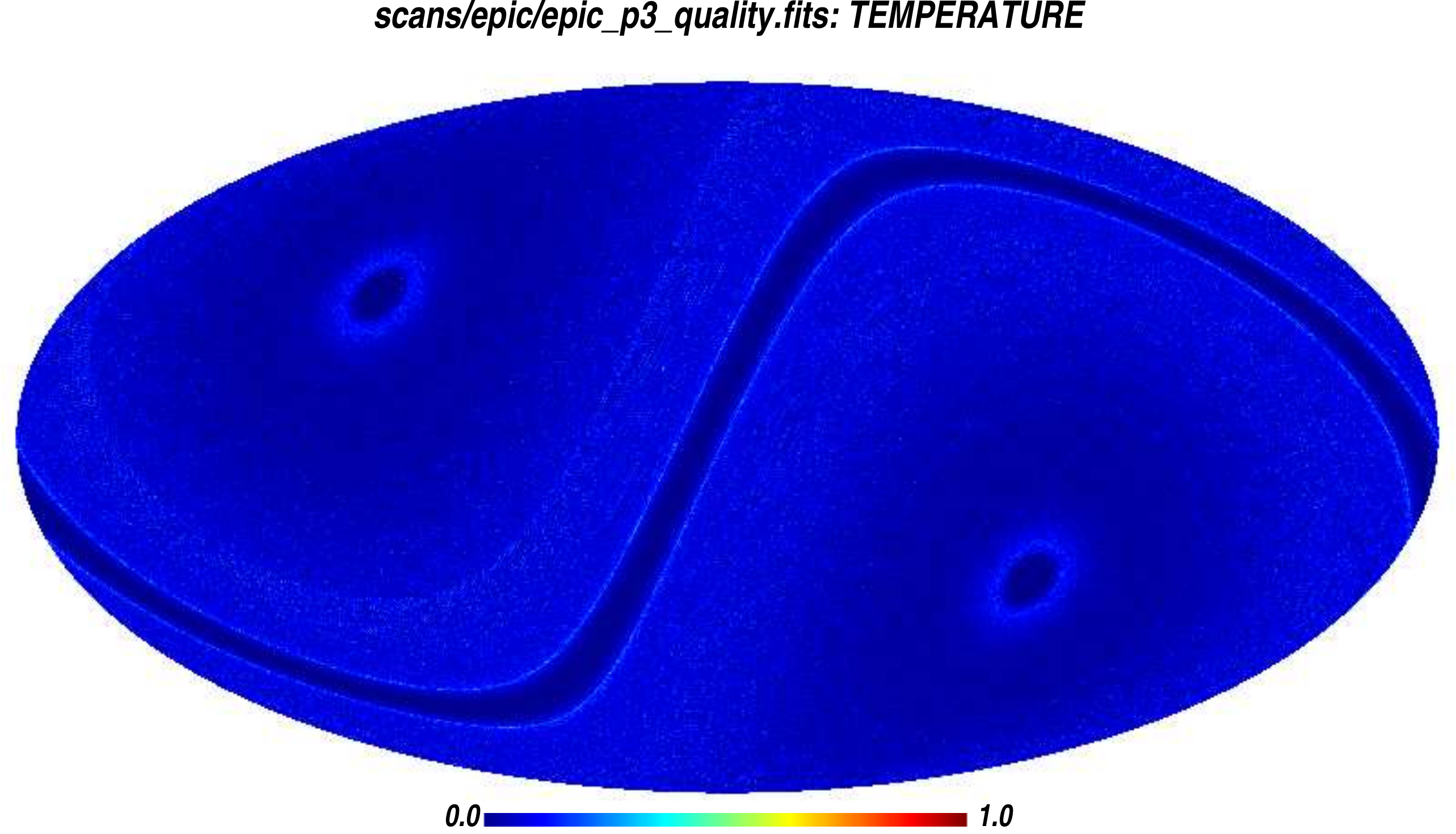}  &
\includegraphics[width=0.33\linewidth, trim=0cm 0cm 0cm 1cm, clip=true]{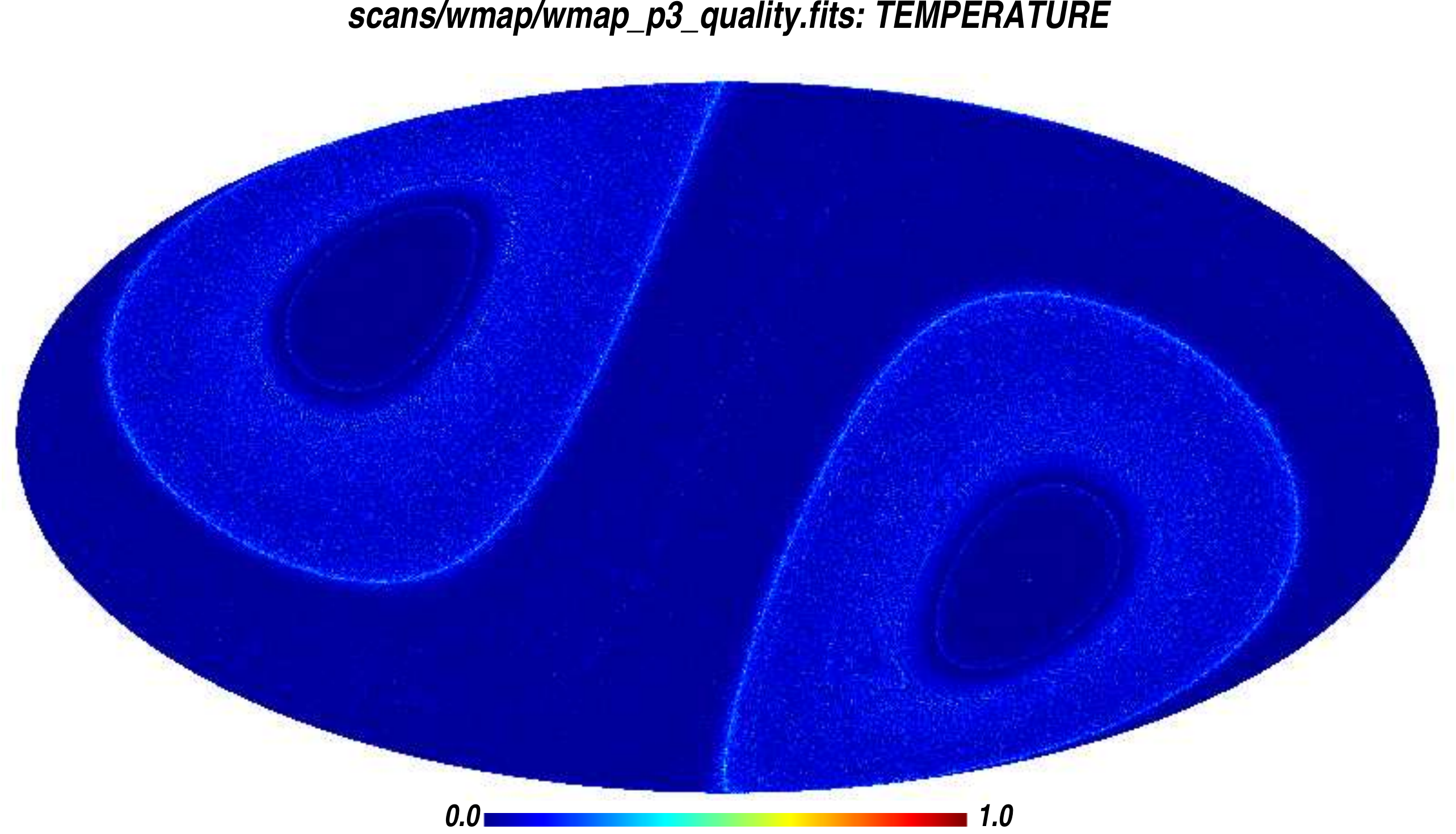}&
\includegraphics[width=0.33\linewidth, trim=0cm 0cm 0cm 1cm, clip=true]{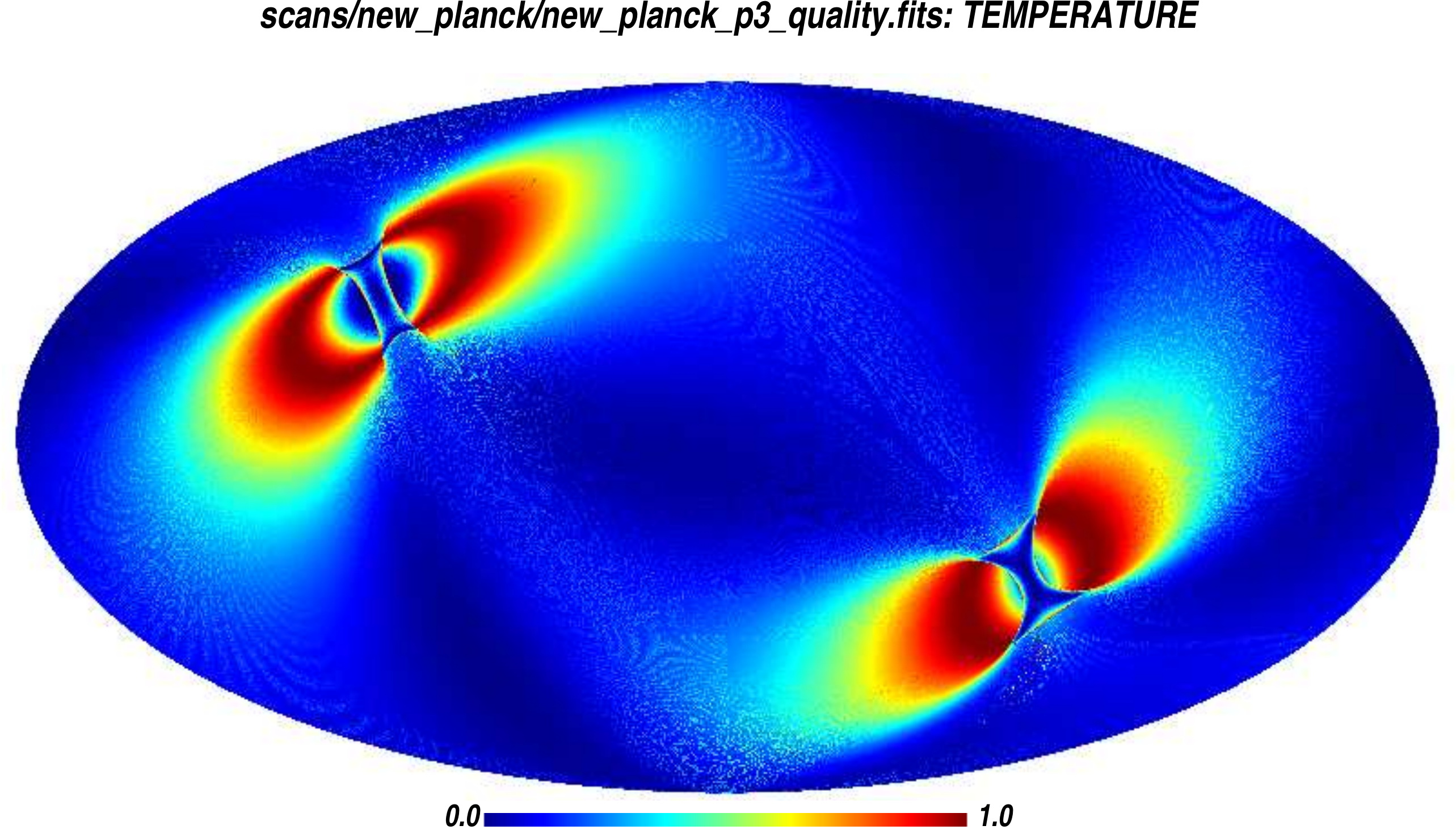}\\
%&$|\tilde{h}_1|^2$&\\
\includegraphics[width=0.33\linewidth, trim=0cm 0cm 0cm 1cm, clip=true]{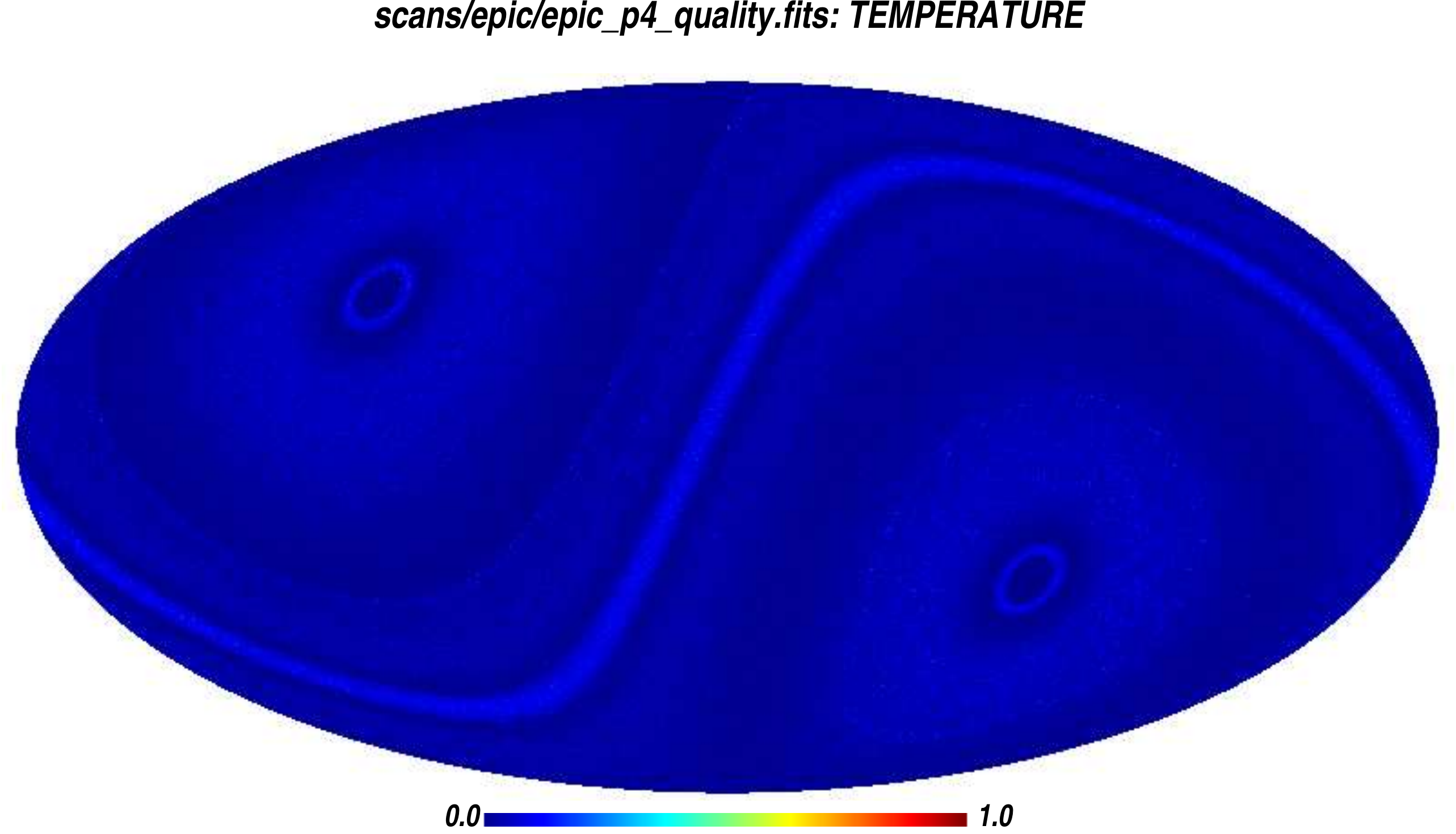}  &
\includegraphics[width=0.33\linewidth, trim=0cm 0cm 0cm 1cm, clip=true]{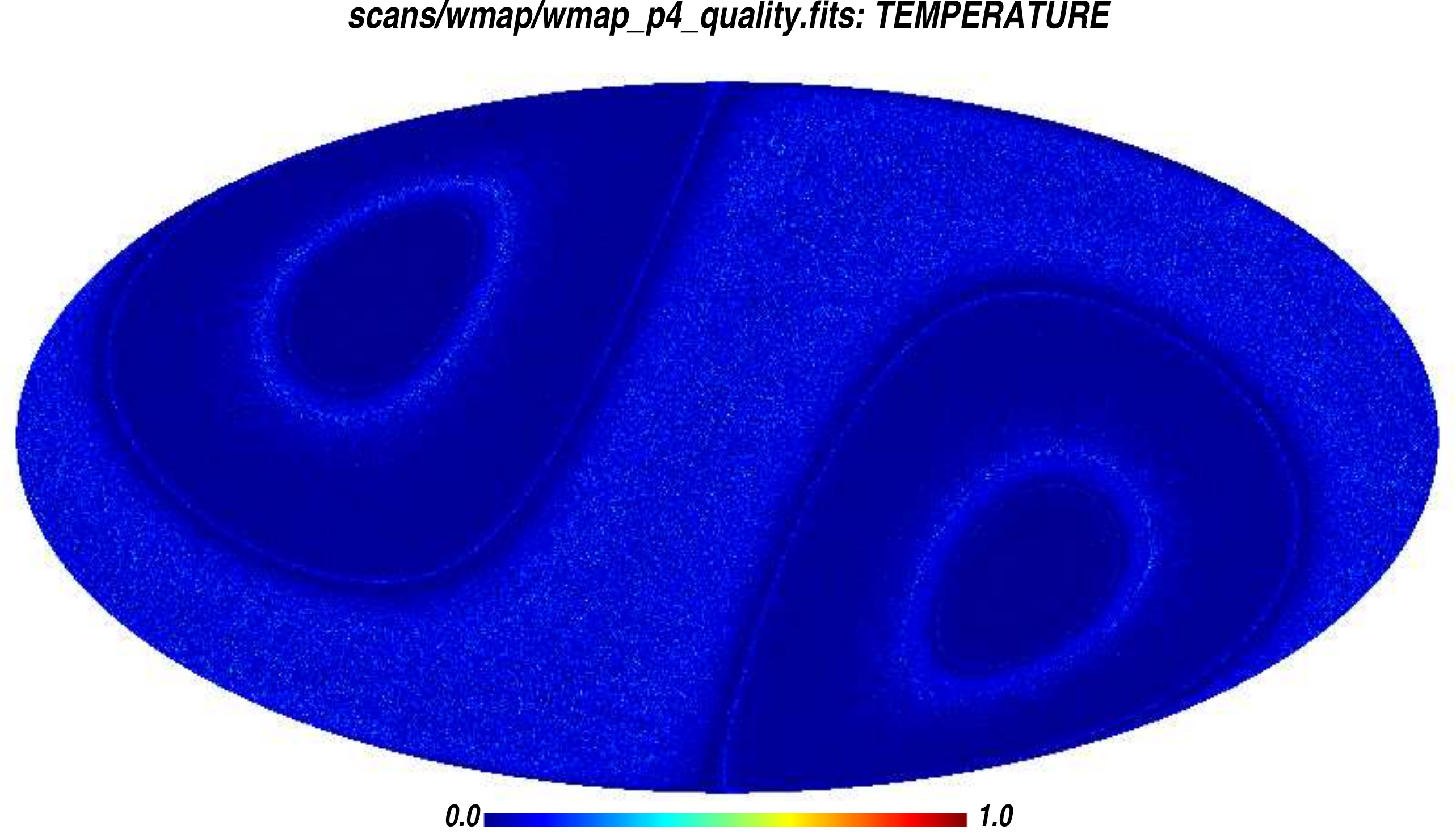}&
\includegraphics[width=0.33\linewidth, trim=0cm 0cm 0cm 1cm, clip=true]{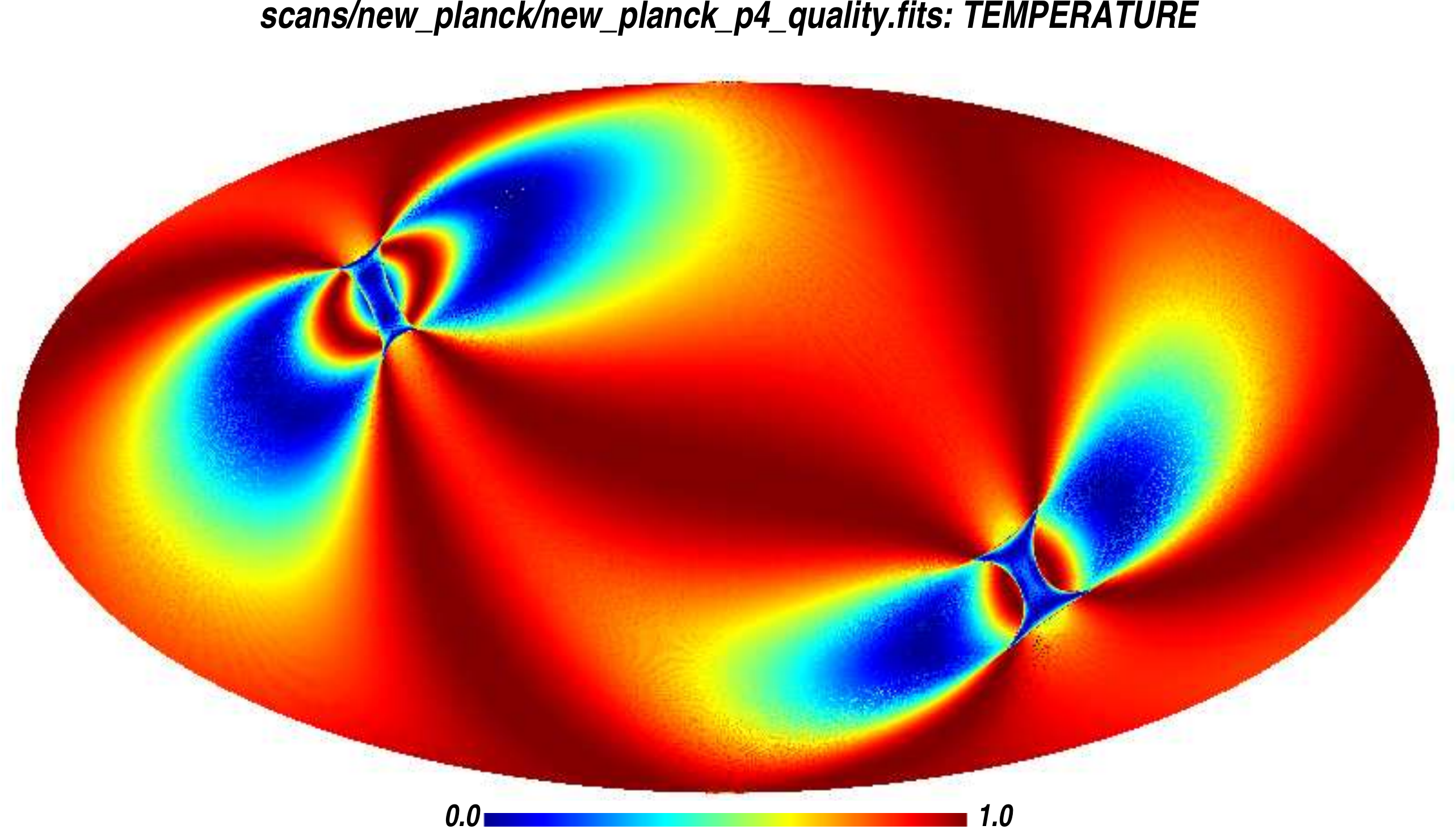}\\

\end{tabular}
\caption{{\it Top row:} The hit maps for the EPIC ({\it left}), WMAP
  ({\it centre}) and {\it Planck} ({\it right}) scan strategies used in the
  simulations at $N_{\rm side}{=}2048$. {\it Lower rows:} Maps of
  $|\tilde{h}_n|^2$, defined in Table~\ref{tab:terms}, for
  $n=\{1,2,3,4\}$, for the different scan strategies. The lower the
  value of $|\tilde{h}_n|^2$ the smaller the temperature to
  polarisation leakage is. Note that in our formalism $\langle
  |\ft{h}_2|^2\rangle = \left\langle \left(\frac{1}{N_{\rm
      hits}}\sum_j\cos(2\psi_j)\right)^2\right\rangle + \left\langle
  \left(\frac{1}{N_{\rm
      hits}}\sum_j\sin(2\psi_j)\right)^2\right\rangle$. This is the
  same quantity as that plotted in Figure~3.5 of \citet{2009arXiv0906.1188B}.}
\label{fig:scan_info}
\end{center}
\end{figure*}

%% file: sections/fig/b_mode_error.tex
% !TEX TS-program = compile

\begin{figure*}
\begin{center}
\begin{tabular}{c c}
\includegraphics[width=0.45\linewidth, trim=0cm 0cm 0cm 0cm, clip=true]{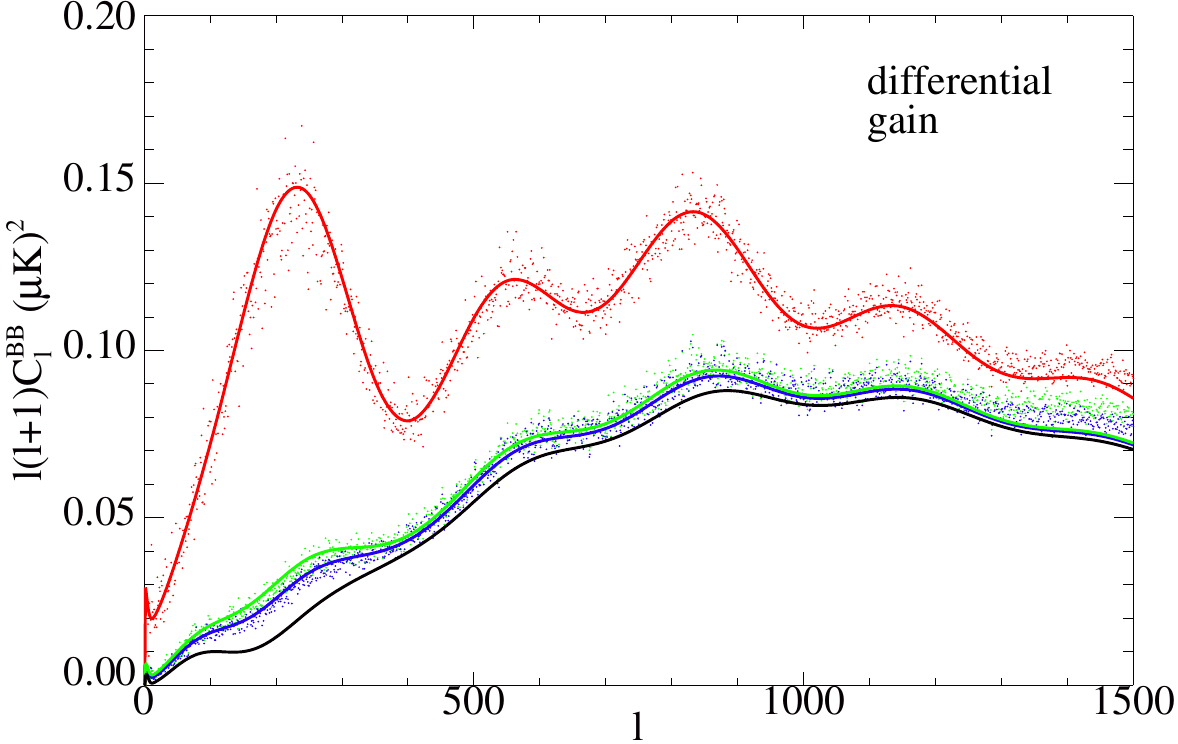}  &
\includegraphics[width=0.45\linewidth, trim=0cm 0cm 0cm 0cm, clip=true]{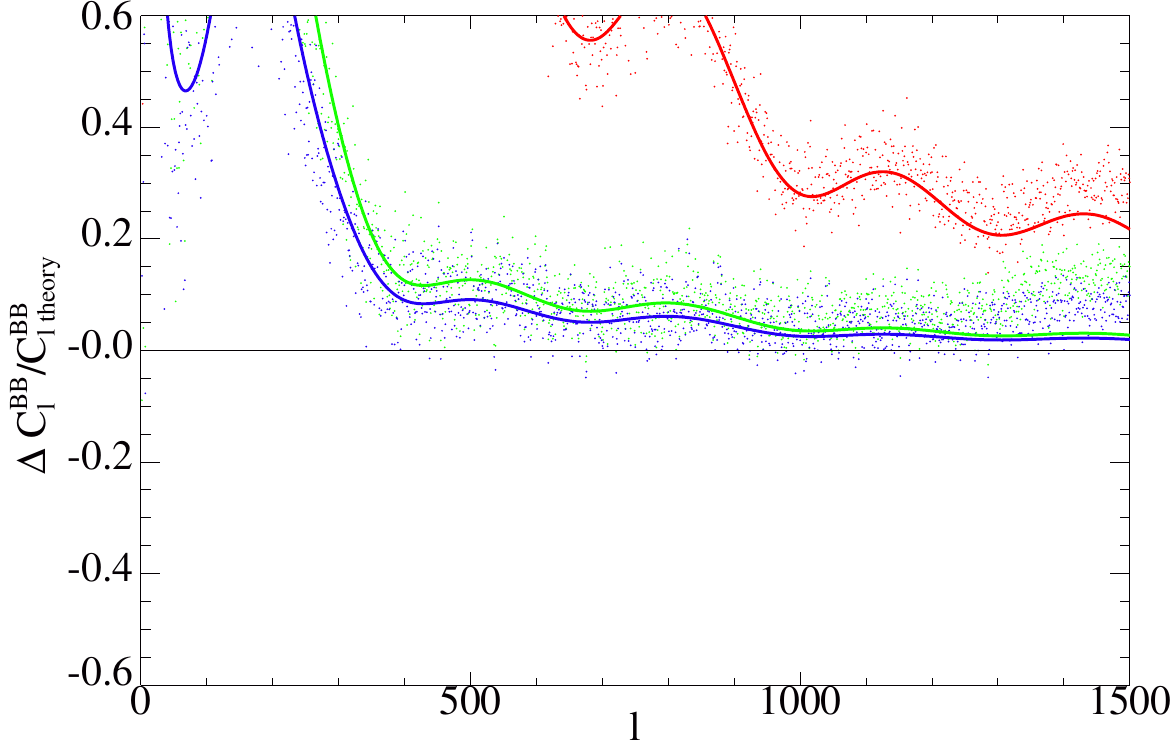}  \\
\includegraphics[width=0.45\linewidth, trim=0cm 0cm 0cm 0cm, clip=true]{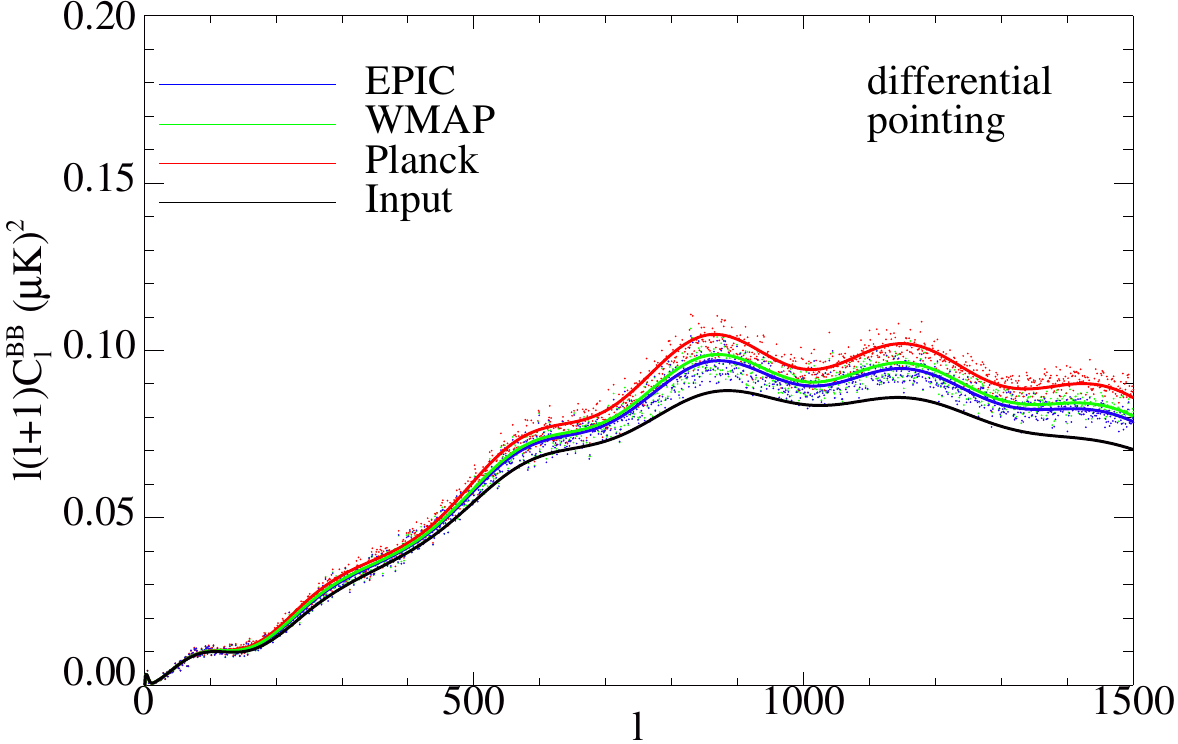}  &
\includegraphics[width=0.45\linewidth, trim=0cm 0cm 0cm 0cm, clip=true]{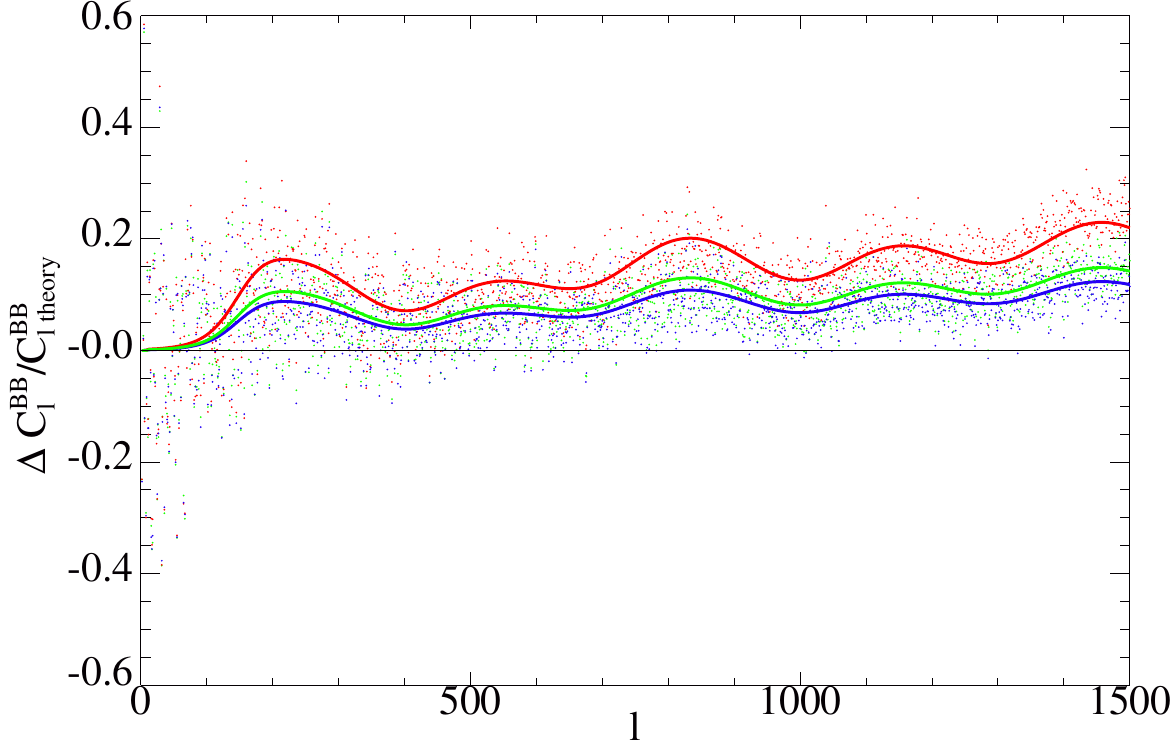}  \\
\includegraphics[width=0.45\linewidth, trim=0cm 0cm 0cm 0cm, clip=true]{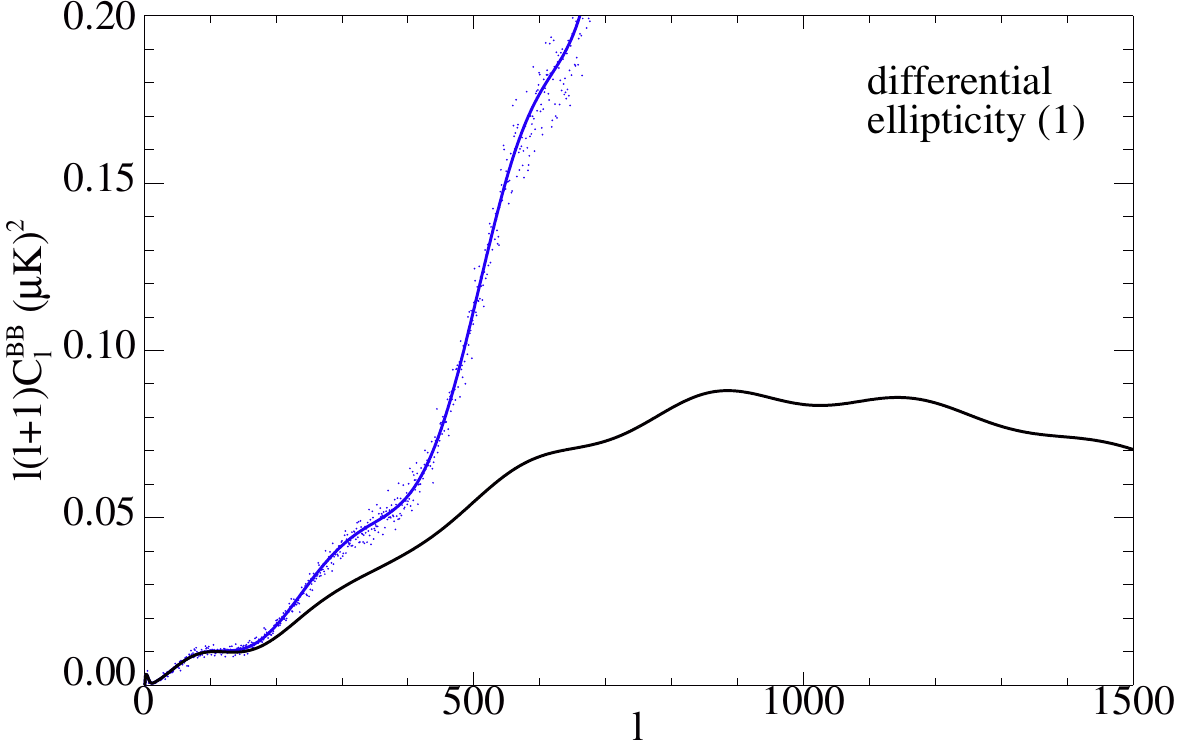}  &
\includegraphics[width=0.45\linewidth, trim=0cm 0cm 0cm 0cm, clip=true]{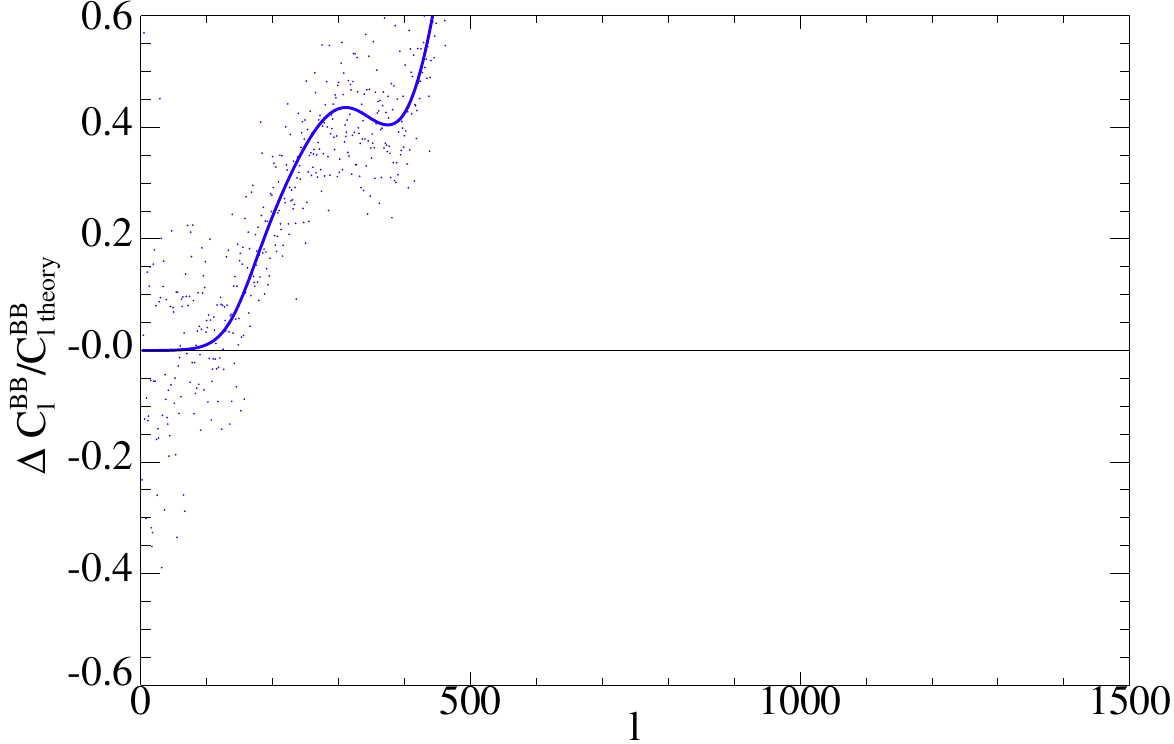}  \\
\includegraphics[width=0.45\linewidth, trim=0cm 0cm 0cm 0cm, clip=true]{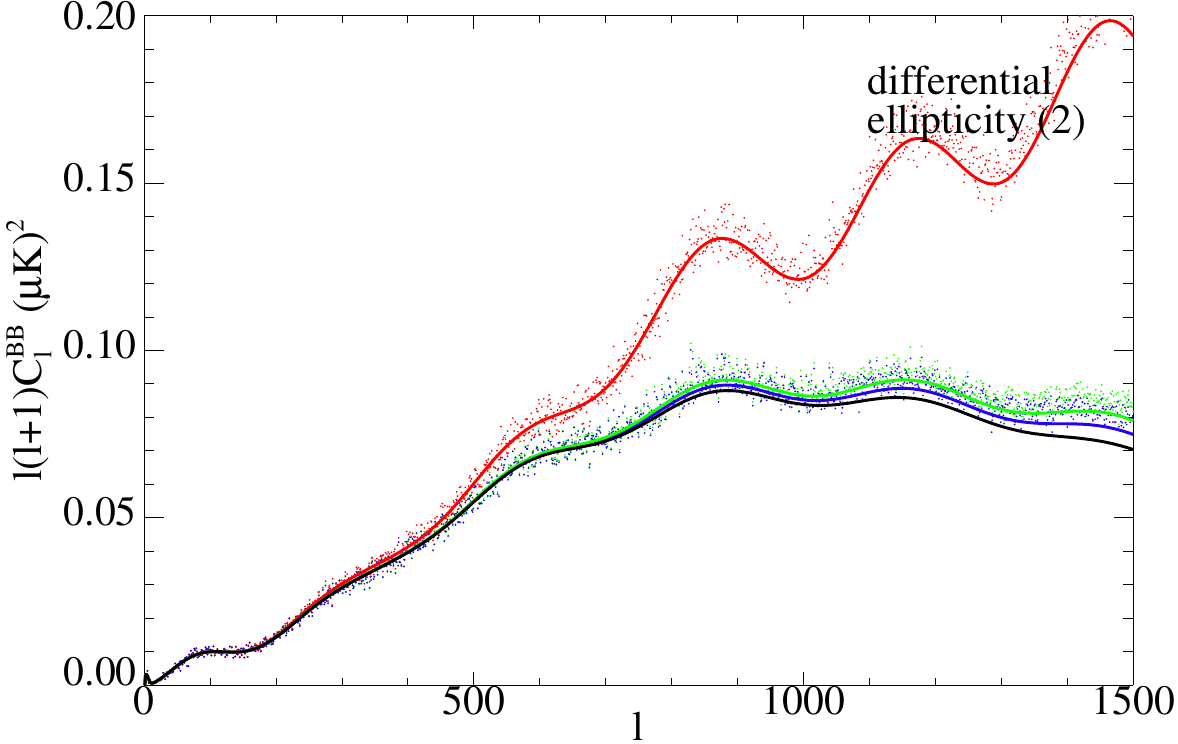}  &
\includegraphics[width=0.45\linewidth, trim=0cm 0cm 0cm 0cm, clip=true]{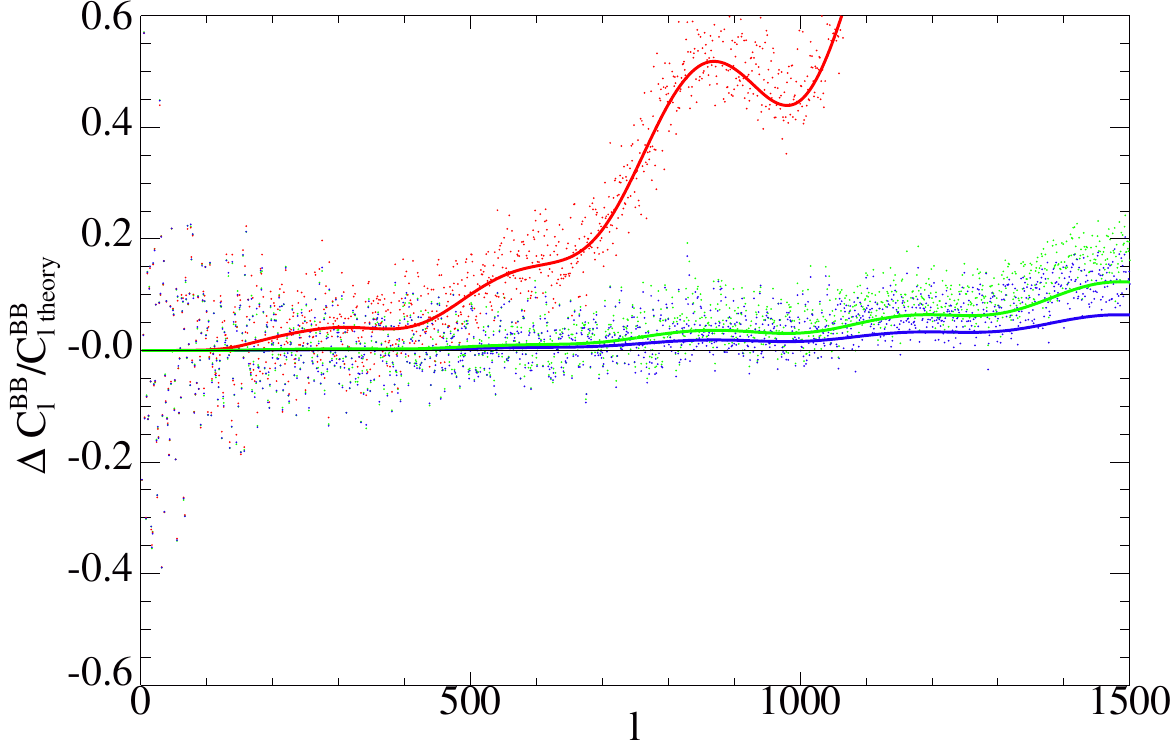}  \\
\end{tabular}

\caption{The recovered $B$-mode power spectrum from simulations
  including systematic effects when using the {\it Planck}, WMAP and
  EPIC scan strategies (shown in red, green and blue
  respectively). The points show the result for one simulation --- see
  Section \ref{sec:sims} for a full description. The left hand panels
  show the recovered power spectra alongside the input spectra (shown as
  the smooth black curves). The right hand panels show the
  fractional bias in the recovered spectra. The simulations included differential gain of 1\%
  (top row), differential pointing of 1.5\% (second row),
  differential ellipticity of 5\% in both detector pairs (third row, to test
  the first term of equation~\eqref{eq:de}), and differential
  ellipticity in only one detector pair (bottom row, to test the second
  term in equation~\eqref{eq:de}). Also plotted are the
  predictions from equations~\eqref{eq:dg}--\eqref{eq:de} for the
  systematic effects showing good agreement with the simulated results. 
  {\color{black}This plot demonstrates the accuracy of the analytic predictions for the 
  systematics. Note that in some cases the levels of the systematics
  were deliberately increased in order to clearly demonstrate the accuracy of the predictions.} A tensor to
  scalar ratio of $r=0.1$ was used in the simulations along with a
  fiducial CMB temperature power spectrum. Note that the spurious
  $B$-mode produced by the effect corresponding to the first term in
  equation~\eqref{eq:de} is independent of the scan strategy. We,
  therefore, plot the result from just one simulation in the third row
  of panels.}

\label{fig:B_rec}
\end{center}
\end{figure*}

%% file: sections/strat_sims.tex
% !TEX TS-program = compile

\section{Scan Strategy Parameter Space}\label{sec:scan_params}

\subsection{Scan strategy parameters}\label{sec:scan_param_constraints}
Having established the accuracy of
equations~\eqref{eq:dg}--\eqref{eq:de} for a representative selection
of scans, we now proceed to use these expressions to quantify the
effectiveness of the scan strategy to mitigate leakage
 as a function of the observational parameters that define it.

The model we adopt to describe a satellite scan uses five
parameters. Firstly, the telescope will spin about the
major axis of the satellite. We denote the time period of this
rotation as $T_{\rm spin}$. The boresight of the telescope will be at an
angle to the spin axis which we call $\beta$. This spin axis is then
allowed to precess around an axis. We choose this precession
axis to be the extended line passing through the Sun and the Earth, 
presuming that the satellite is placed at the second Lagrange point (L2) 
of the Earth-Sun system. This
arrangement therefore allows the telescope to be facing away from the
Sun as much as possible. The angle between the spin axis and the
precession axis is denoted by $\alpha$, and the time period for the
precession is $T_{\rm prec}$. {\color{black}A schematic diagram of this set up is
shown in Fig.~\ref{fig:pram_schematic}.}
Finally, the satellite will sample the
sky at a frequency of $f_{\rm samp}$.

\input{sections/fig/parameter_schematic}

At first glance this seems to suggest that there are five free parameters
to describe the scan strategy. However there are a number of
additional constrains one may wish to enforce. The first and most
obvious constraint is that the telescope must observe the entire
sky. This requires that
\ba
\alpha + \beta > 90\degree.
\ea

 Further constraints arise from considering the spin rate of the
 telescope. One can envisage three potential constraints. Firstly,
 one may wish to ensure that neighbouring rings on the sky are mapped
 sequentially. As the telescope spins, it maps out a ring of radius
 $\beta$ on the sky. The precession of the telescope means that the
 next ring will be displaced from the preceding one. One may then wish
 to ensure that the spatial separation of sequential rings is less
 than some maximum separation which may be chosen to be e.g.~a
 fraction of the beam width, or a fraction of the field of view. Such
 a constraint would then allow for continuous mapping of the
 sky. This requirement places an upper bound on the ratio $T_{\rm
   spin}/T_{\rm prec}$ of
\ba
 \frac{T_{\rm spin}}{T_{\rm prec}} < \frac{\theta_{\bot}}{2\pi\sin\alpha}, 
\label{eq:mapping_con}
 \ea
 where $\theta_\bot$ is the desired separation of sequentially mapped
 rings. 

A second consideration that impacts the choice of $T_{\rm spin}$ is
the potential requirement that the scan speed is fast enough such that
the large scale modes in the sky are not confused with gain drifts in
the detectors creating ``$1/f$ noise''. The noise power spectra of detectors are in general not
white. They can often be modelled by the sum of a white component and
an additional $1/f$ component that becomes important on long time scales.
The transition point is often termed the knee frequency of the
$1/f$ noise, $f_{\rm knee}$. {\color{black}\citet{1998MNRAS.298..445D}
  previously investigated the impact of scan strategies on the resulting noise level of a map in
the presence of $1/f$ noise. Here we place a simple requirement on
$T_{\rm spin}$: we require that a particular scale of
interest on the sky, quantified by $\ell_{\rm min}$, appears in the timestream at
least a factor of $F$ in frequency higher than $f_{\rm knee}$. 
This places an upper bound on the value of $T_{\rm spin}$:
 \ba
T_{\rm spin} < \frac{2\ell_{\rm min}\sin\beta}{Ff_{\rm knee}}. 
\label{eq:f_knee_con}
\ea 
In this work we set $\ell_{\rm min} = 2$, $f_{\rm knee} = 0.01$ Hz (consistent with a slight improvement of the 
{\it Planck} 143 GHz detector, \citealt{2011A&A...536A...6P}), and $F=2$. 

In addition to the effects of $1/f$ noise, bolometers can also suffer
if the signal varies on very short timescales. Bolometers
require a finite amount of time to respond to a change in the incoming
radiation. The response of a bolometer, $d$ to a sudden impulse of power 
as a function of time, $t$, is,
\ba
d(t) \propto e^{-t/t_\star},
\ea
where $t_\star$ is the time constant of the detector \citep{1998MNRAS.298..445D}.
It is desirable, therefore, to ensure that the telescope scans slow enough such
that the telescope pointing only moves a fraction, $p$, of the beam FWHM, 
$\theta_{\rm FWHM}$ in
a time $t_\star$. This places a lower bound on the spin period,
\ba
T_{\rm spin} > \frac{2\pi t_{\star}\sin\beta}{p\theta_{\rm FWHM}}.
\ea
If we set $t_{\star}=1$ ms (consistent with the {\it Planck} CMB channels, 
\citealt{2011A&A...536A...6P}), $p=1/4$ and $\theta_{\rm FWHM}=5.0$ arcmin
then this places a lower bound on $T_{\rm spin}$ that is a factor 10
below the upper bound imposed by the requirement to avoid $1/f$ noise.
}

Finally, we require that the sampling frequency $f_{\rm samp}$ must be
fast enough so that the beam width is fully sampled. We quantify this
by requiring that the telescope must not move further than some
fraction of the beam width, $W$, between samples. This translates to a
lower bound on the sampling frequency of,
\ba 
f_{\rm samp} > \frac{2\pi \sin\beta}{W\theta_{\rm FWHM}T_{\rm spin}},\label{eq:sample_con}
\ea 
In this work we set $W=1/4$.

\subsection{Practical constraints on scan strategies}\label{sec:prac_con}

\input{sections/fig/practical_cons}

In addition to the science-driven requirements detailed above,
one also needs to consider a number of practical constraints which
also limit the possible scan strategy parameter values. {\color{black}The exact
values considered in this section for the practical constraints
are used an as example. The values are reasonable constraints at the time
of writing and may be relaxed in future}.
 
Due to computing considerations, large values of $T_{\rm spin}$ are
easier to implement in our scan-strategy simulations, as the higher
spin periods lead to lower sampling frequencies, this leads to fewer pointings
to calculate. In the following
work we therefore choose $T_{\rm spin}$ to be the largest allowed
value, given the chosen joint constraints on it and the other
parameters. Additionally, given that higher sampling rates are
problematic from the point of view of data transfer considerations, in
all of our simulations we choose the lowest possible value of $f_{\rm  samp}$
 given the constraint of equation~\eqref{eq:sample_con}.

One practical constraint that must be considered when choosing a scan
strategy is the required fuel to maintain the precession. The required
torque to maintain a gyroscopic precession is
\ba
\tau = I_{zz} \omega_{\rm prec}\omega_{\rm spin} \sin \alpha,
\ea
where $I_{zz}$ is the moment of inertia of the telescope about the
spin axis, $\omega_{\rm prec}=2\pi/T_{\rm prec}$ is the angular
velocity of the precession and $\omega_{\rm
  spin}=2\pi/T_{\rm spin}$ is the angular velocity of the telescope
spin {\color{black}\citep{feynmanlectures1,WMAP_driving}.} 
This leads to a total velocity impulsion of
\ba
\Delta v &=& \int dt \frac{\tau}{M_{\rm sat}R_{\rm lever}}\\
&=& \frac{\tau T_{\rm mission}}{M_{\rm sat}R_{\rm lever}},
\ea
where $M_{\rm sat}$ is the mass of the satellite, $R_{\rm lever}$ is
the distance of the rocket from the spin axis and $T_{\rm mission}$ is
the lifetime of the mission. In the upper panels of Fig.~\ref{fig:phy_con} we plot the
required impulsion for a telescope using typical values for an
experiment, $I_{zz}=2000~{\rm kg\,m^2}$, $M_{\rm  sat}=2000~{\rm kg}$,
 $R_{\rm lever}=1.5~{\rm m}$ and $T_{\rm  mission}=4~{\rm yrs}$.
 We show the results for a range of $T_{\rm  prec}$
 and $\alpha$ values and we have set $\alpha + \beta = 95\degree$ in all cases. 

We present results on the required impulsion for two cases. The first is the case where
$T_{\rm spin}$ is chosen such that it meets the requirement on
continuous mapping (equation~\ref{eq:mapping_con}) where we set 
$\theta_\bot=3~{\rm arcmin}$. The second is for the case where $T_{\rm
  spin}$ is chosen according to the requirement to limit the impact of
low frequency $1/f$ detector noise (equation
\ref{eq:f_knee_con}). For this latter case, we set $\ell_{\rm min} =
2$, $F=2$ and $f_{\rm knee}=0.01~{\rm Hz}$.

The second major practical concern is the implications that different
scan strategies have for the required data rate. Higher sampling
frequencies obviously require a faster data transfer to the
earth. The required data rate per detector is proportional to the
sampling frequency and hence,
\ba
{\rm data rate} = N_{\rm b} f_{\rm samp}, 
\ea
where $N_{\rm b}$ is the number of bits per sample, which we
choose to be 8. In the lower panels of
Fig.~\ref{fig:phy_con} we present the required data rate per detector
as a function of $T_{\rm prec}$ and $\alpha$. As for the investigation
of the required impulsion discussed above, we present our
results for two possible ways of choosing $T_{\rm spin}$, according
to either equation~\eqref{eq:mapping_con} or equation~\eqref{eq:f_knee_con}.

For comparison we also plot contours of ``reasonable values'' for the impulsion
and data rate. We plot the contours of impulsion requirements for 
$\Delta v =$ 48, 113, 290 ms$^{-1}$. These correspond to the possible impulsion
values after having obtained ``small'', ``medium'' and ``large''
Lissajou orbits centred around L2. This assumes that the satellite
will have a fixed amount of fuel for both
orbit injection and to drive the scan. We have here subtracted the fuel
required to achieve the respective orbits.
For the data rate plots we show a contour of data rate per detector
corresponding to a total data rate of 20 Mbps for 4800 detectors.

The telescope pointing with respect to the Sun is also of significant 
importance. To minimise far sidelobe pick up of the Sun the
telescope must never be pointed too close to the Sun. For this 
reason we have set $\alpha + \beta = 95\degree$. This means our
telescope never points closer than $85\degree$. However,
the Sun's influence can also place constraints on $\alpha$ itself. As $\alpha$ is
the angle between the spin axis and the Sun-Earth line, large values
of $\alpha$ mean the Sun will be shining on the side of the telescope
at a more acute angle. Therefore, large values of $\alpha$ must 
also be accommodated with more effective heat shielding. Solar panels 
must be located on the warm service module rather than on the cold 
payload. A high value of the precession angle $\alpha$ results in less 
efficient orientation of the solar panels w.r.t. the Sun.
Therefore, a more sophisticated set up for the solar panels
may be required.

The data transfer antenna sets a limit on the maximum precession
angle allowed. The earth aspect angle $\theta_{\rm earth}$ cannot exceed
a certain value to allow the antenna to point towards the earth. Here we set
$\theta_{\rm earth} \leq 62\degree$ as currently available antenna
choices set this limit.
If the limit was changed the possible scan strategies would increase.
This aspect angle is increased by both the precession angle and the
position of the satellite,

\ba
\theta_{\rm earth} &=& \alpha + \theta_{\rm orbit},\label{eq:aspect_angle}\\
&\le& 62\degree,\\
\theta_{\rm orbit} &=& \arctan\left(\frac{R^{\rm max}_{\rm orbit}}{R_{\rm L2}}\right),
\ea
where $R_{\rm orbit}$ is the radius of the satellite around L2 and
$R_{\rm L2}$ is the distance of L2 from the earth. Therefore,
if the satellite is placed in a smaller orbit around L2 then
a larger precession angle $\alpha$ will be possible. However,
the fuel required to place the satellite in this smaller orbit will
be larger which leaves us less fuel to drive the scan strategy. 

The results from this assessment of practical considerations suggest
that a large fraction of parameter space is difficult to acheive in practice.
The reader should therefore bear in mind Fig. \ref{fig:phy_con} when
interpreting the results, displayed in Fig.~\ref{fig:b_mode_grid},
regarding the impact of different scan strategies on mitigating
systematics which we now go on to discuss.

%\subsection{Error on the $B$-mode polarisation power spectrum}
\subsection{Error on the main science goals}\label{sec:science_goals}

The analysis of Section~\ref{sec:ana} provides a quick but accurate method to
predict the error on the recovered $B$-mode polarisation power
spectrum due to certain systematic errors and given certain features of
the scan strategy. We have additionally developed a fast scan strategy
simulation code that calculates the pointing and orientation of the
telescope with respect to North for a given set of scan strategy
parameters over 1 year of observations. This pointing information
can then be used to calculate the $\langle |\ft{h}_n|^2 \rangle$
values for that particular set of scan strategy parameters. The
calculated values can then be used to predict the error on the
recovered $B$-mode power spectrum using equations
(\ref{eq:dg})--(\ref{eq:de}).

In Fig.~\ref{fig:h_num} we plot the $\langle |\ft{h}_n|^2 \rangle$
values as a function of the scan strategy parameters, $\alpha$ and
$T_{\rm prec}$. In all cases, we have overplotted the positions of the
{\it Planck}, WMAP, and proposed EPIC scan strategies. We have also
overplotted an example LiteBIRD scan strategy on the grid for
reference. It should be noted that the baseline LiteBIRD design
includes a rotating
HWP in the optical chain which will improve both the ability to mitigate systematic effects
and the noise properties of the experiment \citep{2014SPIE.9143E..1FM}. In implementing the scan strategies for each point in the
$\{\alpha$--$T_{\rm prec}\}$ parameter space, we have made a number
of choices for the values of the other observational parameters. These
choices were motivated by the considerations outlined in
Section~\ref{sec:scan_param_constraints}. Firstly, we have set $\alpha
+ \beta =95\degree$. We have checked that this choice has little
impact on the results within $\pm5\degree$. 
%It simply translates the shape of the plots.
We therefore choose the angle sum to be similar to other scan
strategies in the literature. In particular, both the {\it Planck} and
WMAP scan strategies have $\alpha + \beta = 92.5\degree$ whilst
the proposed EPIC scan strategy has the sum equalling $95\degree$. 

We chose the value of $T_{\rm spin}$ such that it satisfies the
requirement of equation~\eqref{eq:f_knee_con}. As before, we set
$\ell_{\rm min}=2$, $F=2$ and $f_{\rm knee}=0.01~{\rm Hz}$.
{\color{black}When the constraint of equation~\eqref{eq:mapping_con} was used, for 
 $T_{\rm prec}\lesssim 10^2$ hrs, the
$\langle |\ft{h}_n|^2 \rangle$ values showed no discernible
change.}
The fact that the $h$-values were unchanged is to be expected because $T_{\rm spin}
\ll T_{\rm prec}$ in both cases. Finally, we have chosen $f_{\rm samp}$ such
that it fulfils the requirement of equation~\eqref{eq:sample_con}, for a
$\theta_{\rm FWHM} = 5$ arcmin beam width.

From equations (\ref{eq:dg})--(\ref{eq:de}) we can see that
the lower the values of $\langle|\tilde{h}_n|^2\rangle$ the smaller
the temperature to polarisation leakage is for a particular systematic
error in the experiment. $\langle|\tilde{h}_2|^2\rangle$ is important
for mitigating differential gain, $\langle|\tilde{h}_1|^2\rangle$ and
$\langle|\tilde{h}_3|^2\rangle$ are important for mitigating
differential pointing, and finally $\langle|\tilde{h}_4|^2\rangle$ is
important for mitigating the difference between the differential
ellipticity. The results of Fig.~\ref{fig:h_num} show that the
choice of $T_{\rm prec}$ has little impact on the
$\langle|\tilde{h}_n|^2\rangle$ values unless $T_{\rm prec} \gtrsim
20~{\rm hours}$. This can be understood by considering the other
timescales in the problem. With $T_{\rm prec} \lesssim 20~{\rm hours}$ then
$T_{\rm spin} \ll T_{\rm prec} \ll 1~{\rm year}$ meaning that 
this would have little effect on the quality of the scan
strategy. {\color{black}If the value of $T_{\rm prec}$ is too large then the 
scan strategy cannot observe the entire sky in 1 year. This region of
parameter space is shown in white in Fig.~\ref{fig:h_num}.}

Fig.~\ref{fig:h_num} does show, however, that the precession angle,
$\alpha$, has a significant impact on the quality of the scan
strategy. A smaller boresight angle ($\beta$) will
result in the satellite scanning in smaller circles. Given the
constraint $\alpha + \beta = 95\degree$, a smaller value of $\beta$
corresponds to a larger precession angle $\alpha$. Scanning in smaller
circles generally creates a larger range of orientations for each
pixel and thus improves the quality of the scan strategy by lowering the
$\langle|\tilde{h}_n|^2\rangle$ values, where $n$ is even. When $n$ is odd 
scanning in larger circles results in a small range of
orientation angles. However six months later the scan is flipped,
creating a symmetry that renders the odd terms close to zero. This
symmetry is enhanced when the range of orientation angles is small,
creating a deep valley in the $\langle|\tilde{h}_1|^2\rangle$ and
$\langle|\tilde{h}_3|^2\rangle$ values at $\alpha\approx5\degree$.

\input{sections/fig/scan_quality}

Using equations~(\ref{eq:dg})--(\ref{eq:de}) we can translate the
$\langle |\ft{h}_n|^2 \rangle$ surfaces of Fig.~\ref{fig:h_num} into
estimates of the resulting error on the recovered $B$-mode power
spectrum. To do this, we must make some assumptions regarding the
levels of the systematics to include in the
calculations. {\color{black}Apart from differential gain, we use
typical values for these types of systematics as found in the BICEP2
experiment instrumental systematics paper
\citep{2015arXiv150200608B}. For differential gain, we note that satellite-based
missions can use the CMB dipole to calibrate the bolometers, and can
therefore achieve much lower levels of differential gain, of order 0.2\% \citep{2015arXiv150201587P}.
Specifically, we choose a differential
gain of 0.2\%, a differential pointing of 2\% of the beam and a differential
ellipticity of 5\% corresponding to $q=1.05$ (see equation~\ref{eq:gaus_beam}).

With these calculated errors on the recovered power spectrum, given a
scan strategy, we then go on to estimate the error on the recovered
cosmological parameters.  We focus on two key parameters: the
tensor-to-scalar ratio $r$, and $A^{BB}_{\rm lens}$, which
parametrises the amplitude of the $B$-mode lensing signal, as used in
\citet{2014ApJ...794..171T}. In order to do this we calculate the
error on two important $C_{\ell}^{BB}$ multipole bins. Our first bin
covers the multipole range $2 \le \ell \le 201$. We use the error on
this bin (which we call $\Delta C_{b=1}^{BB}$) to assess an
experiment's ability to recover an inflationary $B$-mode signal on
large scales. We calculate the amplitude of the predicted signal in
this bin (which we call $C_{b=1}^{BB}$) for a cosmology where $r=1$
and in the absence of lensing $B$-modes. We
assume the latter have been removed to make a best estimate of $r$. As
the amplitude of the signal within this bin is simply scaled by the
value of $r$, we can therefore find the error on the recovered value
of $r$, which we call $\Delta r$,
\ba
\Delta r = \frac{\Delta C_{b=1}^{BB}}{C_{b=1}^{BB}}.
\ea
The other bin we consider covers the
range $801 \le \ell \le 1000$ in order to assess our ability to
recover the lensing $B$-mode power spectrum. We follow a similar
procedure: we calculate the error on this bin ($\Delta C_{b=2}^{BB}$) and the predicted value
for $A^{BB}_{\rm lens}=1$ ($C_{b=2}^{BB}$). As the parameter simply scales the
$B$-mode power spectrum the error on $A^{BB}_{\rm lens}$ will be,
\ba
\Delta A^{BB}_{\rm lens} = \frac{\Delta C_{b=2}^{BB}}{C_{b=2}^{BB}}.
\ea
The results are presented
in Fig.~\ref{fig:b_mode_grid}. This figure clearly shows, as described
before, that the dominant parameter
 in mitigating temperature-to-$B$-mode
polarisation leakage is the precession angle $\alpha$, once $T_{\rm prec} \ll 1$ year:
 as $\alpha$ increases the spurious signal induced in the $B$-mode polarisation
power spectrum reduces dramatically.

It should be noted that our method for predicting the
temperature-to-$B$-mode leakage provides us with a worse case scenario
as we have only considered an experiment with two pairs of
detectors. In a real experiment, the leakage from each detector pair
will be different and in certain situations the leakage from different
detector pairs could be completely uncorrelated. This would result in
the overall leakage averaging to a lower value. The differential gain
between detector pairs is unlikely to be correlated. However, the
effects of differential pointing and differential ellipticity could
conceivably be a function of the position of the detector pair in the
focal plane and/or imperfections in the primary lens or mirror of the
telescope. Such a scenario would result in the leakage from different
detector pairs being correlated and hence not averaging to an overall
lower value.

\input{sections/fig/b_mode_error_grid}

\subsection{Long Time Scale Drifts}

Before we can choose a scan strategy it is imperative that we consider
other problematic consequences a particular scan strategy can have on
the final data analysis. One potential complication is the need to
filter the TOD in order to remove long time scale drifts of the
detectors. These long timescale drifts are a result of the $1/f$ noise
considered in Section \ref{sec:scan_param_constraints}. In order to
minimise the impact of $1/f$ noise, we required that the scan speed of
the telescope should be fast enough such that the largest angular
scale of interest ($\ell=2$) appears in the TOD at a time scale twice
that of $f_{\rm knee}$. This choice could be problematic, as large
values of $\alpha$ can lead to the telescope observing a relatively
small patch of the sky for long time periods even if, over that patch,
it is scanning quickly.

In data analysis there is often a step where long timescale drifts are
removed from the TOD, in order to suppress the effects of $1/f$ noise
(stripes) on the resulting map. This typically involves the use of
de-striping algorithms, many of which filter the TOD with a high pass
filter removing long timescales
\citep{1998A&AS..127..555D,2010MNRAS.407.1387S,2010ApJS..187..212C}.
While this filtering reduces the long timescale noise in the map, it
can also remove some of the true sky signal. Filtering of true sky
signal in this way is something which we would like to avoid and/or
minimize.

To fully assess the impact that the scan strategy choice has on TOD
filtering requirements in the presence of $1/f$ noise would require
full TOD simulations including realistic $1/f$ noise and the
subsequent application of a de-striping algorithm to the simulated
data. One would then need to analyse the resulting maps to investigate
which scans required the least amount of filtering in order to meet
the science goals. The results of this exercise would depend strongly
on the choice, and implementation of, the de-striping algorithm, so
ideally one would use many and compare the results. Such an analysis is
beyond the scope of the current work. Additionally there
could well be developments in de-striping algorithms before the data
is collected and analysed and one would ideally use those, as yet
undeveloped, superior algorithms in such an investigation.

We can however obtain some insight on the effect that the scan strategy has on
the resulting filtering requirements by implementing a naive TOD
filtering for each scan. To do this, for each scan, we have performed signal only
simulations of the TOD from one detector, which is only sensitive to
temperature fluctuations. We have then filtered the TOD by removing the mean from 3 min sections
of the TOD. We then make maps of the temperature sky using these TODs, where the 
input sky model only contains CMB fluctuations. The power spectrum of
the reconstructed maps are calculated and we assess the result of the
filtering by calculating,
\ba
F_\ell = \frac{C_{\ell}^{T~{\rm meas}}}{C_\ell^{T~{\rm true}}} \label{eq:f_ell_def},
\ea
where $C_{\ell}^{T~{\rm meas}}$ is the power spectrum of the map after
filtering and $C_{\ell}^{T~{\rm true}}$ is the power spectrum of the
map before filtering.  $F_\ell$ ranges from 0 to 1 --- the higher the
value, the less signal has been removed.  Fig. \ref{fig:fl_av_grid}
shows the average of $F_\ell$ over the range $2\le\ell\le25$.  We
choose to plot this range as the large scale power is the most
affected by this type of filtering.  Figure \ref{fig:fl_av_grid} shows
the regions of scan strategy space that should be avoided in order to
minimize loss of signal due to filtering. Unless
$T_{\rm prec}\lesssim1$ hrs then $\alpha \lesssim 60\degree$ is
necessary to avoid filtering the TOD to a significant level.

\input{sections/fig/filtered_grid}

%% file: sections/fig/parameter_schematic.tex
% !TEX TS-program = compile

\begin{figure}
\begin{center}
\begin{tabular}{c}
~\\
~\\
\includegraphics[width=0.8\linewidth, trim=5cm 10cm 11.7cm 4cm, clip=true]{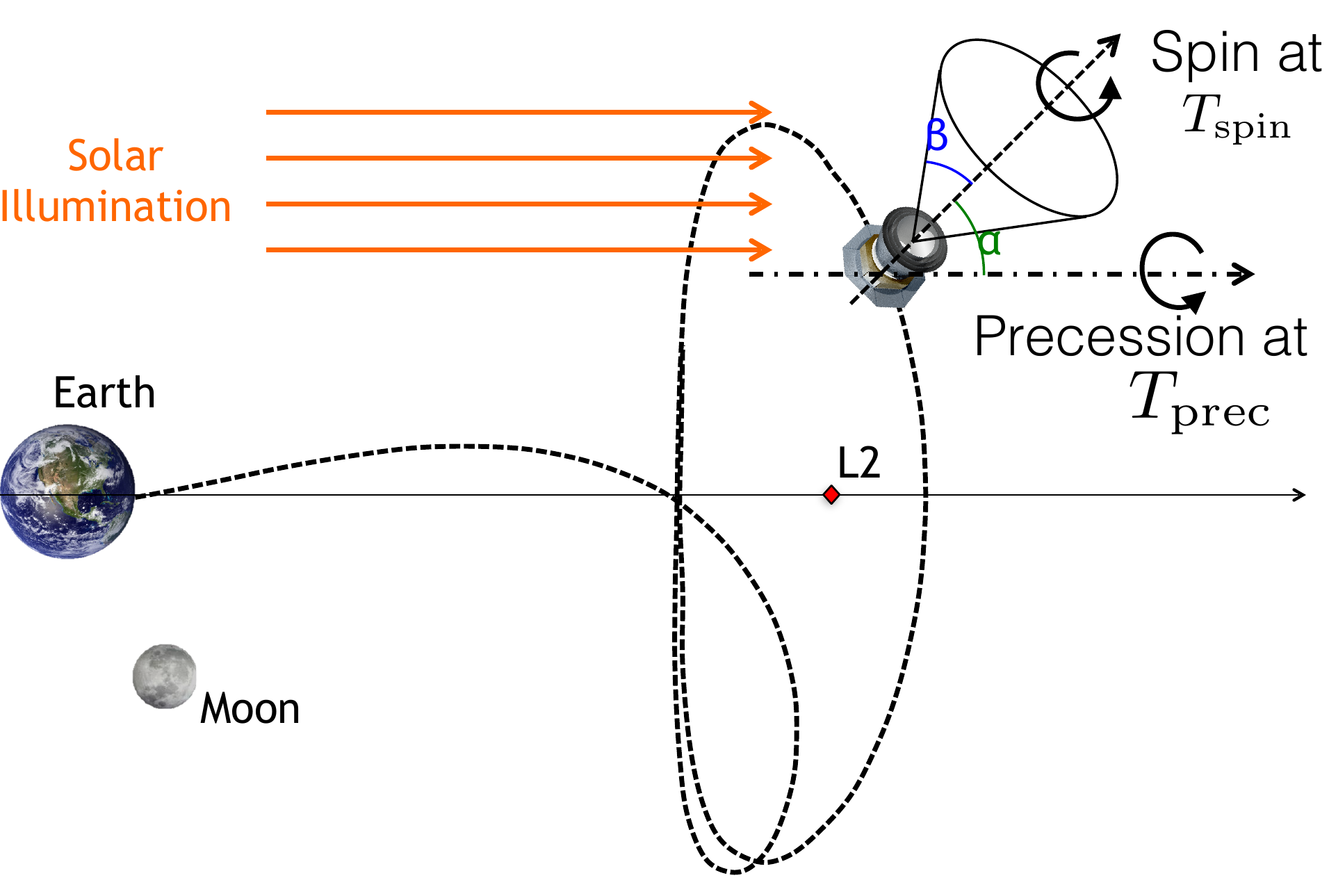}\\
\end{tabular}
\caption{{\color{black}Schematic diagram describing the observational parameters used to define the scan strategies.}}
\label{fig:pram_schematic}
\end{center}
\end{figure}

%% file: sections/fig/practical_cons.tex
% !TEX TS-program = compile

\begin{figure*}
\begin{center}
\begin{tabular}{c}
~\\
~\\
\includegraphics[width=\linewidth, trim=0cm 0cm 0cm 0cm, clip=true]{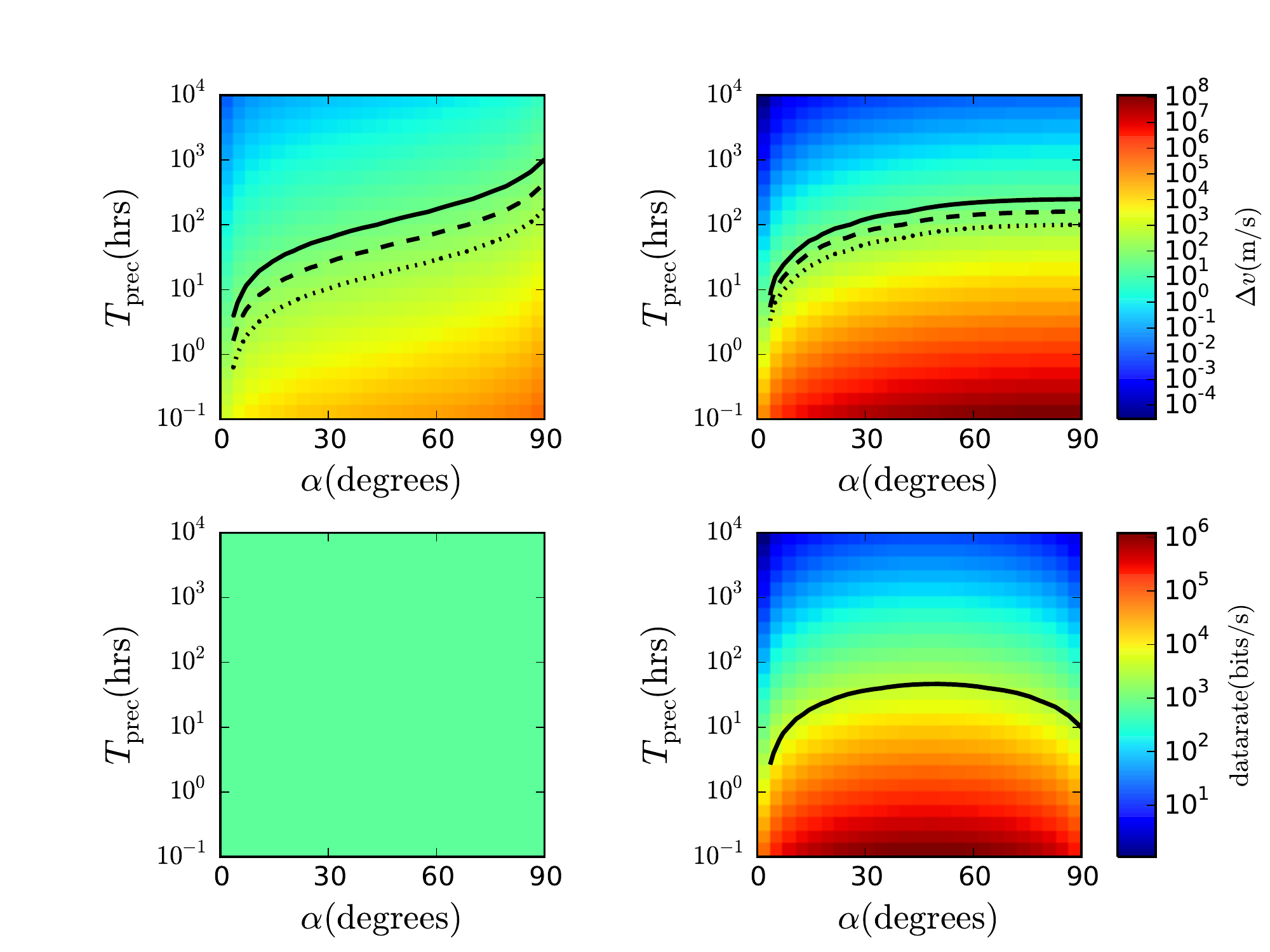}\\
\end{tabular}
\caption{{\it Upper panels:} We plot the required impulsion to
  maintain the scan strategies as a function of the scan
  parameters. In the left panel, we have used the $T_{\rm spin}$
  constraint from equation~\eqref{eq:f_knee_con}. The panel on the
  right shows the result when the constraint from
  equation~\eqref{eq:mapping_con} is used. Overplotted are contours of
  $\Delta v = 48, 113, 290~{\rm m/s}$ to show reasonable impulsion
  values. These correspond to the possible impulsion
values after having acheived a small, medium and large L2 orbit (shown
with solid, dashed and dotted lines respectively).
 {\it Lower panels:} We plot the data rate per detector
  requirements given the required sampling frequency. We have used the
  $T_{\rm spin}$ constraint from equation~\eqref{eq:f_knee_con} on the
  left and the constraint from equation~\eqref{eq:mapping_con} on the
  right. See Section \ref{sec:prac_con} for details. Overplotted is a contour of
  constant data rate corresponding to a total data rate of 20 Mbps for 4800 detectors.
  As the  $T_{\rm spin}$ constraint from equation~\eqref{eq:f_knee_con} requires
  all the scan strategies to scan at the same speed, and the sampling frequency is set by the 
  scan speed, the data rate is therefore constant for all values of $\alpha$ and $T_{\rm prec}$.}
\label{fig:phy_con}
\end{center}
\end{figure*}

%% file: sections/fig/scan_quality.tex
% !TEX TS-program = compile

\begin{figure*}
\begin{center}
\begin{tabular}{c}
~\\
~\\
\includegraphics[width=\linewidth, trim=0cm 0cm 0cm 0cm, clip=true]{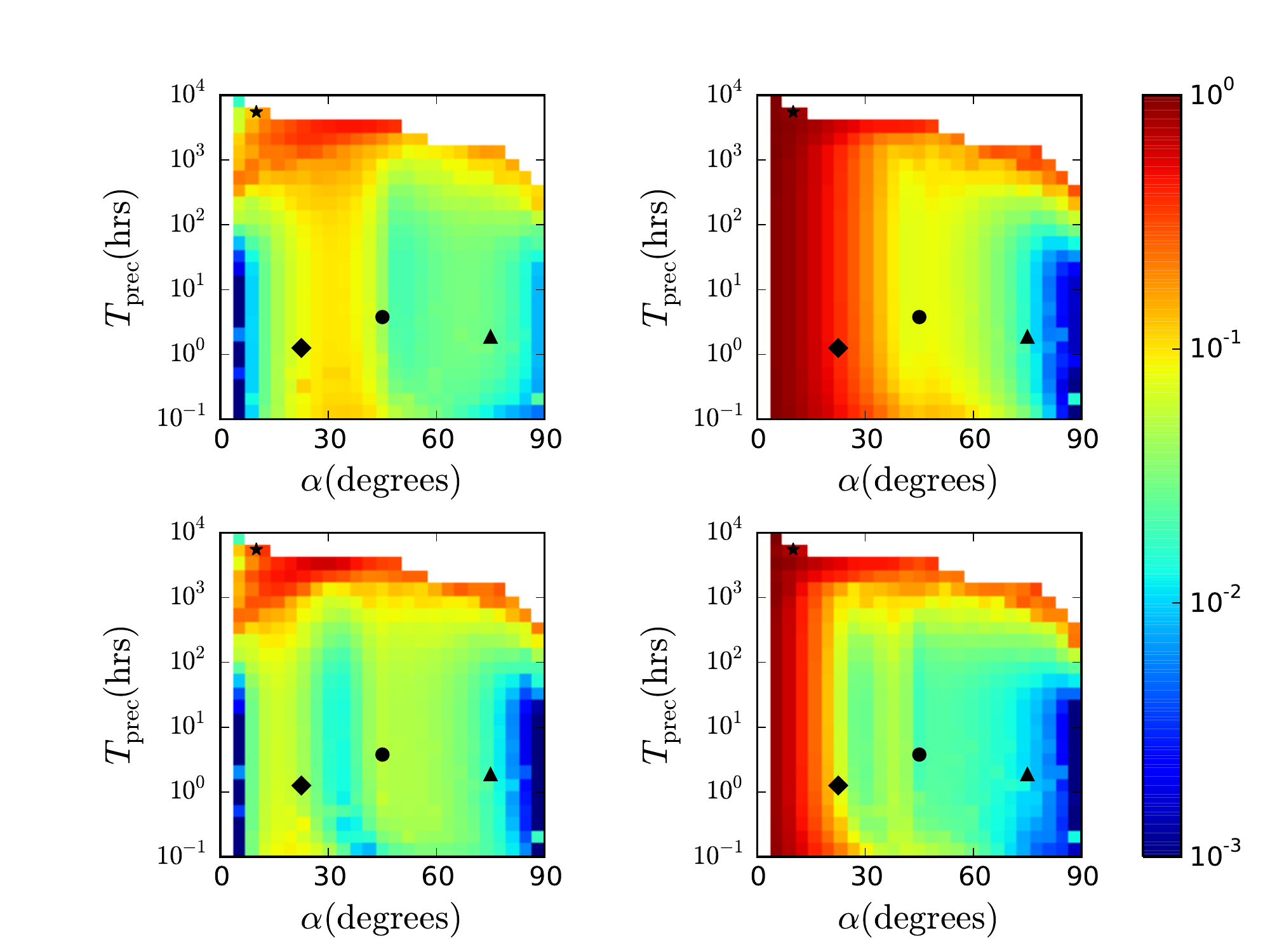}\\
\end{tabular}
\caption{We plot the values of $\langle|\tilde{h}_n|^2\rangle$, used
  in equations~(\ref{eq:dg})--(\ref{eq:de}) as a function of the scan
  strategy parameters. The $\langle|\tilde{h}_n|^2\rangle$ values for
  $n=\{1,2,3,4\}$ are displayed in the top left, top right, lower left
  and lower right panels respectively. The lower the value of
  $\langle|\tilde{h}_n|^2\rangle$ the smaller the temperature to
  polarisation leakage is for a particular systematic error in the
  experiment. $\langle|\tilde{h}_2|^2\rangle$ is important for
  mitigating differential gain, $\langle|\tilde{h}_1|^2\rangle$ and
  $\langle|\tilde{h}_3|^2\rangle$ are important for mitigating
  differential pointing, and finally $\langle|\tilde{h}_4|^2\rangle$
  is important for mitigating the difference between the differential
  ellipticity. {\color{black}White regions show areas where the entire sky
is not observed in 1 year.}
In each panel, we indicate the positions of the
  {\it Planck}, WMAP, EPIC and LiteBIRD scan strategies with a star,
  diamond, circle and triangle, respectively.}
\label{fig:h_num}
\end{center}
\end{figure*}

%% file: sections/fig/b_mode_error_grid.tex
% !TEX TS-program = compile

\begin{figure*}
\begin{center}
\begin{tabular}{c c}
~\\
~\\
\includegraphics[width=0.5\linewidth, trim=0cm 0cm 0cm 0cm, clip=true]{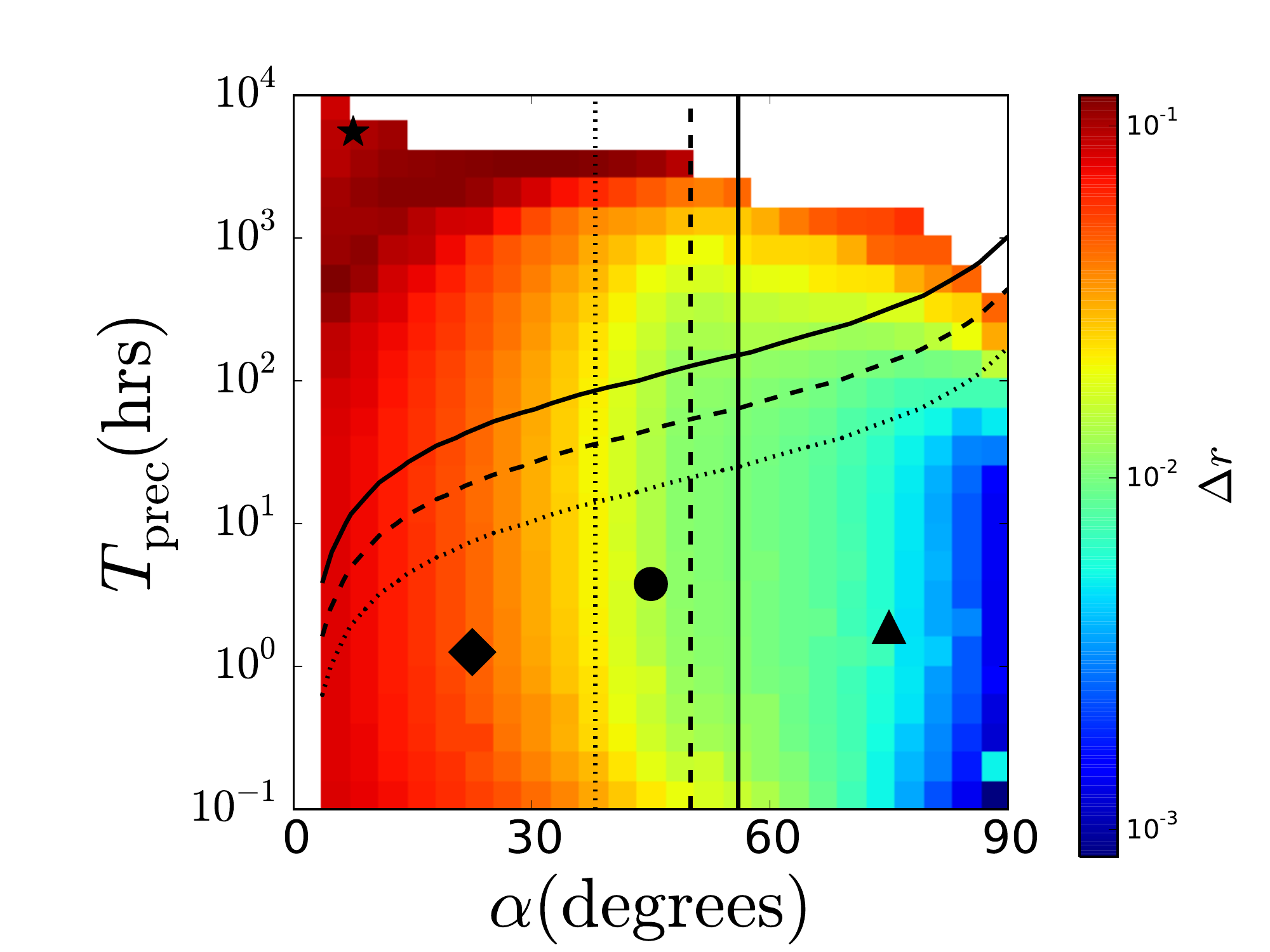}  &
\includegraphics[width=0.5\linewidth, trim=0cm 0cm 0cm 0cm, clip=true]{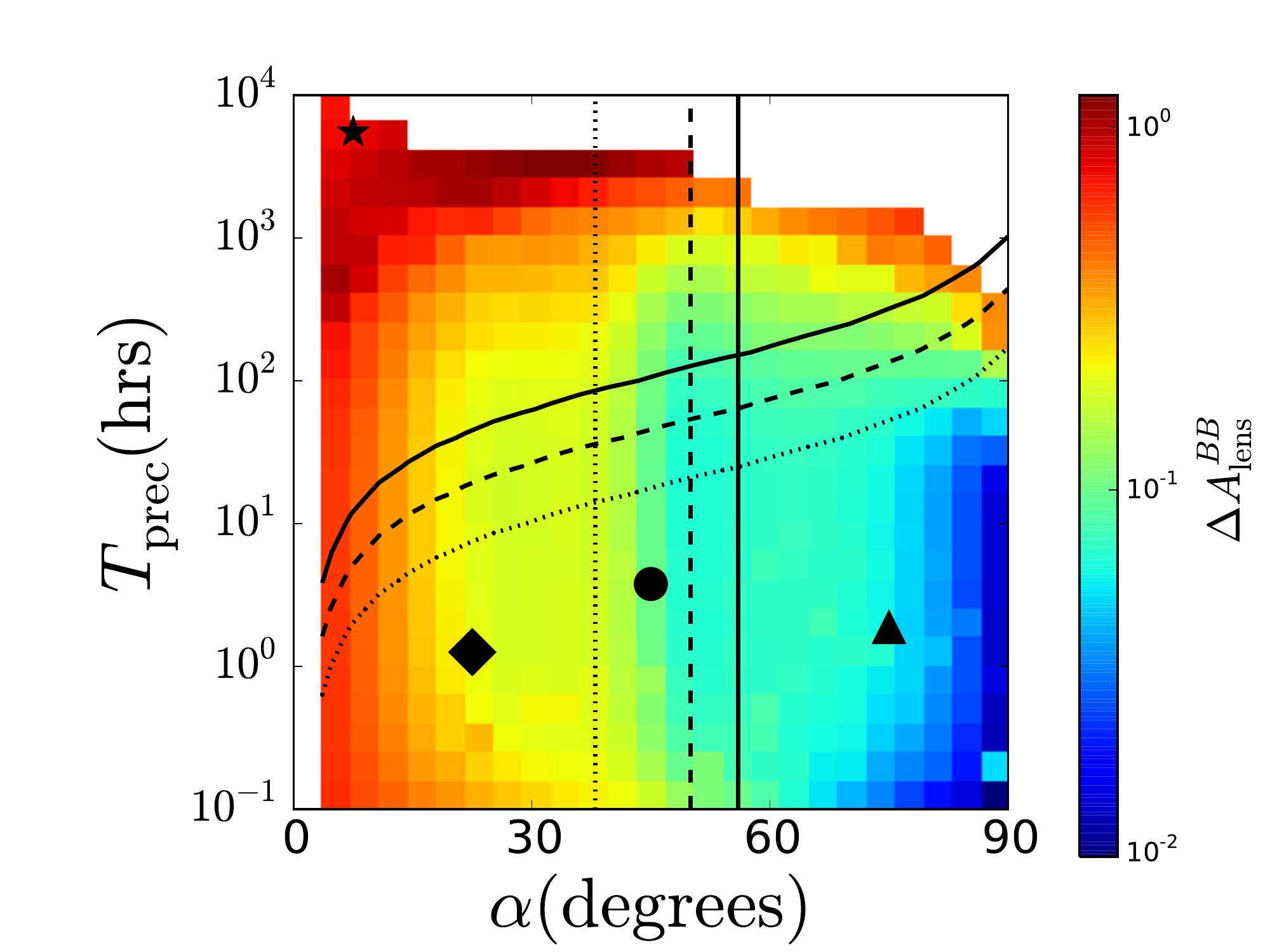}\\

\end{tabular}
\caption{{\it Left panel:} The error on the tensor to scalar ratio $r$
  as a function of the scan strategy parameters (see Section
  \ref{sec:science_goals} for details). A differential gain of 0.2\%,
  a differential pointing of 2\% and a differential ellipticity in one
  detector of 5\% aligned with the polarisation sensitivity was
  assumed. {\it Right panel:} The same as the left panel but for the
  lensing amplitude parameter $A_{\rm lens}^{BB}$.  Note that the
  morphology of the two plots are the same but the amplitude is
  different. White regions show areas where the entire sky is not
  observed in 1 year.  In each panel, we indicate the positions of the
  {\it Planck}, WMAP, EPIC and LiteBIRD scan strategies with a star,
  diamond, circle and triangle, respectively. The overplotted lines
  give an indication of the likely restrictions on scan strategy
  parameter space due to the maximum fuel capacity of the satellite
  and the maximum aspect angle of the satellite with respect to the
  Earth, which is limited by data transmission requirements. Regions
  of parameter space below the curved lines require more fuel to drive
  the scan strategy than what is likely to be available after
  injection into orbit around L2. The solid, dashed and dotted lines
  correspond to injection into small, medium and large Lissajou L2
  orbits respectively (see Section~\ref{sec:scan_investigation} for
  details). The vertical lines indicate the likely maximum possible
  values for $\alpha$ in order to meet the data transfer requirements
  on the satellite aspect angle. Once again, the solid, dashed and
  dotted lines correspond to small, medium and large Lissajou L2
  orbits respectively.}
\label{fig:b_mode_grid}
\end{center}
\end{figure*}

%% file: sections/fig/filtered_grid.tex
% !TEX TS-program = compile

\begin{figure}
\begin{center}
\begin{tabular}{c}

\includegraphics[width=\linewidth, trim=0cm 0cm 0cm 0cm, clip=true]{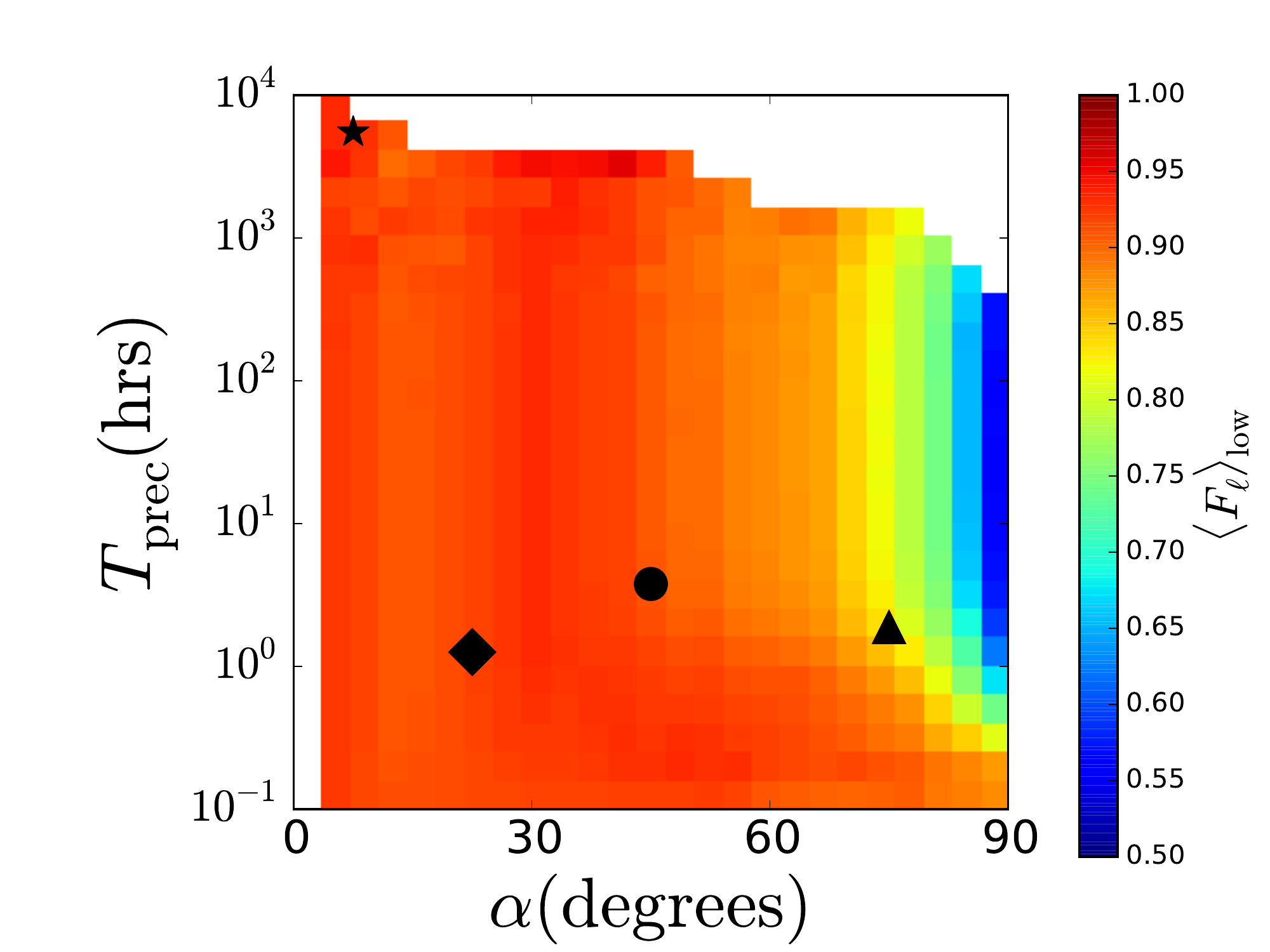}  \\

\end{tabular}
\caption{The average of $F_\ell$ (see equation \ref{eq:f_ell_def}),
  over a multipole range $2\le\ell\le 25$ for a naively filtered
  timestream for a range of scan strategies. The simulated TOD was
  filtered by removing the mean signal from 3 minute sections of
  TOD. A binned temperature map was constructed from this filtered
  timestream and its power spectrum calculated. This procedure naively
  models the effect of a de-striping algorithm filtering the
  cosmological signal. We indicate the positions of the
  {\it Planck}, WMAP, EPIC and LiteBIRD scan strategies with a star,
  diamond, circle and triangle, respectively.}
\label{fig:fl_av_grid}
\end{center}
\end{figure}

%% file: sections/scan_investigation.tex
% !TEX TS-program = compile

%\section{Chosen Scan Investigation}\label{sec:scan_investigation}
\section{Choosing a scan strategy}\label{sec:scan_investigation}

{\color{black}We are now in a position, as an example, to suggest a scan strategy for 
the improved COrE+ ESA M5 mission call.} 
We aim to choose the scan strategy that will mitigate systematic effects 
most effectively given the practical constraints on possible scans outlined
in Section \ref{sec:prac_con}. We examine three potential L2 orbits for
the satellite.

\input{sections/tables/orbits_fuel_req}

As described in Section \ref{sec:prac_con} there is a maximum aspect
angle that the satellite can have with respect to the Earth, of
62$\degree$, set by the antenna transmitting data to the ground. The
larger the orbit, the smaller the scan precession must be (see
equation~\ref{eq:aspect_angle}). However, there is a trade off to be
made with the fuel requirement that injection into the satellite orbit
demands, as this leaves less fuel to drive the scan
strategy. Table~\ref{tab:orbits_fuel_req} shows the impulsion
requirements to enter into three Lissajous orbits which we label as
``large'', ``medium'' and ``small''. We also show the remaining
impulsion that would then be available to drive the scan assuming the
satellite had a total impulsion of 380 ${\rm ms^{-1}}$ to
use. Table~\ref{tab:orbits_fuel_req} also shows the aspect
angle of the satellite with respect to the Earth $\theta_{\rm orbit}$
and therefore, the maximum possible precession angle $\alpha$.
 
The results of our systematic effect mitigation investigation are
summarized in Fig.~\ref{fig:b_mode_grid}. In this figure, we have
overplotted the constraints on the scan strategy due to the fuel and
data transmission considerations discussed above. For each orbit
(large, medium and small) we plot two lines which are dotted, dashed
and solid respectively. The vertical line corresponds to the maximum
precession angle for that orbit. The curved line is a line of constant
impulsion which we must stay above. The result of this analysis shows
that the best scan strategy for each orbit occurs at the limit of both
constraints.  The medium and small orbits allow much better scans
strategies than the large orbit, with the medium orbit being slightly
better than the small orbit.

We show explicitly the optimal scan strategy of each orbit in Table \ref{tab:core_scan_params}.
For the medium orbit we suggest another option where we increase the boresight angle. 
In Figure \ref{fig:core_scans_time_int} and \ref{fig:core_scans_h_2}
 we plot the time integration and $\langle |h_2|^2\rangle$ of the scan strategies
presented in Table \ref{tab:core_scan_params} for a range of mission lengths.

\input{sections/tables/core_scan_parameters}

\input{sections/fig/time_investigation}
\input{sections/fig/time_investigation2}

Fig.~\ref{fig:core_scans_time_int} shows that all the scans can create
a relatively even coverage over the sky after a year of scanning. It
should be noted that in the "small orbit" scenario we have too little
fuel to drive a fast precession period.  This slower precession has
caused a less smooth time integration map. On shorter timescales the
two ``medium orbit" scans can make half sky maps while the ``small
orbit" and ``large orbit" scans require the Earth to orbit around the
Sun to observe a full half sky. The ability to make half sky maps on
short timescales -- on the order of tens of days -- would be very
beneficial as this would allow time dependent systematic effects to be
investigated more easily.  In Fig.~\ref{fig:core_scans_h_2} we plot
the $\langle |h_2|^2\rangle$ value of chosen scans. This allows us to
see how well the scans would allow us to make maps with single
detectors. After a year all scans have a very good polarisation angle
coverage and therefore single detector polarisation maps will be
possible.  On shorter time scales, however, this will not be
possible. The orbit of the Earth around the sun is required to allow
the telescope to observe large regions of the sky at many crossing
angles. The second scan at the ``medium orbit" where we set the
boresight angle $\beta=50\degree$ is intended to improve the
polarisation angle coverage on short time scales. Figure
\ref{fig:core_scans_h_2} does show a small improvement. The ability of
the two ``medium orbits'' to make half sky maps on short time scales,
and to produce excellent polarisation angle coverage over a year, make
them both suitable choices for the {\color{black} improved COrE+ scan strategy.}

%% file: sections/tables/orbits_fuel_req.tex
% !TEX TS-program = compile

\begin{table*}
  \centering
\begin{tabular}{|l|p{3cm}|p {3cm}|p {3cm}|p {3cm}l}%\label{table:terms}
  \hline
{ \bf Orbit around L$_2$} & {\bf Impulsion to reach orbit} ($\Delta v_{\rm orb}$) & {\bf Max scan impulsion possible} ($\Delta v_{\rm prec}$)& {\bf Aspect angle of orbit} ($\theta_{\rm orbit}$) & {\bf Max precession angle possible} ($\alpha$)\\ \hline
Large Lissajous & 90 ${\rm ms^{-1}}$ & 290 ${\rm ms^{-1}}$ & 24$\degree$ & 38$\degree$ \\ \hline
Medium Lissajous & 267 ${\rm ms^{-1}}$ & 113 ${\rm ms^{-1}}$ & 12$\degree$ & 50$\degree$ \\ \hline
Small Lissajous & 332 ${\rm ms^{-1}}$ & 48 ${\rm ms^{-1}}$ & 6$\degree$ & 56$\degree$\\ \hline
\end{tabular}
  \caption{Table showing three possible orbits around L$_2$ and the required impulsion. We also show the implications of such orbits on the remaining fuel to drive the scan strategy and the possible precession angles given the orbit, see Section \ref{sec:prac_con}.}
  \label{tab:orbits_fuel_req}
\end{table*}

%% file: sections/tables/core_scan_parameters.tex
% !TEX TS-program = compile

\begin{table*}
  \centering
\begin{tabular}{|c|c|c|c|c|}%\label{table:terms}
  \hline
{ \bf Scan } &  { \bf Boresight angle} ($\beta$) & {\bf Precession angle }($\alpha$)& {\bf Spin period} $(T_{\rm spin})$&{\bf Precession period} $(T_{\rm prec})$\\ \hline
Large Lissajous & 57$\degree$ & 38$\degree$ & 166 s & 15 hrs\\ \hline
Medium Lissajous & 45$\degree$ & 50$\degree$ & 142 s & 40 hrs \\ \hline
Medium Lissajous ($\beta=50\degree$) & 50$\degree$ & 50$\degree$ & 142s & 40 hrs \\ \hline
Small Lissajous  & 39$\degree$ & 56$\degree$ & 125 s & 130 hrs \\ \hline
\end{tabular}
  \caption{Parameters of the optimal scans at the different L$_2$
    orbits (see Section~\ref{sec:scan_investigation} for details).}
  \label{tab:core_scan_params}
\end{table*}

%% file: sections/fig/time_investigation.tex
% !TEX TS-program = compile

\begin{figure*}
\begin{center}
\begin{tabular}{c c c c}
~\\
~\\
Large & Medium & Medium ($\beta=50\degree$) & Small \\
%&Hit Map&\\0.25
\includegraphics[width=0.25\linewidth, trim=0cm 1cm 0cm 0cm, clip=true, angle=180]{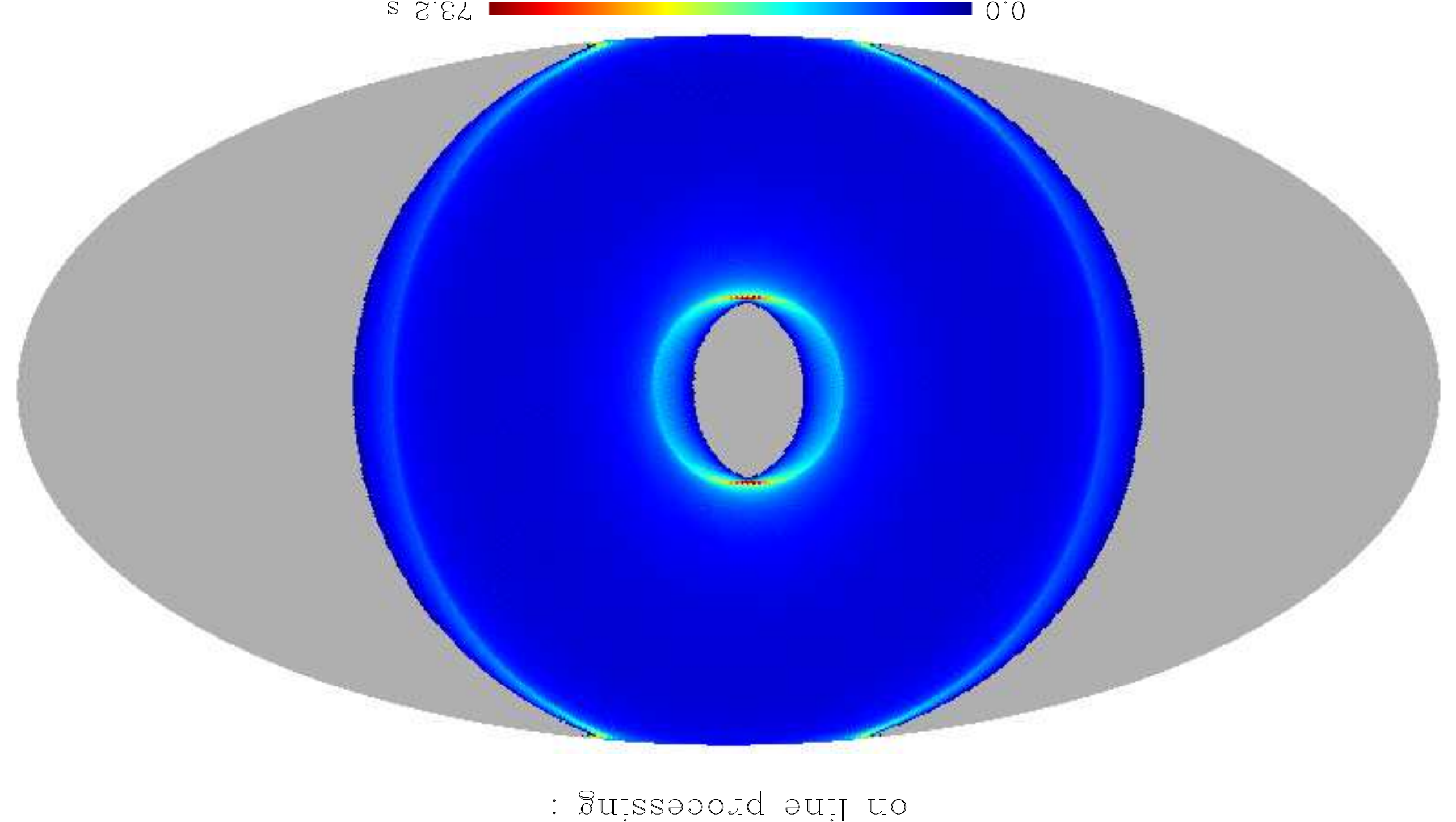}  &
\includegraphics[width=0.25\linewidth, trim=0cm 1cm 0cm 0cm, clip=true, angle=180]{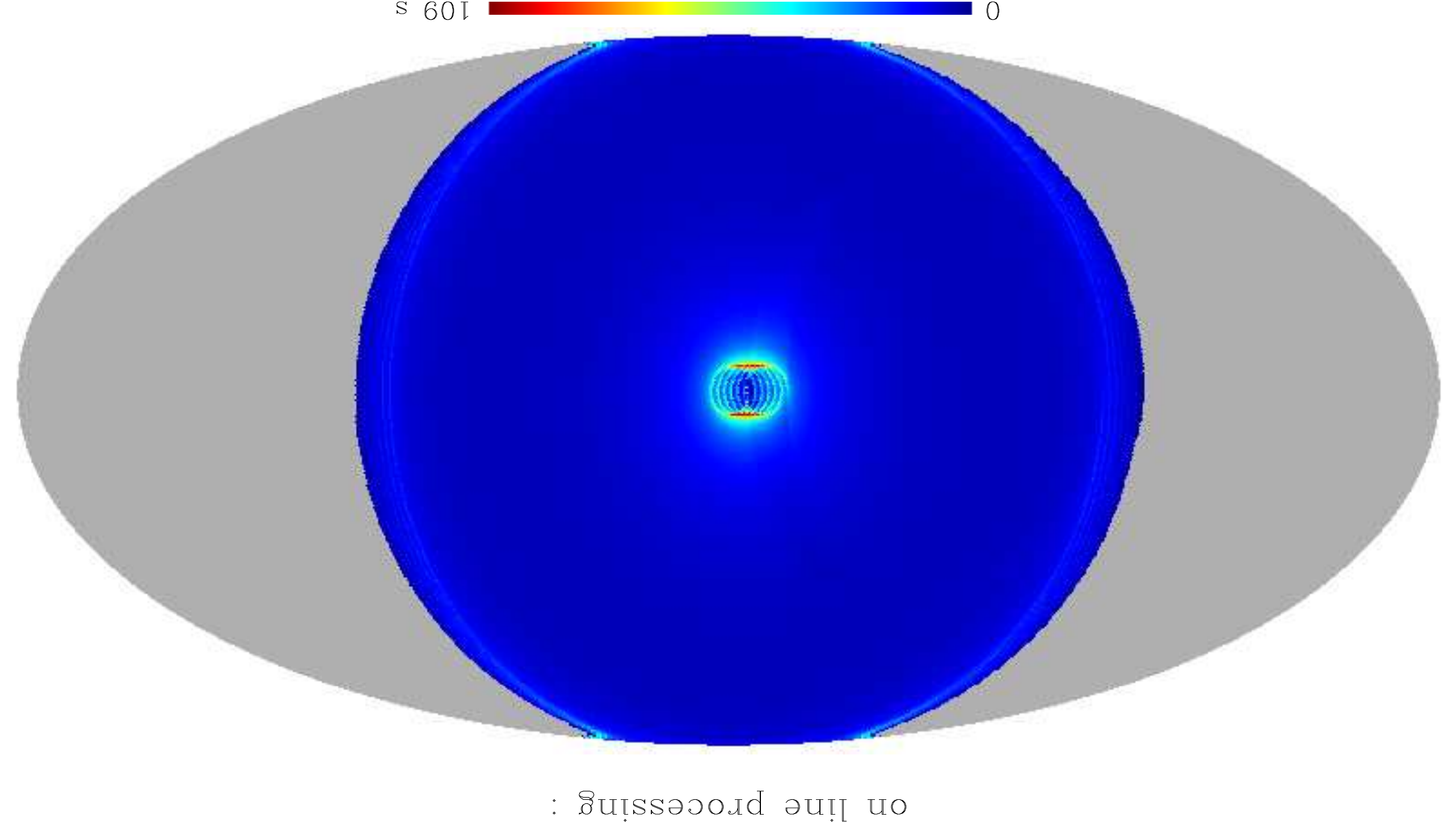}&
\includegraphics[width=0.25\linewidth, trim=0cm 1cm 0cm 0cm, clip=true, angle=180]{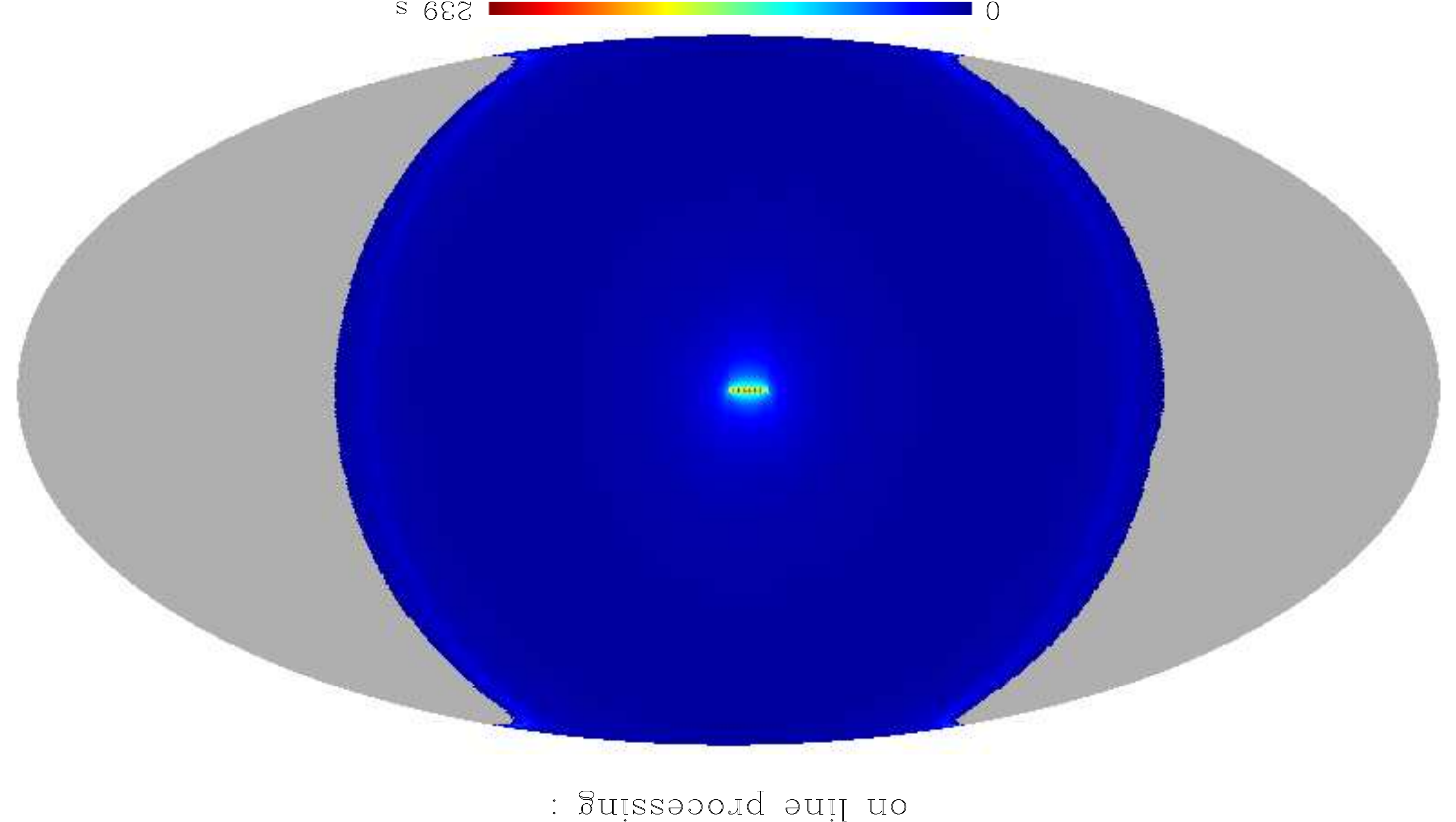}&
\includegraphics[width=0.25\linewidth, trim=0cm 1cm 0cm 0cm, clip=true, angle=180]{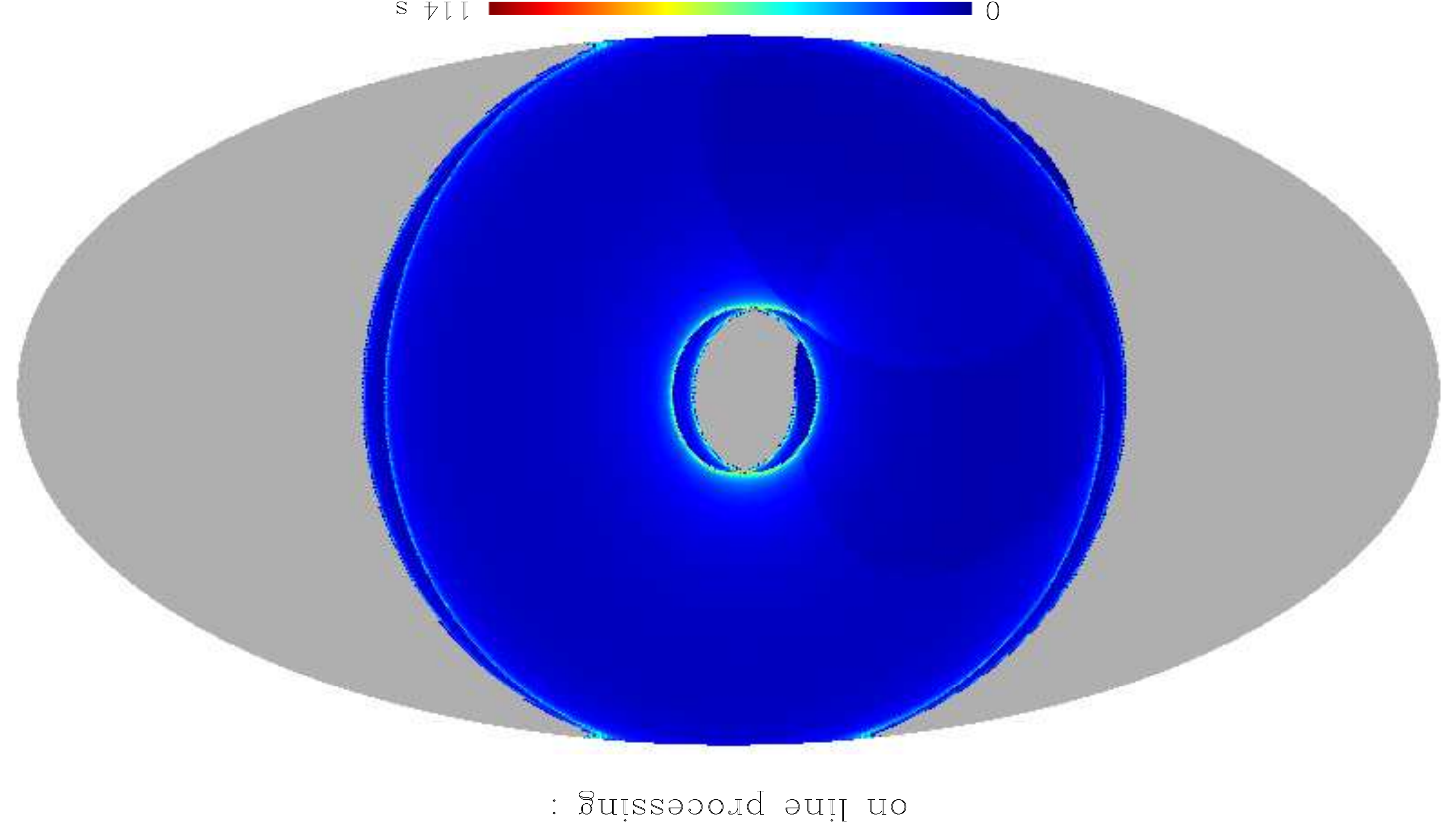}\\

\includegraphics[width=0.25\linewidth, trim=0cm 1cm 0cm 0cm, clip=true, angle=180]{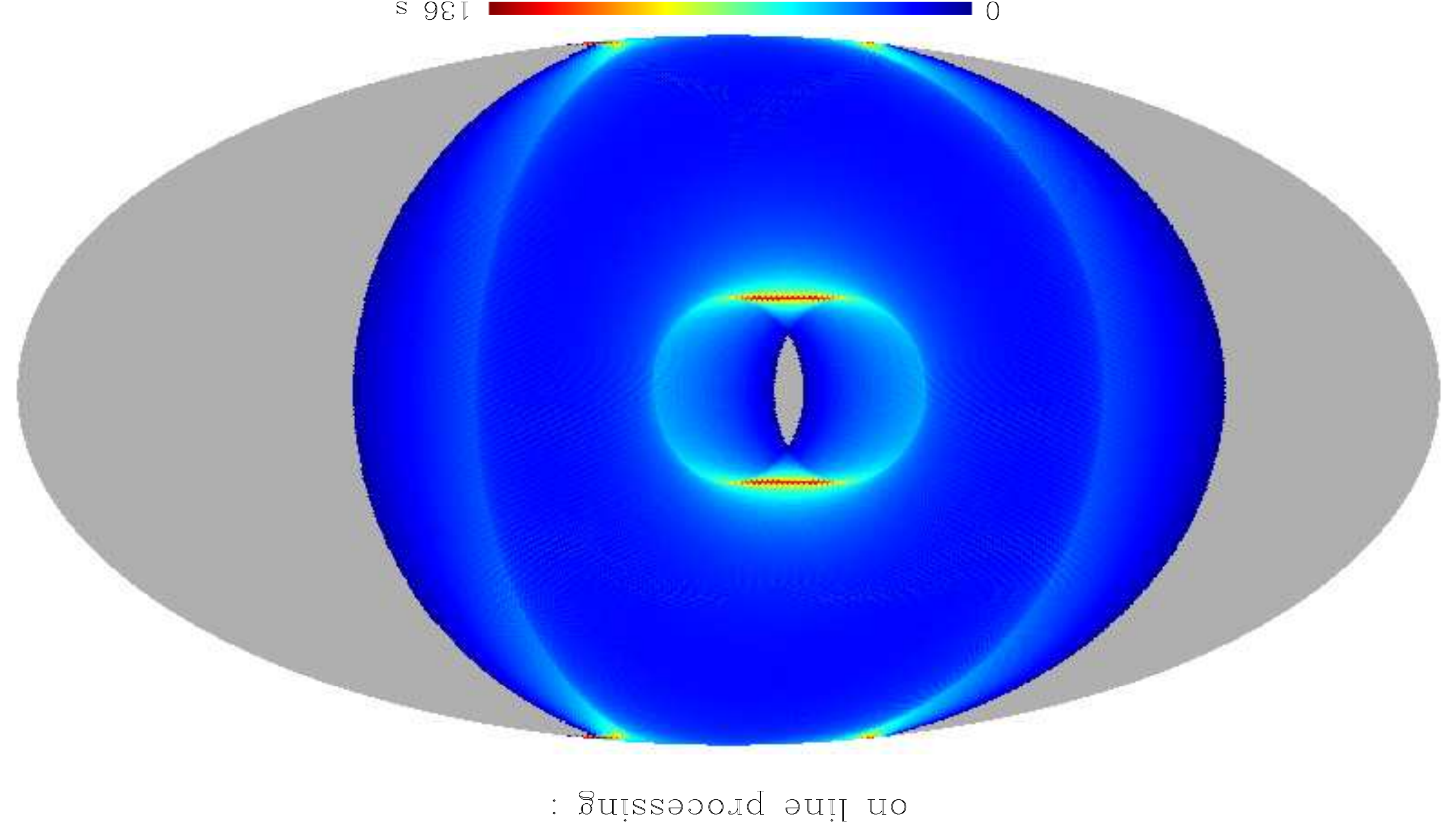}  &
\includegraphics[width=0.25\linewidth, trim=0cm 1cm 0cm 0cm, clip=true, angle=180]{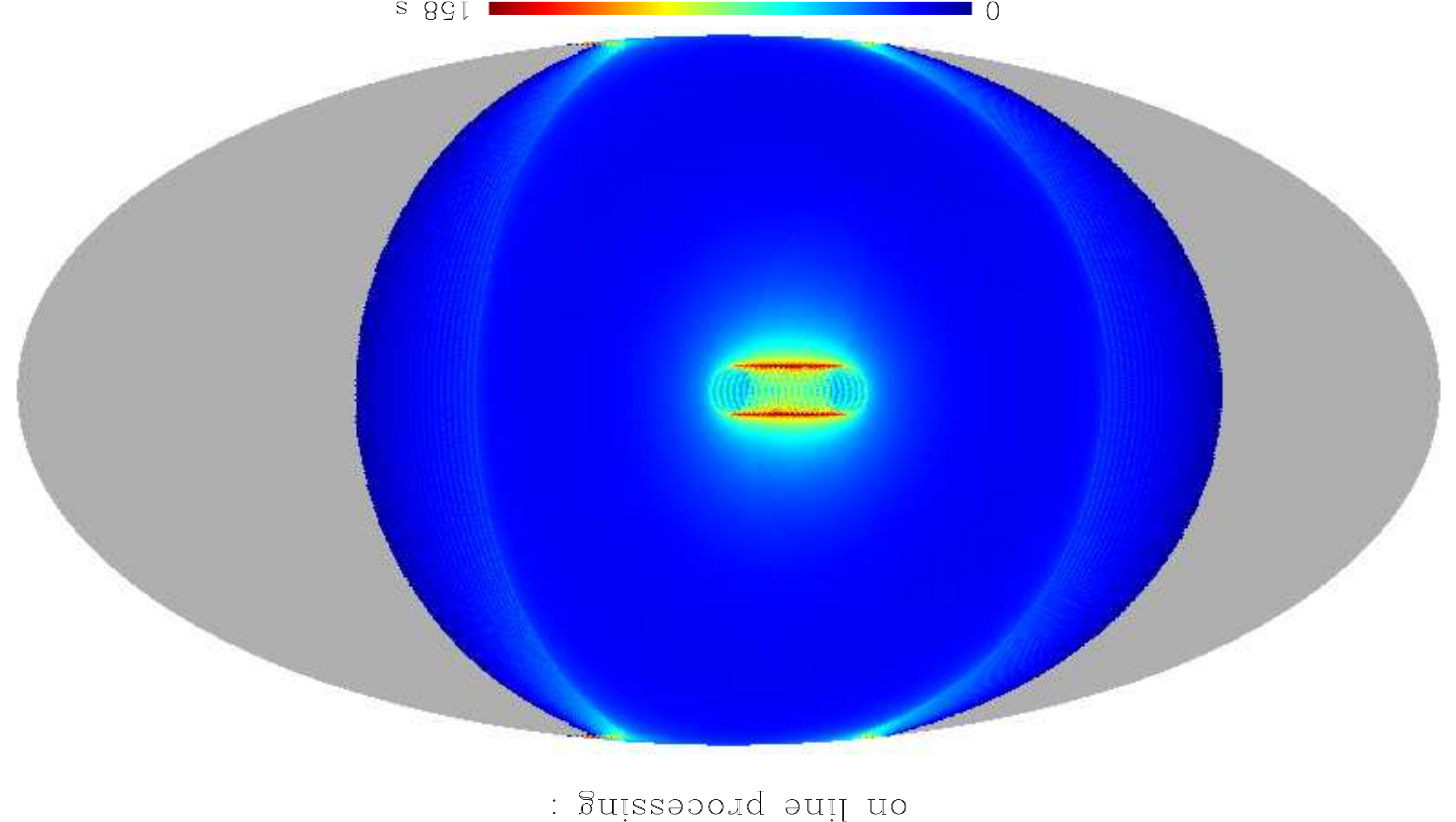}&
\includegraphics[width=0.25\linewidth, trim=0cm 1cm 0cm 0cm, clip=true, angle=180]{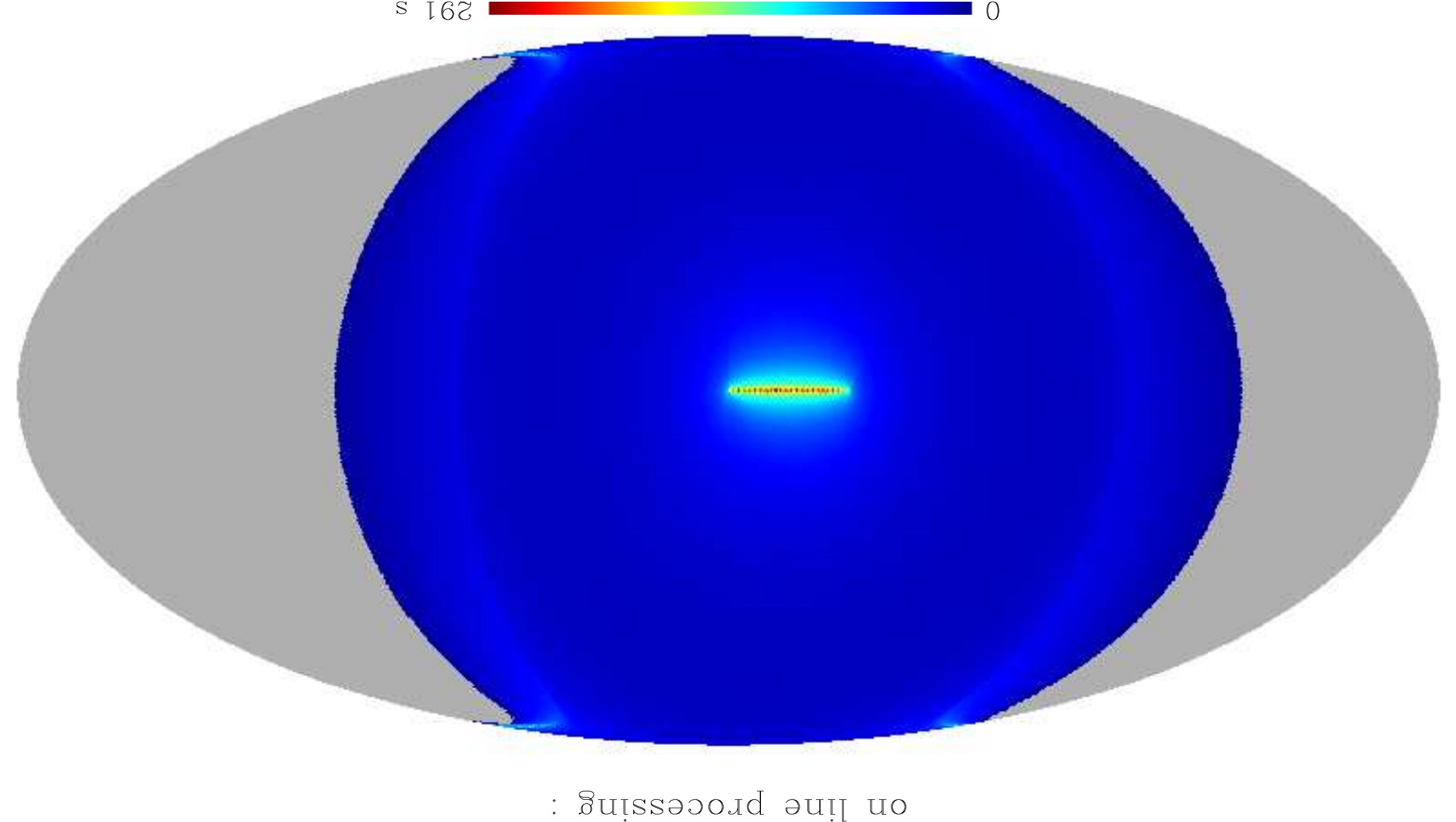}&
\includegraphics[width=0.25\linewidth, trim=0cm 1cm 0cm 0cm, clip=true, angle=180]{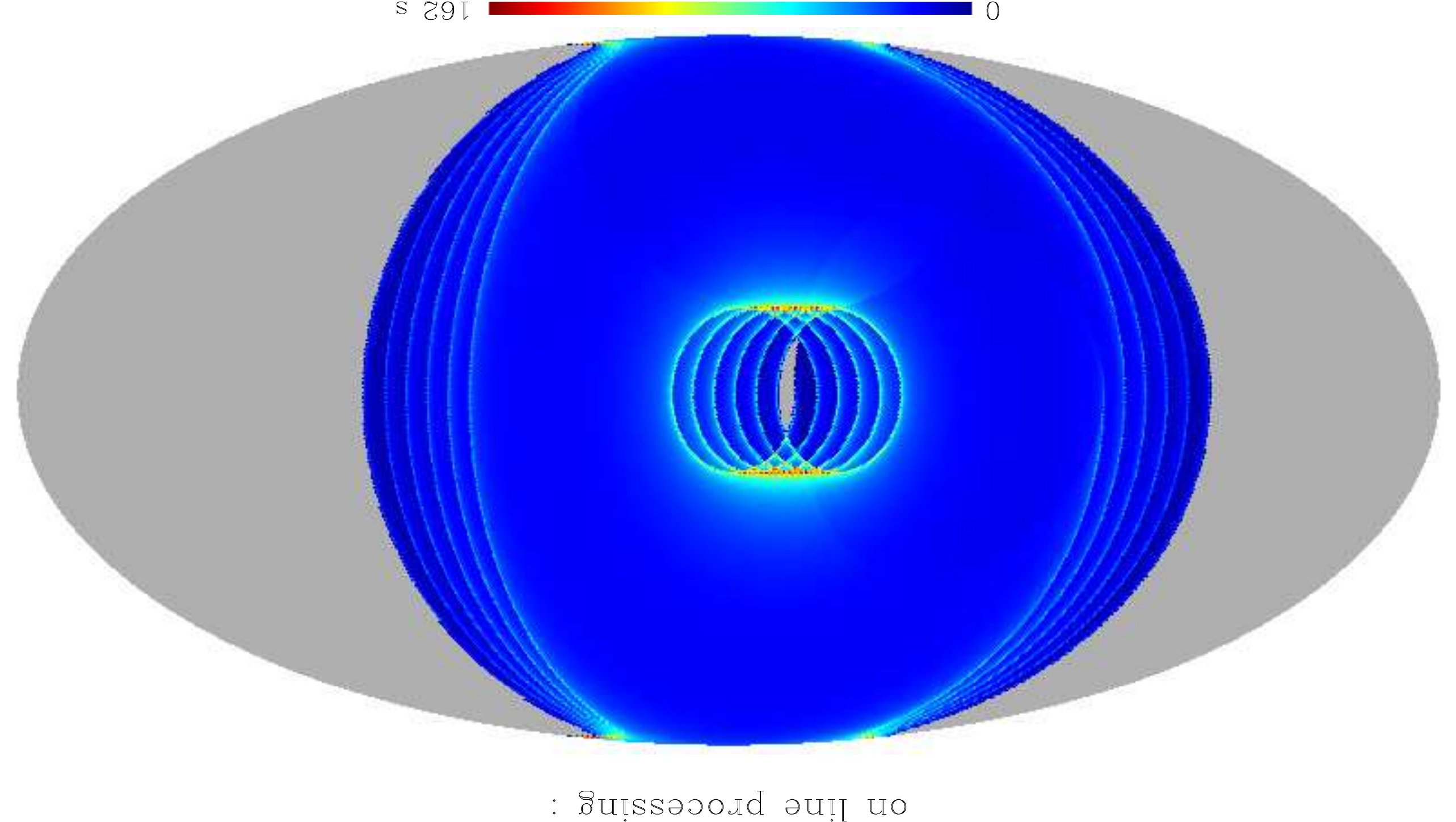}\\

\includegraphics[width=0.25\linewidth, trim=0cm 1cm 0cm 0cm, clip=true, angle=180]{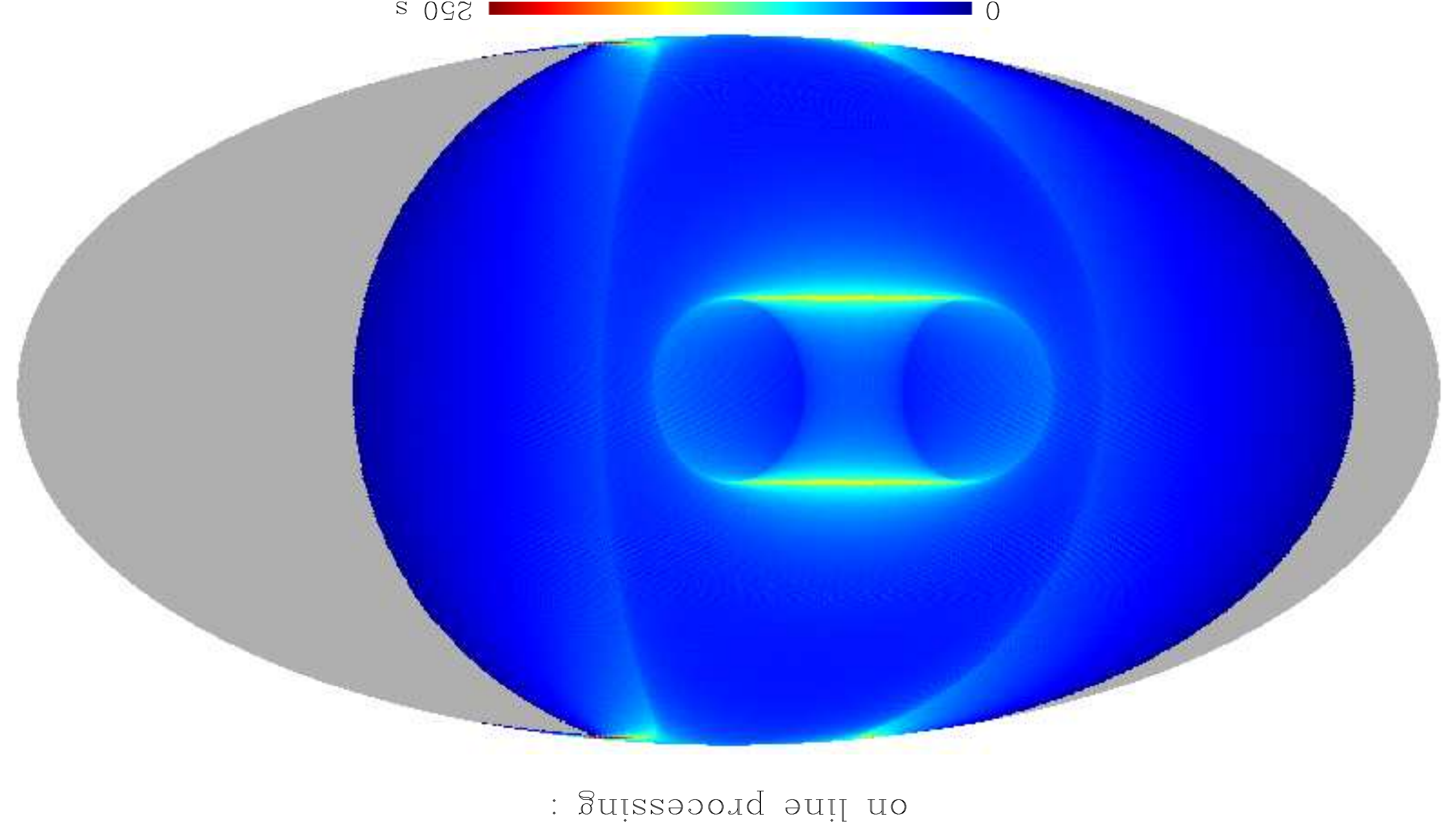}  &
\includegraphics[width=0.25\linewidth, trim=0cm 1cm 0cm 0cm, clip=true, angle=180]{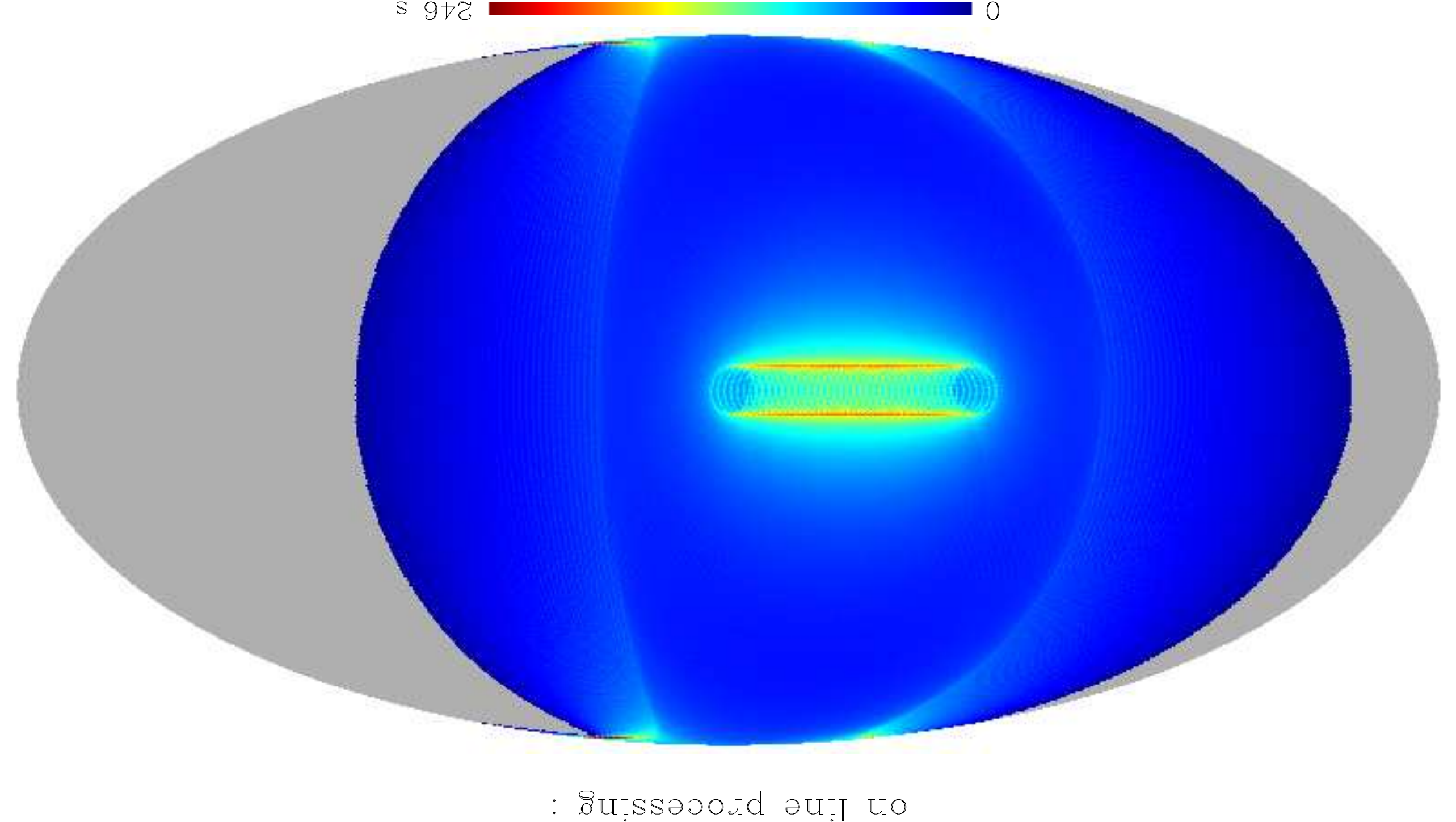}&
\includegraphics[width=0.25\linewidth, trim=0cm 1cm 0cm 0cm, clip=true, angle=180]{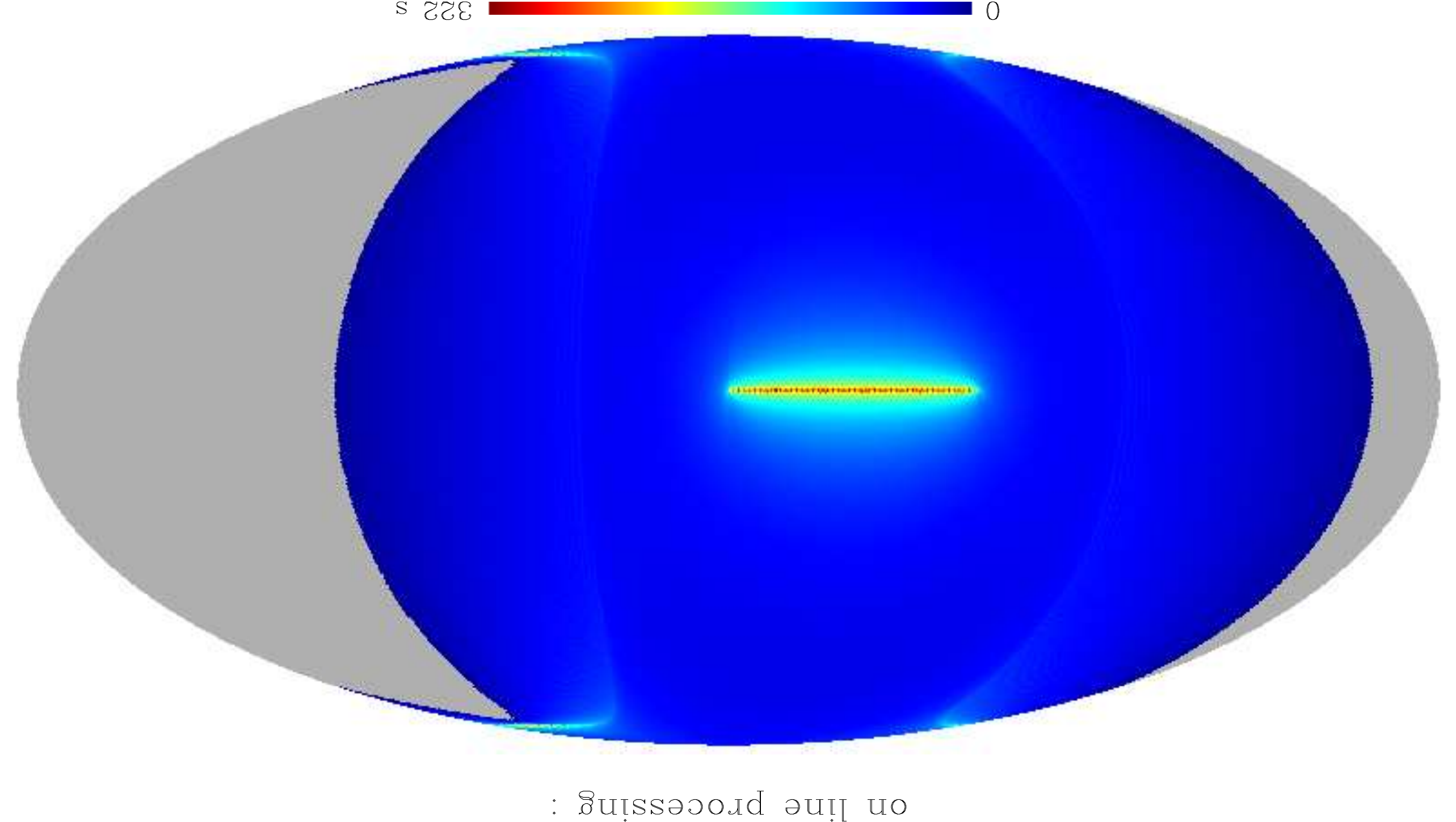}&
\includegraphics[width=0.25\linewidth, trim=0cm 1cm 0cm 0cm, clip=true, angle=180]{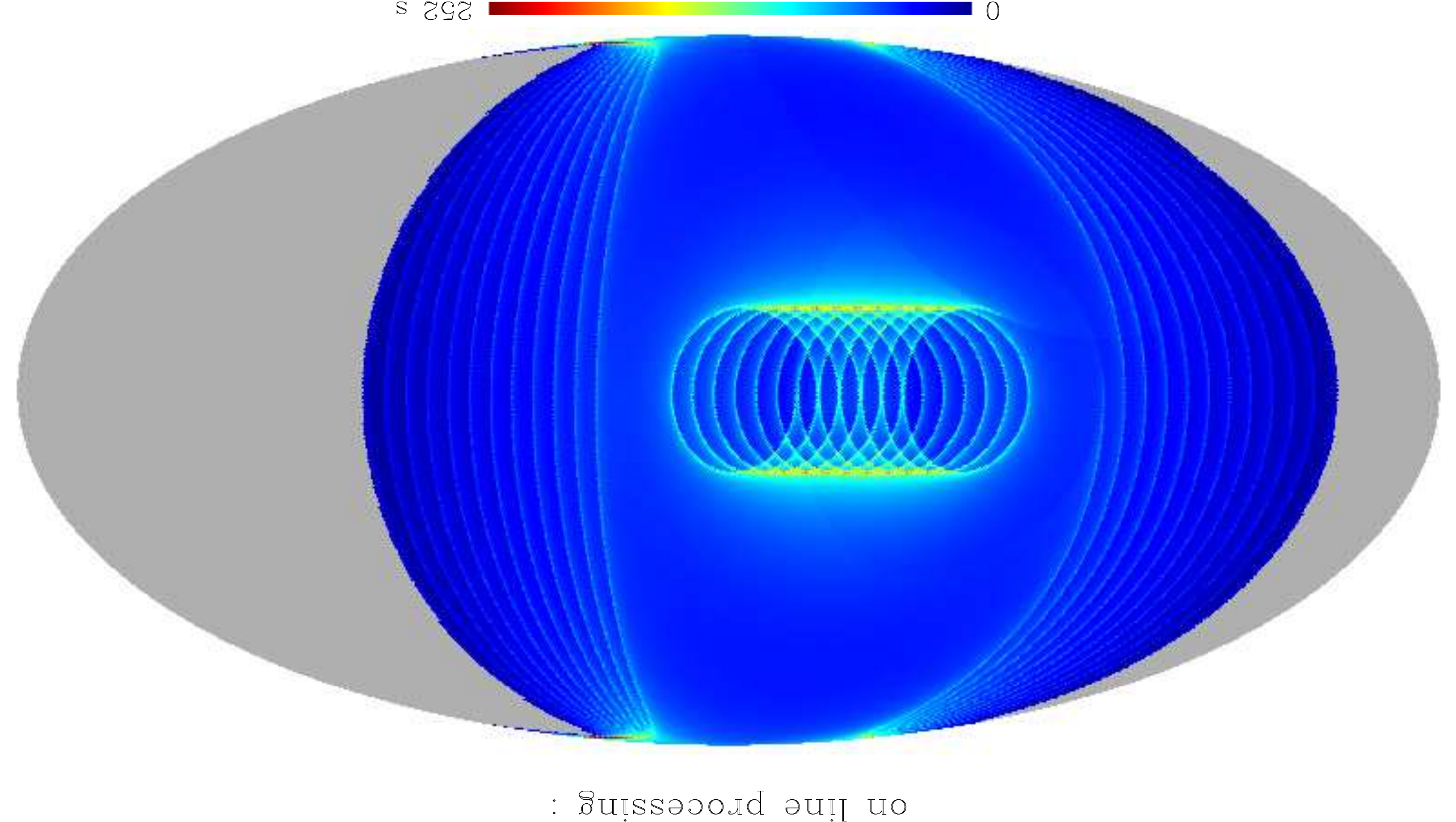}\\

\includegraphics[width=0.25\linewidth, trim=0cm 1cm 0cm 0cm, clip=true, angle=180]{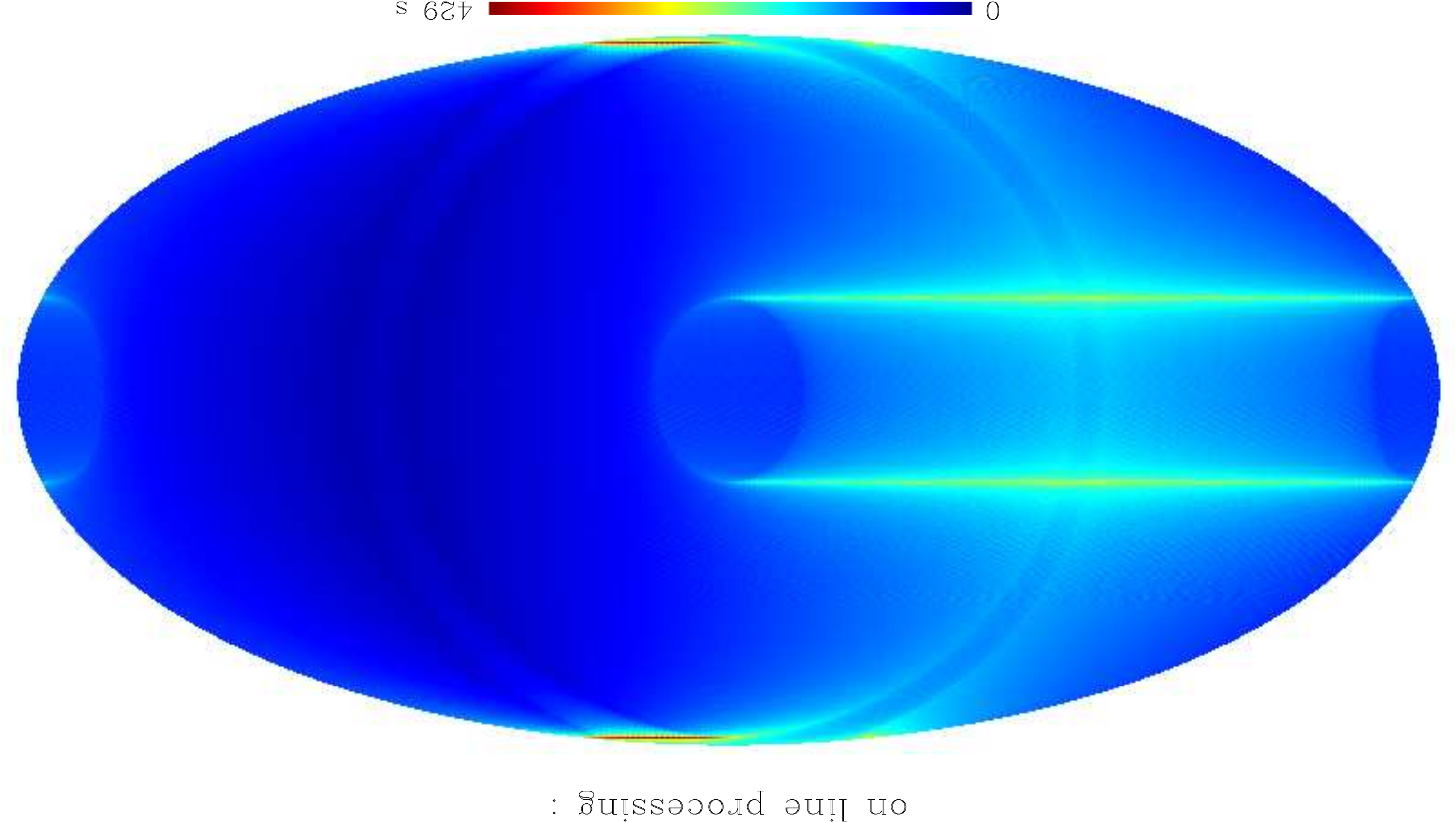}  &
\includegraphics[width=0.25\linewidth, trim=0cm 1cm 0cm 0cm, clip=true, angle=180]{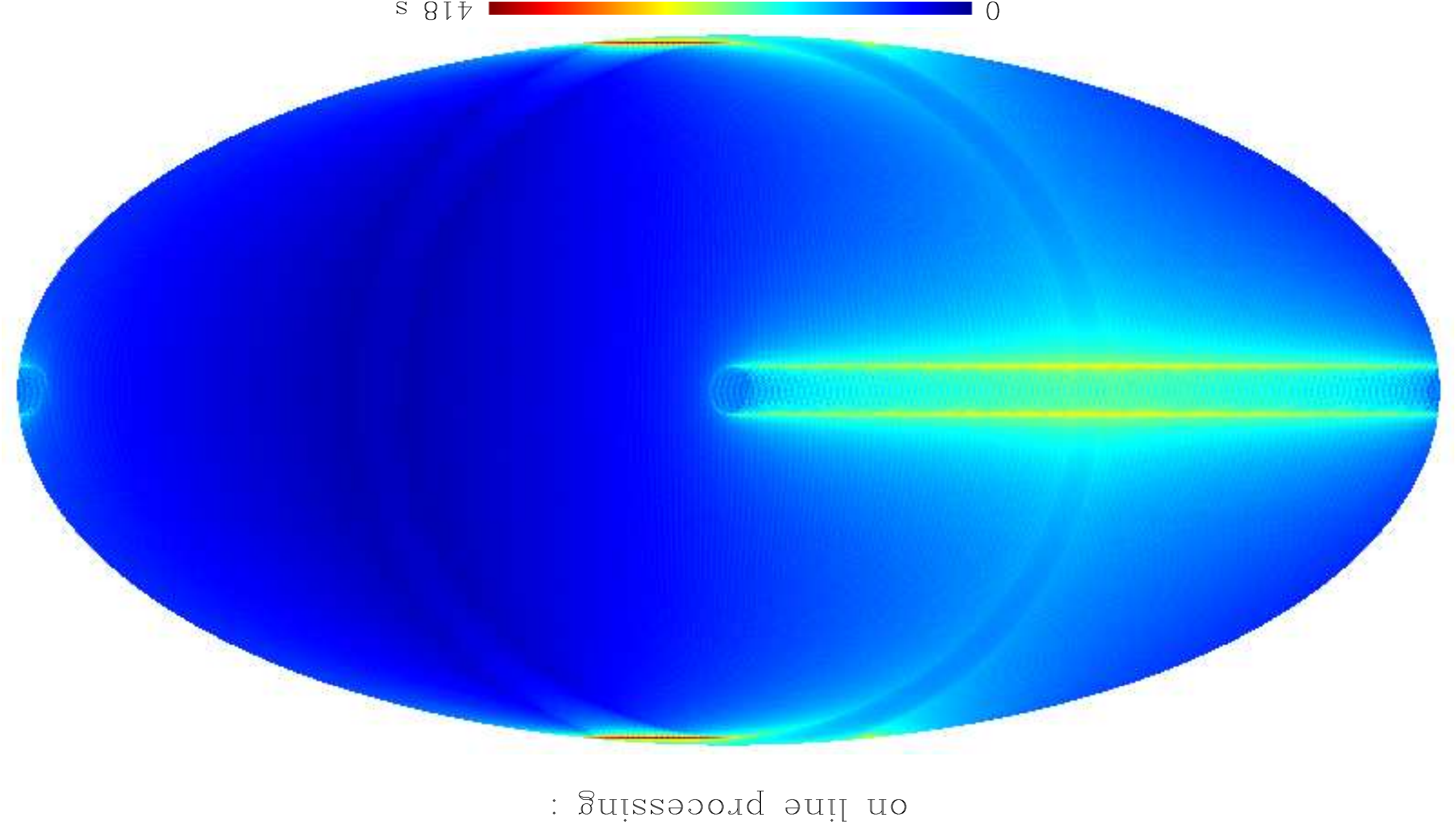}&
\includegraphics[width=0.25\linewidth, trim=0cm 1cm 0cm 0cm, clip=true, angle=180]{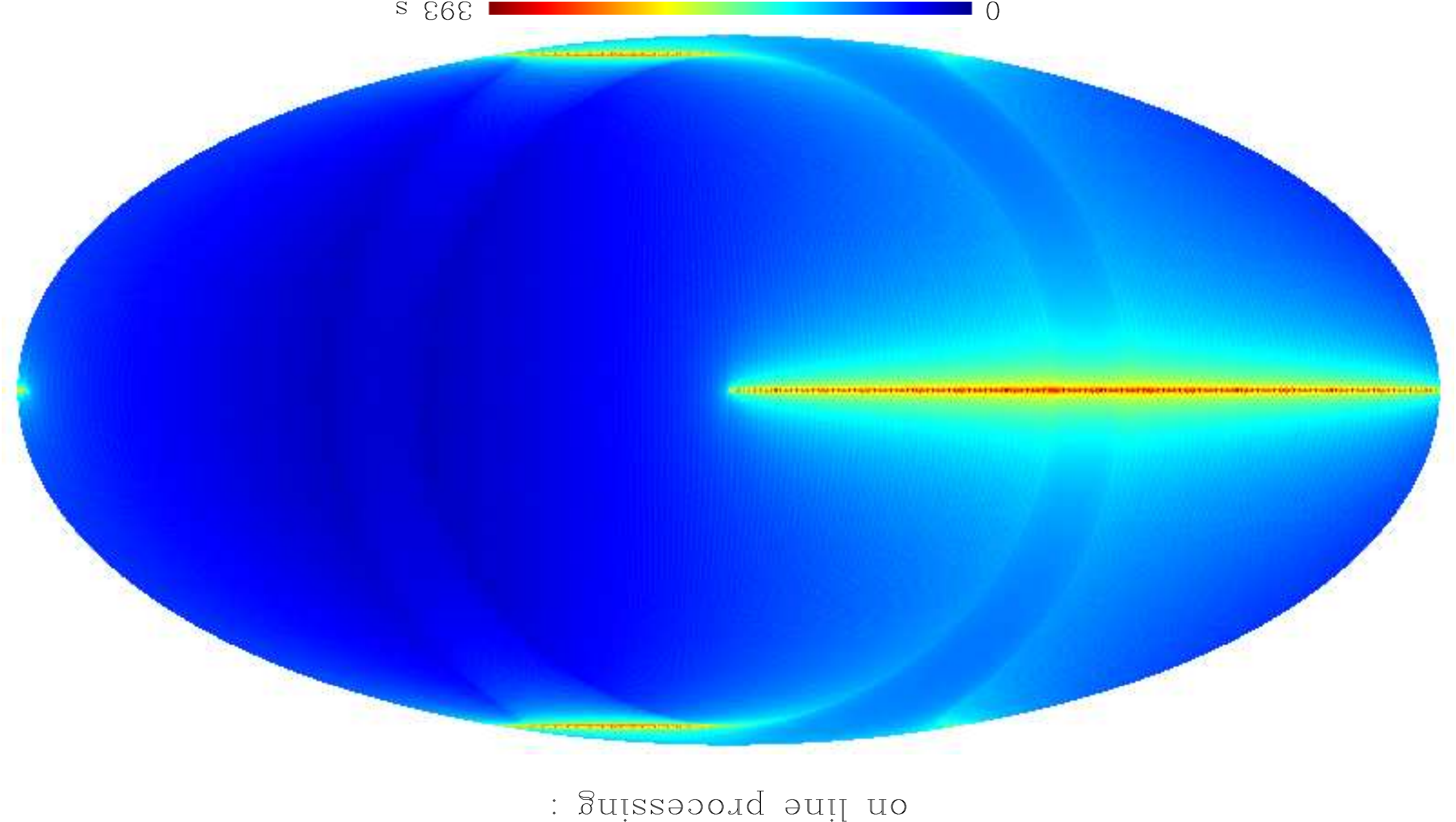}&
\includegraphics[width=0.25\linewidth, trim=0cm 1cm 0cm 0cm, clip=true, angle=180]{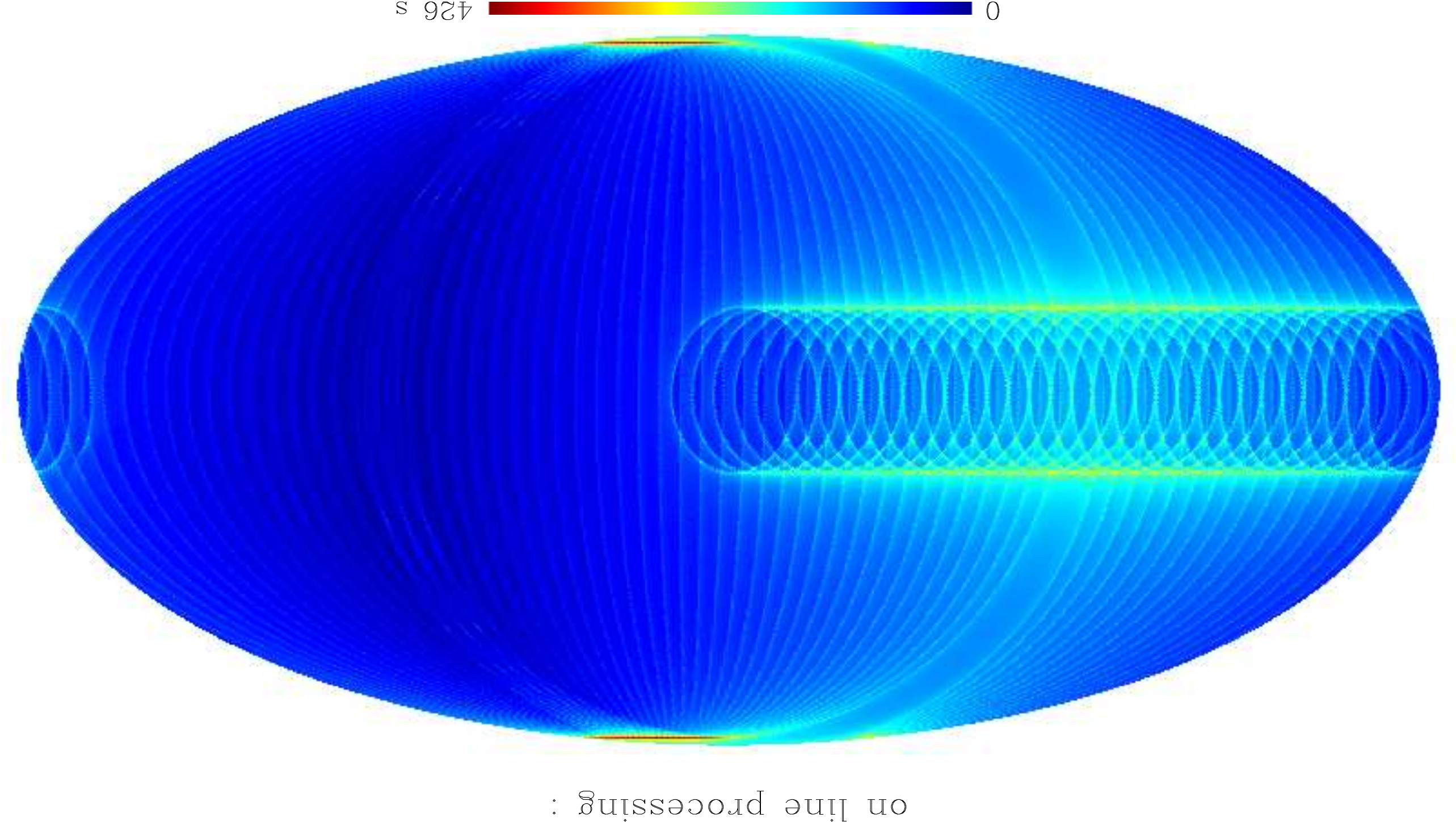}\\

\includegraphics[width=0.25\linewidth, trim=0cm 1cm 0cm 0cm, clip=true, angle=180]{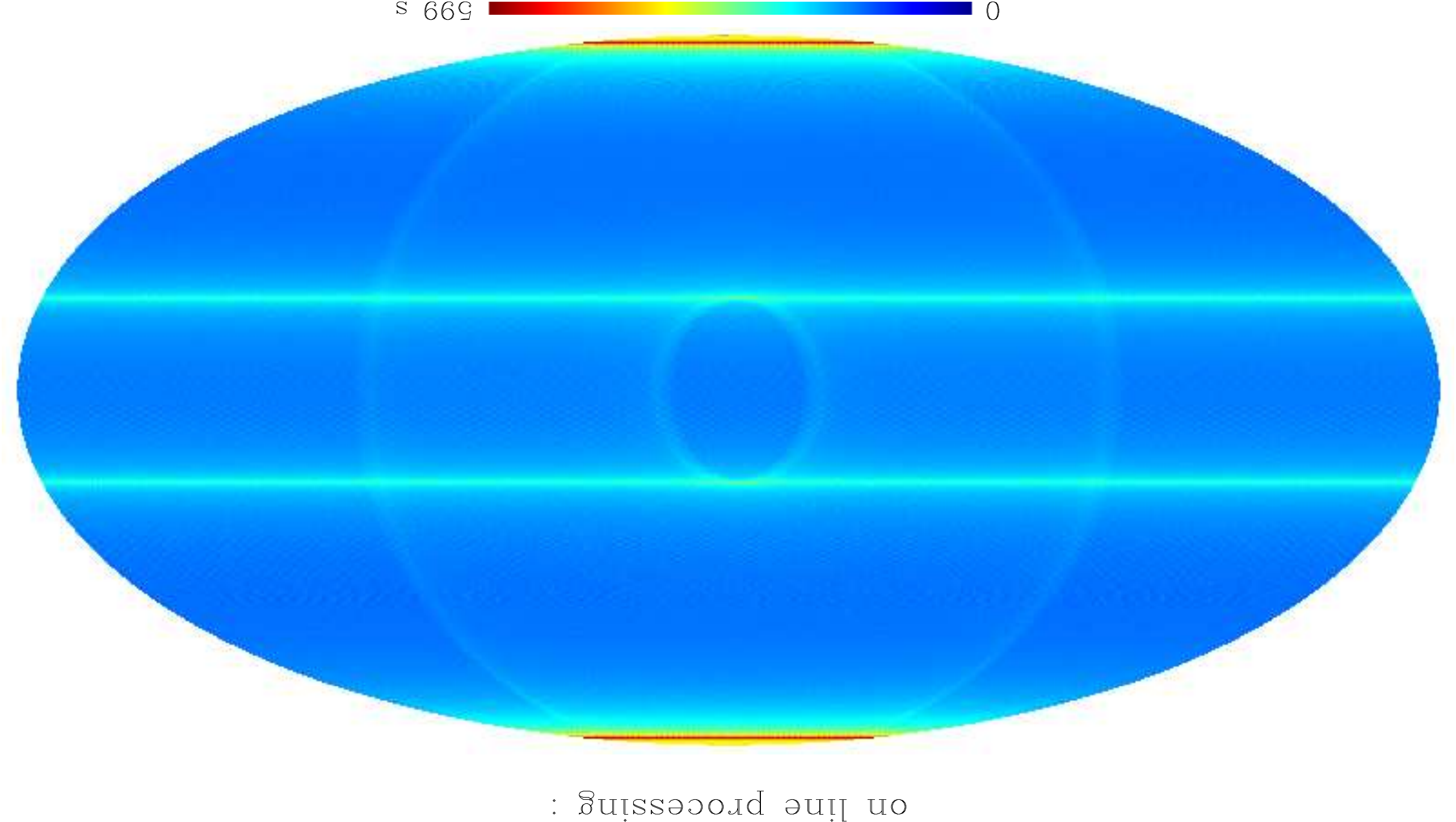}  &
\includegraphics[width=0.25\linewidth, trim=0cm 1cm 0cm 0cm, clip=true, angle=180]{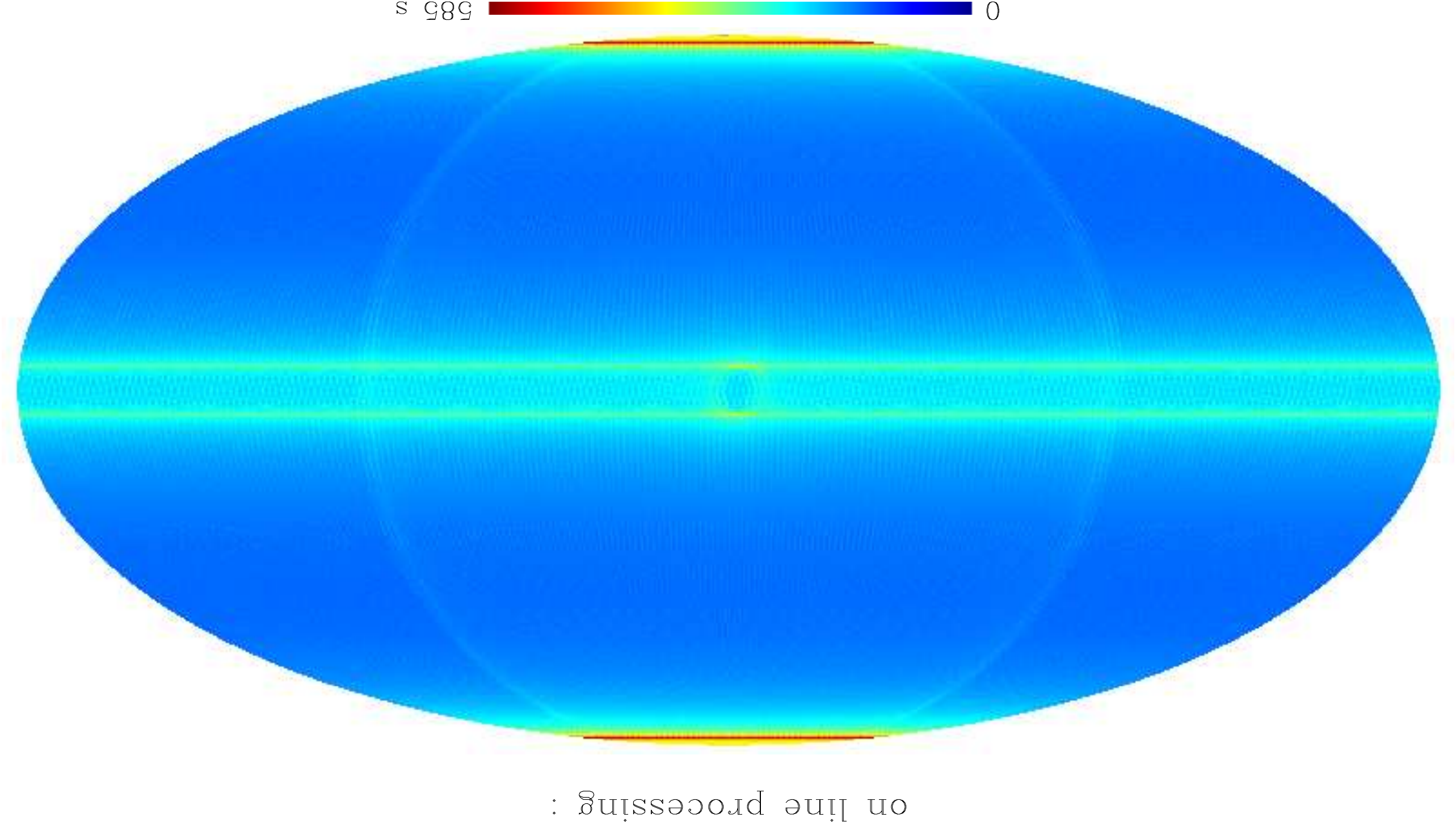}&
\includegraphics[width=0.25\linewidth, trim=0cm 1cm 0cm 0cm, clip=true, angle=180]{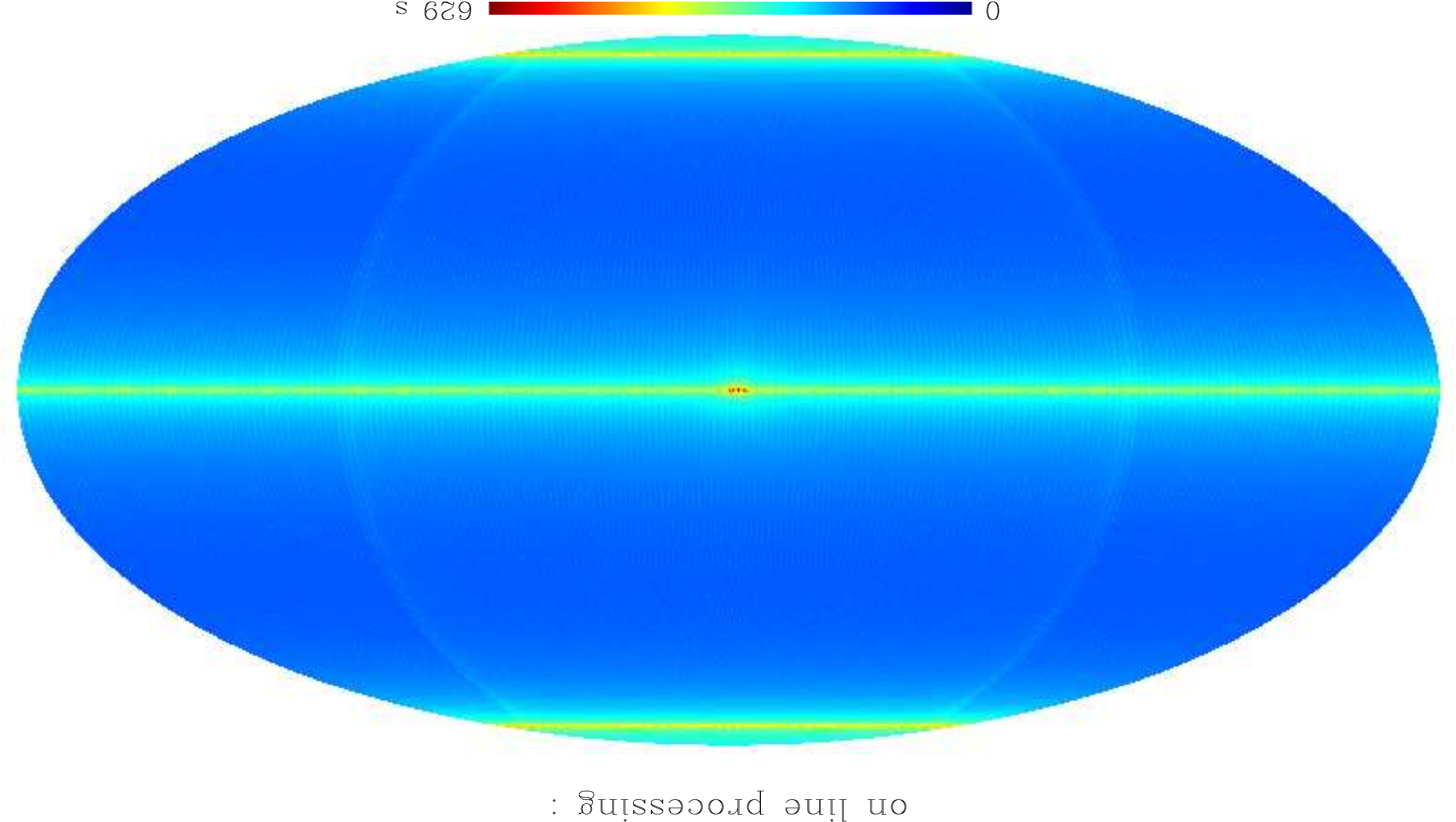}&
\includegraphics[width=0.25\linewidth, trim=0cm 1cm 0cm 0cm, clip=true, angle=180]{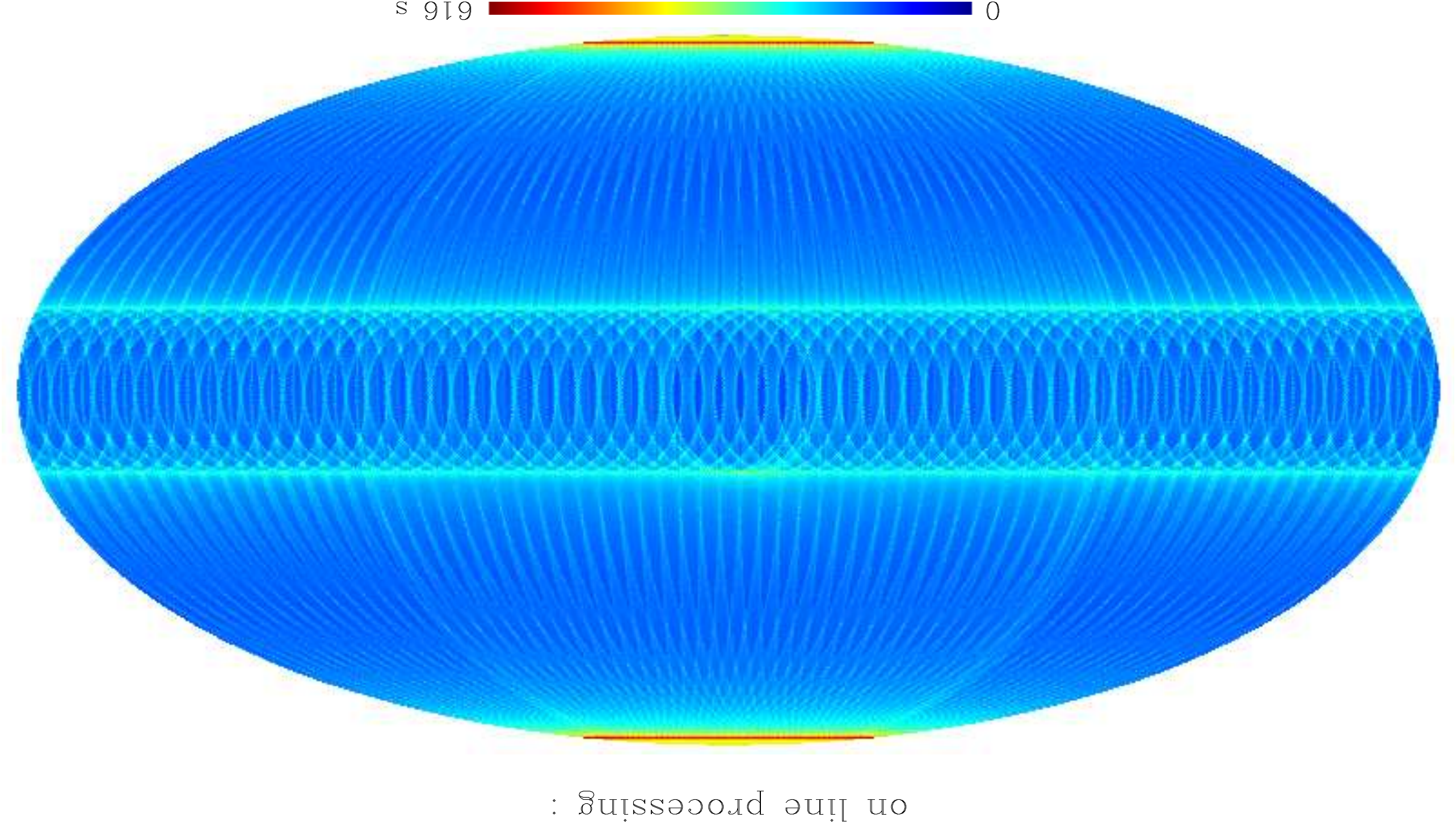}\\
\end{tabular}
\caption{The time integration of each scan listed in
  Table~\ref{tab:core_scan_params} on a map of $N_{\rm side} =$
  128. The rows correspond to the scan running for 10 days, 1 month, 2
  months, 6 months and 1 year, from top to bottom respectively.}
\label{fig:core_scans_time_int}
\end{center}
\end{figure*}

%% file: sections/fig/time_investigation2.tex
% !TEX TS-program = compile

\begin{figure*}
\begin{center}
\begin{tabular}{c c c c}
~\\
~\\
Large & Medium & Medium ($\beta=50\degree$) & Small \\
%&h2 Map&\\0.25
\includegraphics[width=0.25\linewidth, trim=0cm 1cm 0cm 0cm, clip=true, angle=180]{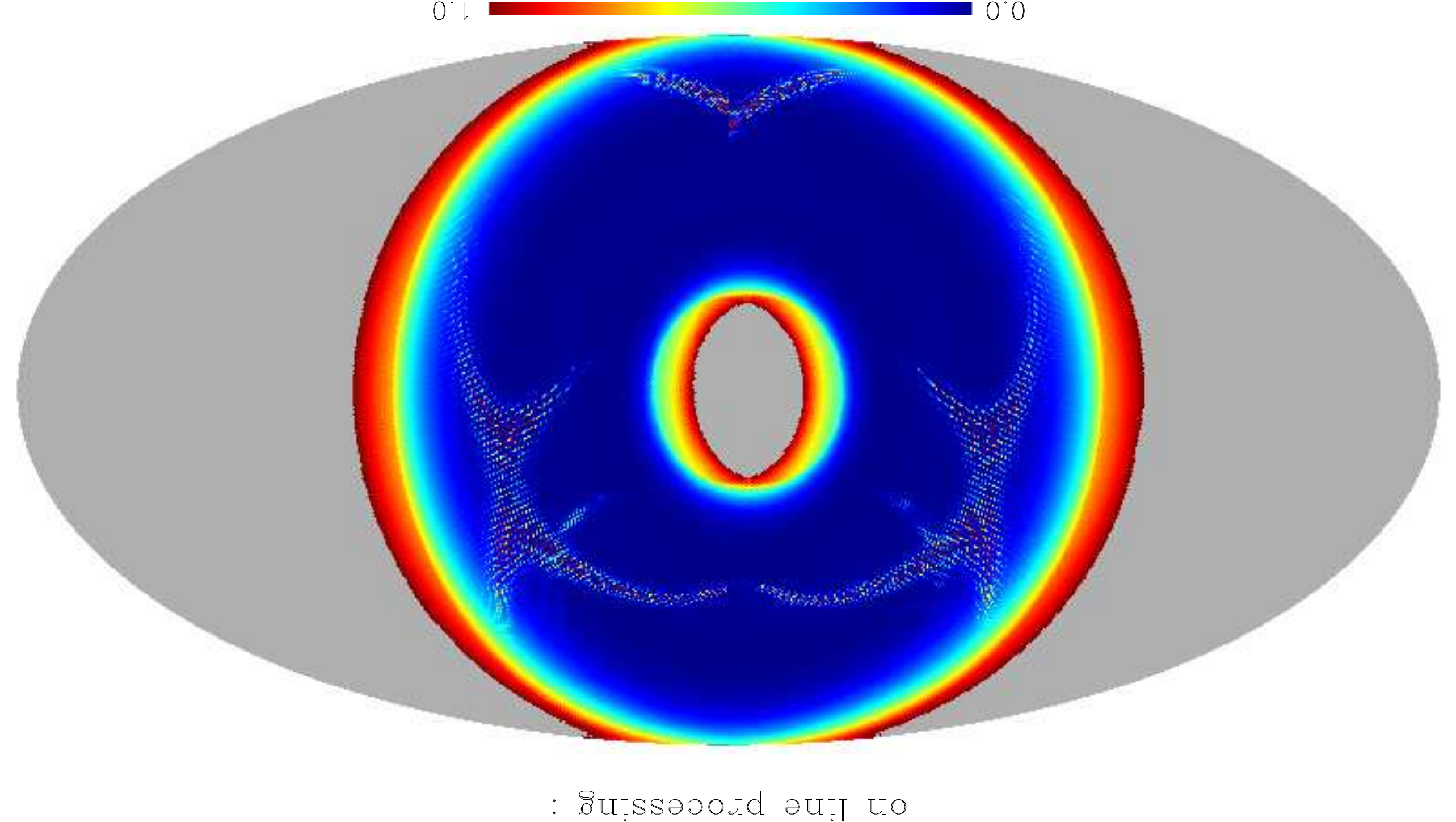}  &
\includegraphics[width=0.25\linewidth, trim=0cm 1cm 0cm 0cm, clip=true, angle=180]{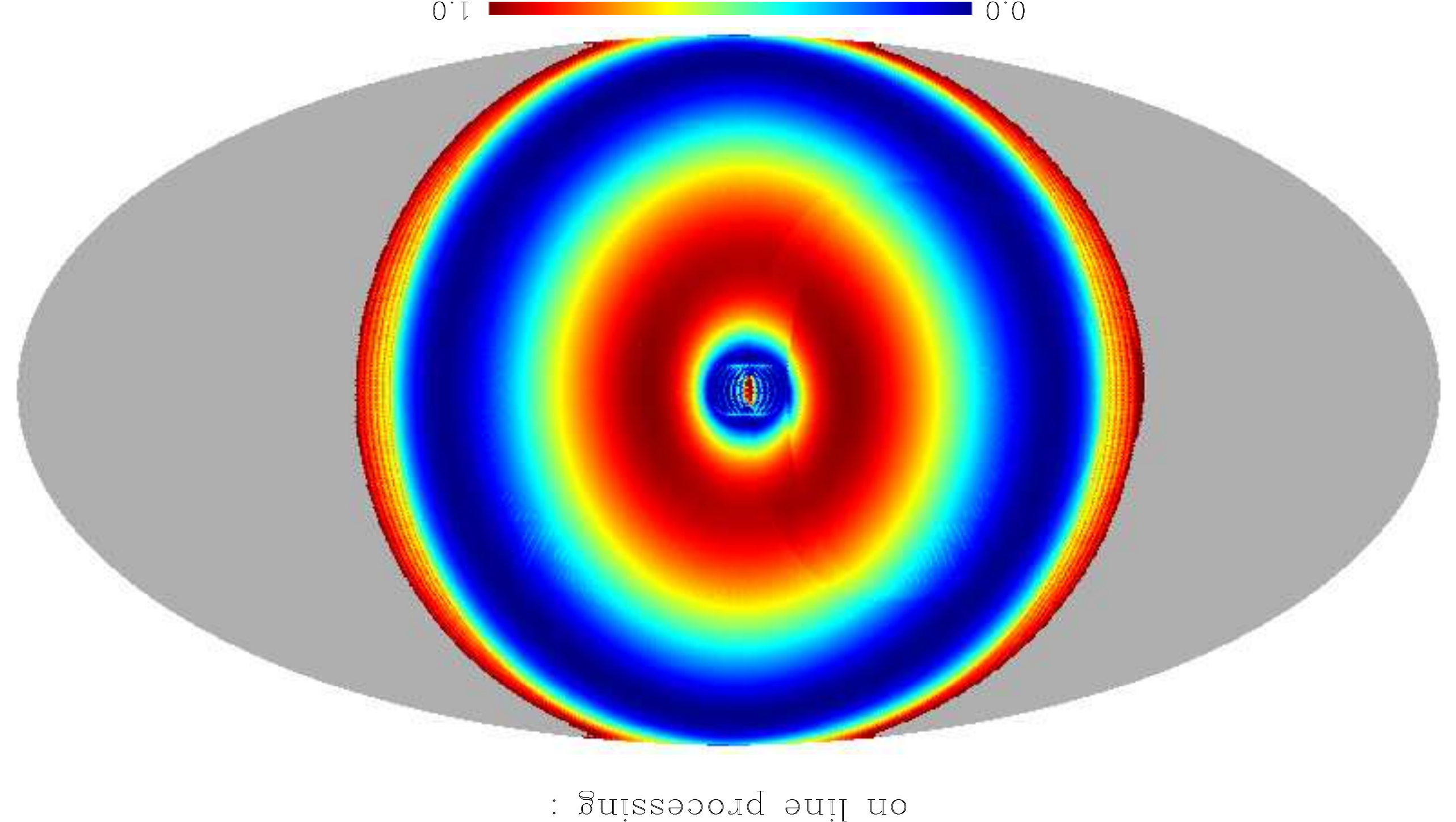}&
\includegraphics[width=0.25\linewidth, trim=0cm 1cm 0cm 0cm, clip=true, angle=180]{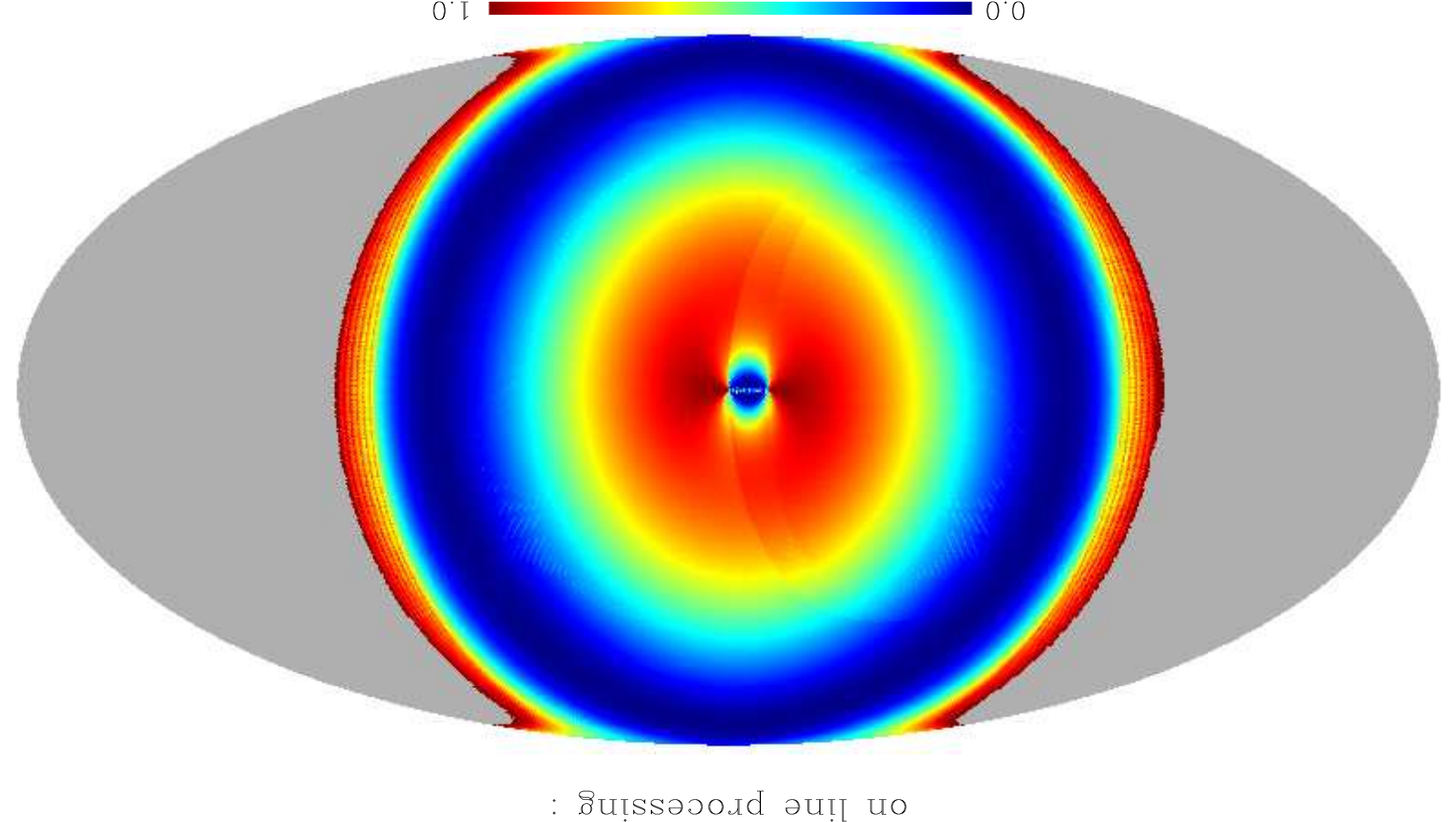}&
\includegraphics[width=0.25\linewidth, trim=0cm 1cm 0cm 0cm, clip=true, angle=180]{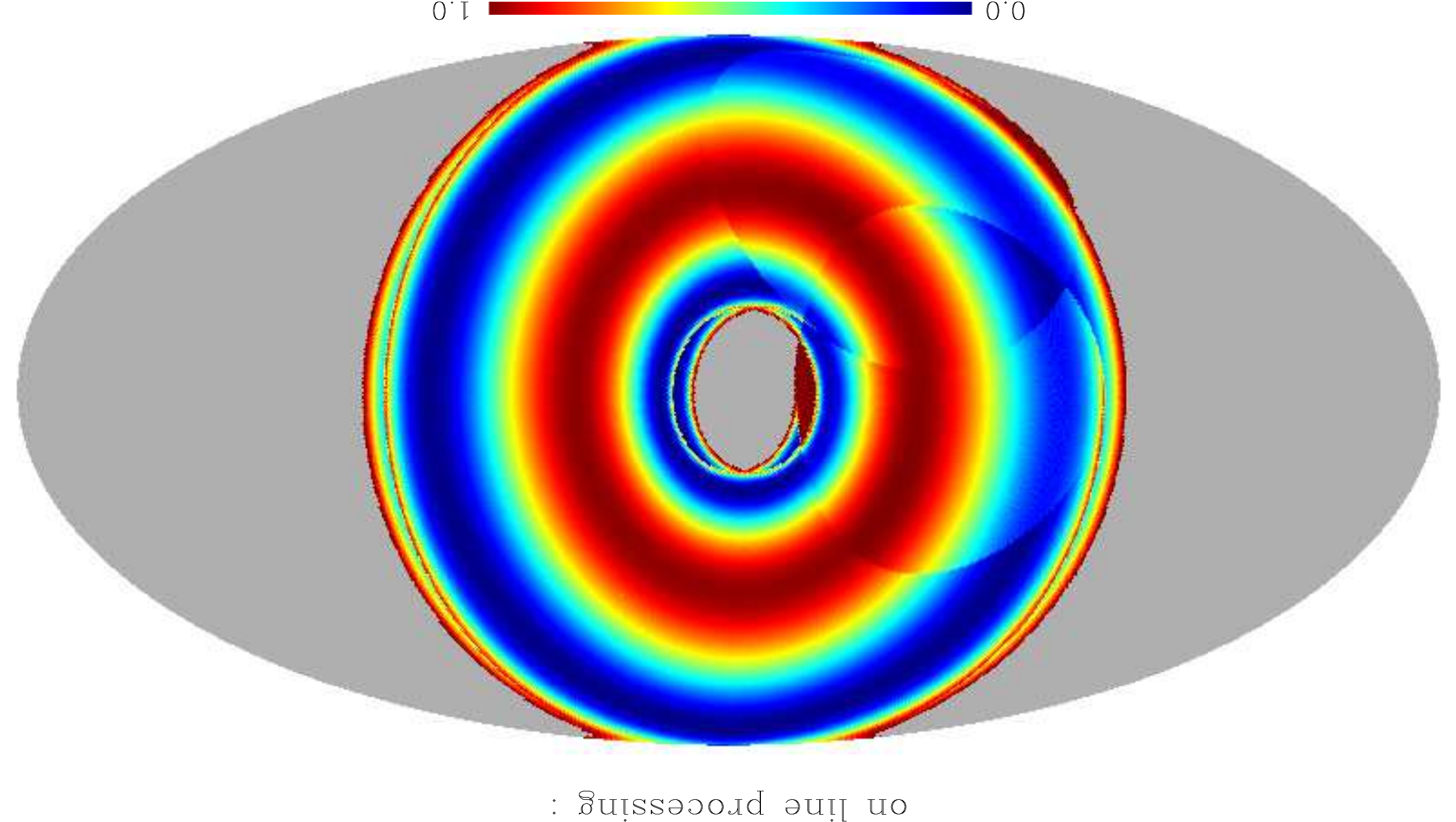}\\

\includegraphics[width=0.25\linewidth, trim=0cm 1cm 0cm 0cm, clip=true, angle=180]{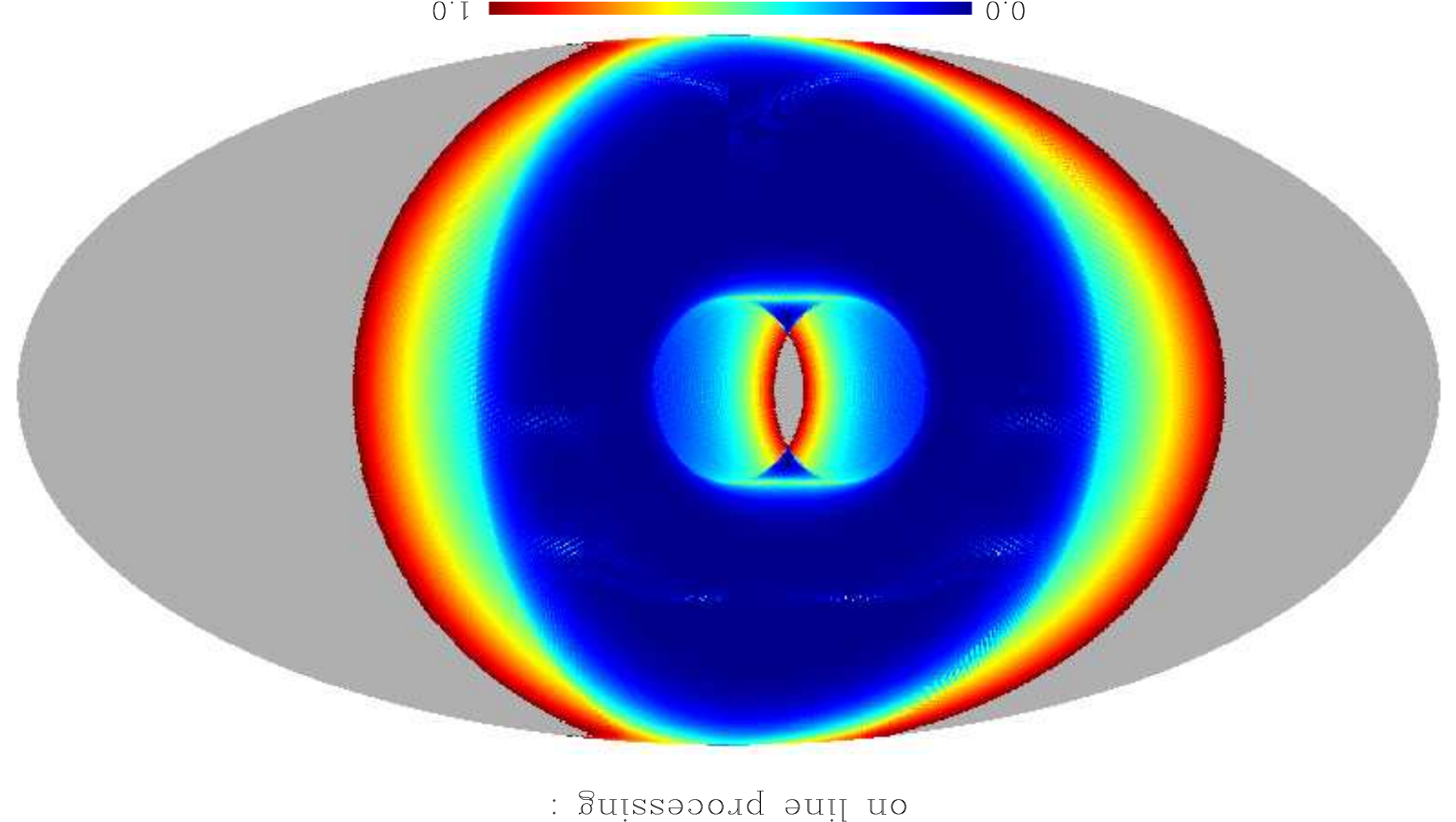}  &
\includegraphics[width=0.25\linewidth, trim=0cm 1cm 0cm 0cm, clip=true, angle=180]{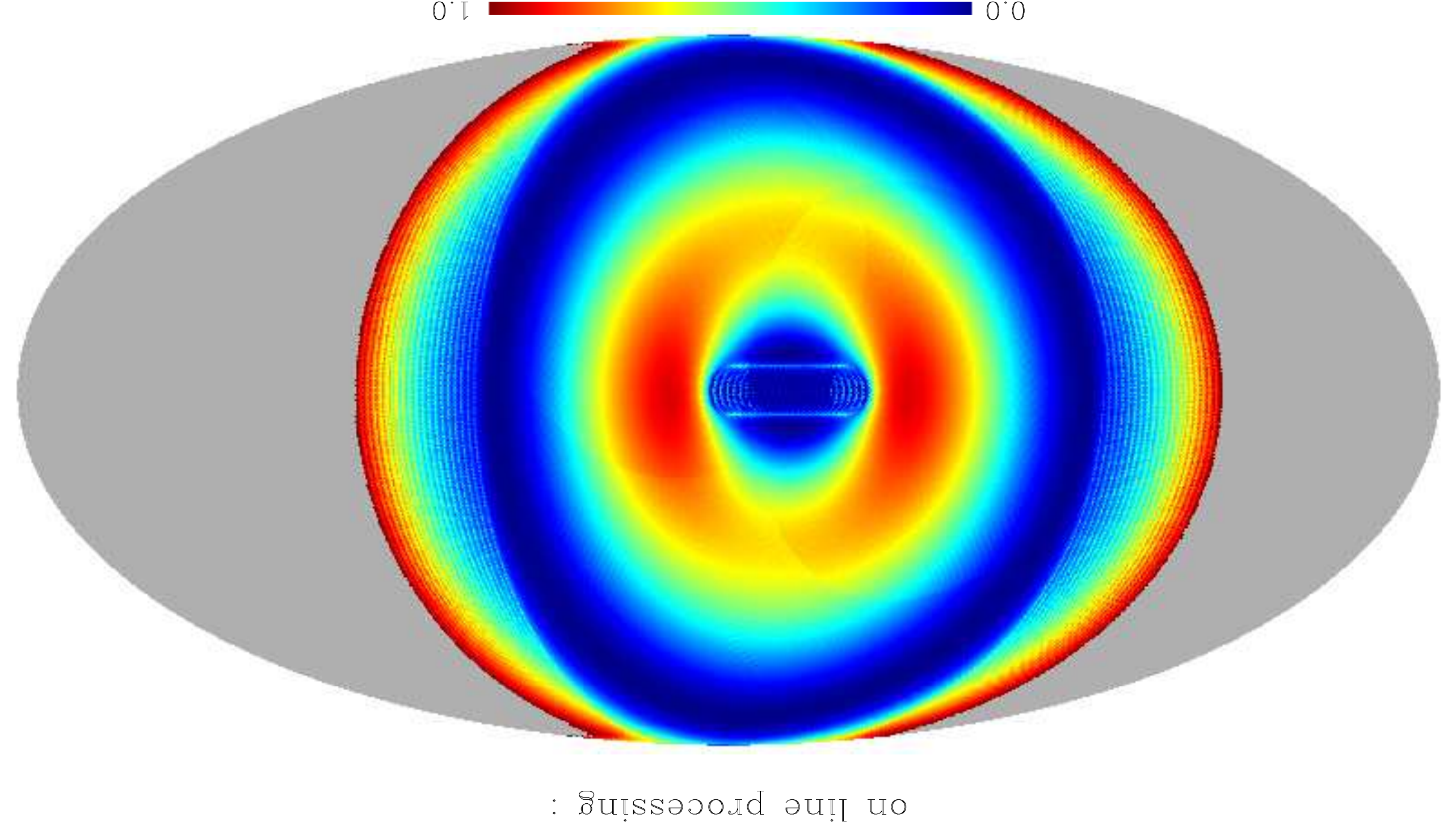}&
\includegraphics[width=0.25\linewidth, trim=0cm 1cm 0cm 0cm, clip=true, angle=180]{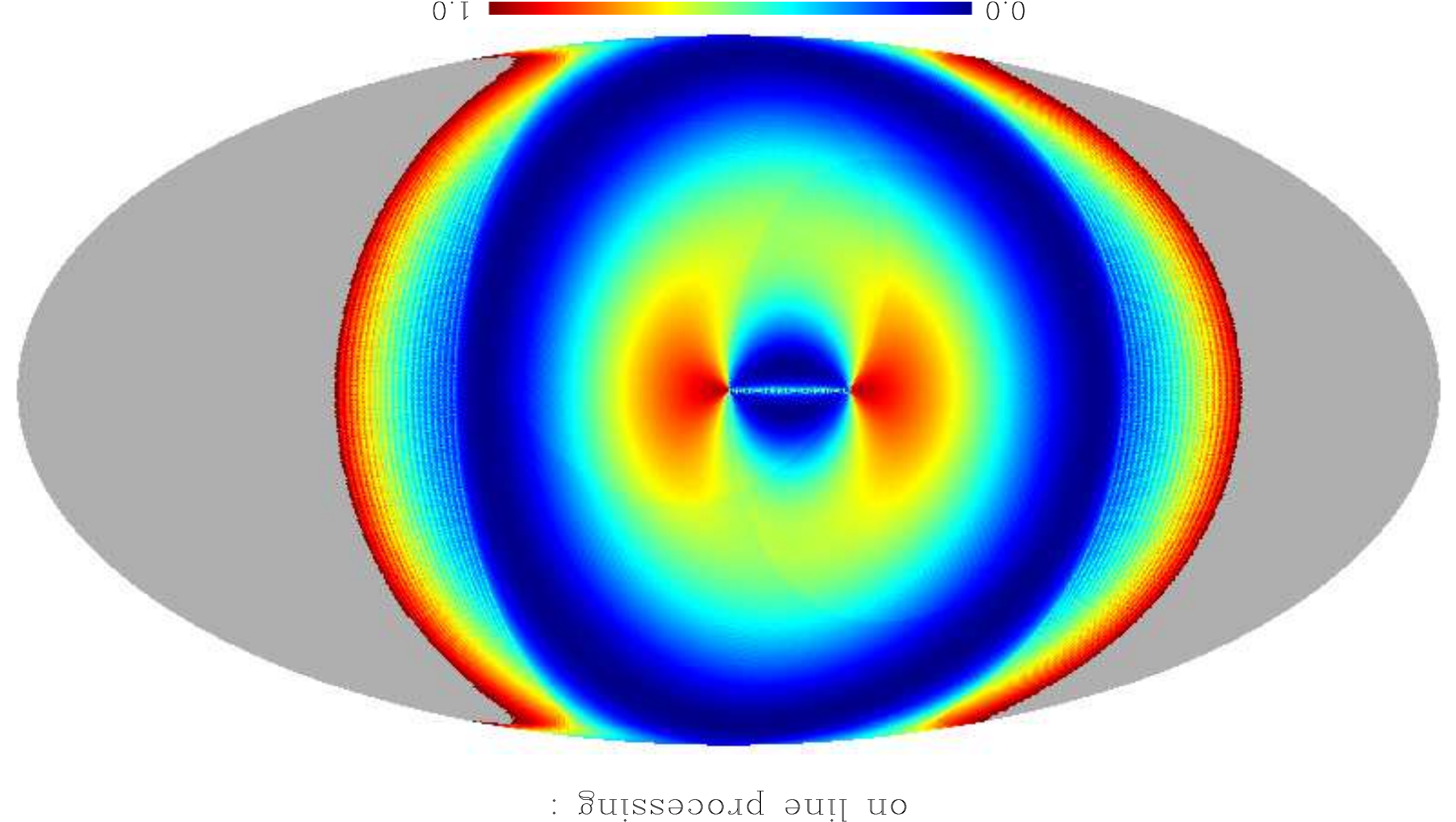}&
\includegraphics[width=0.25\linewidth, trim=0cm 1cm 0cm 0cm, clip=true, angle=180]{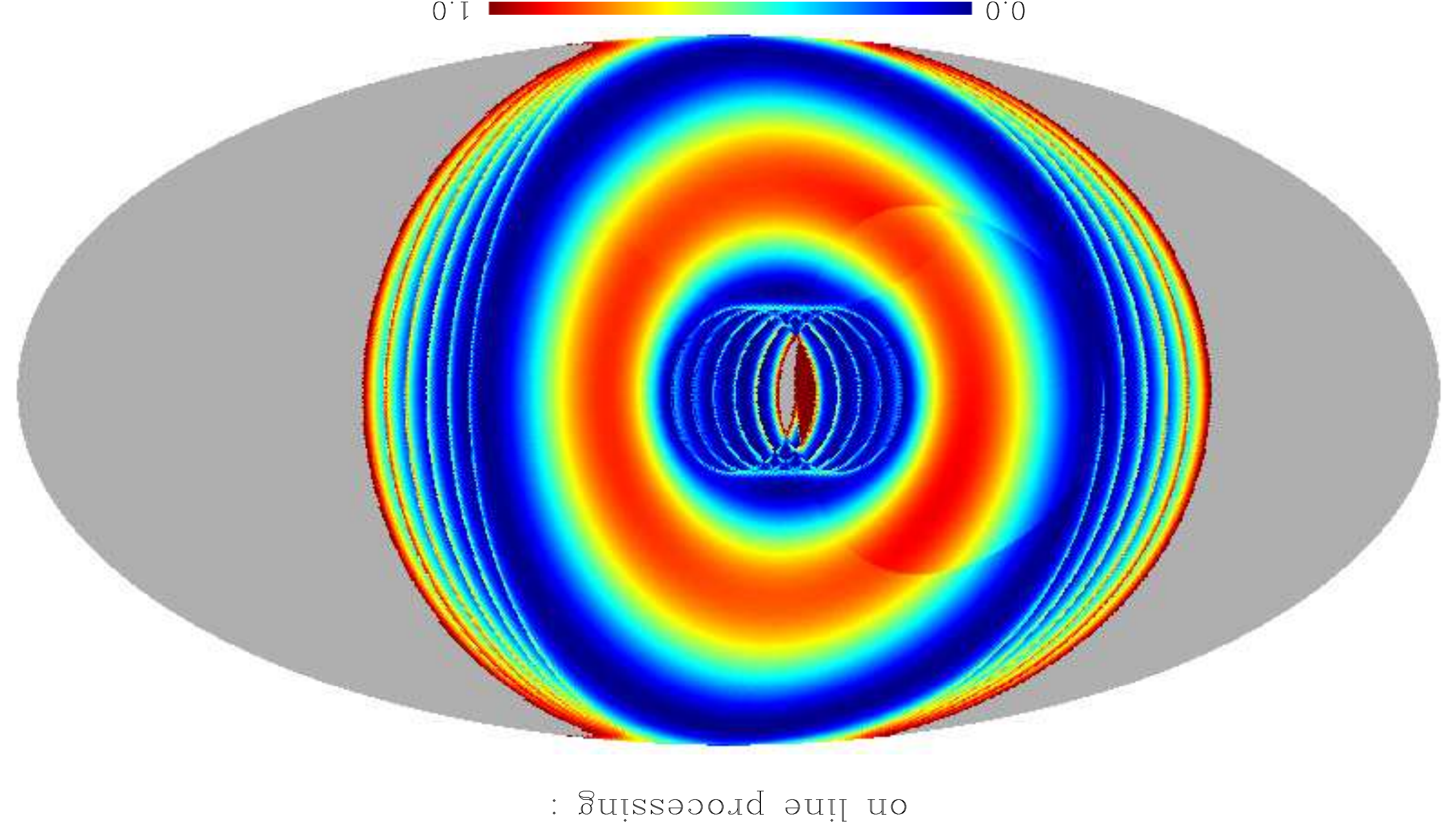}\\

\includegraphics[width=0.25\linewidth, trim=0cm 1cm 0cm 0cm, clip=true, angle=180]{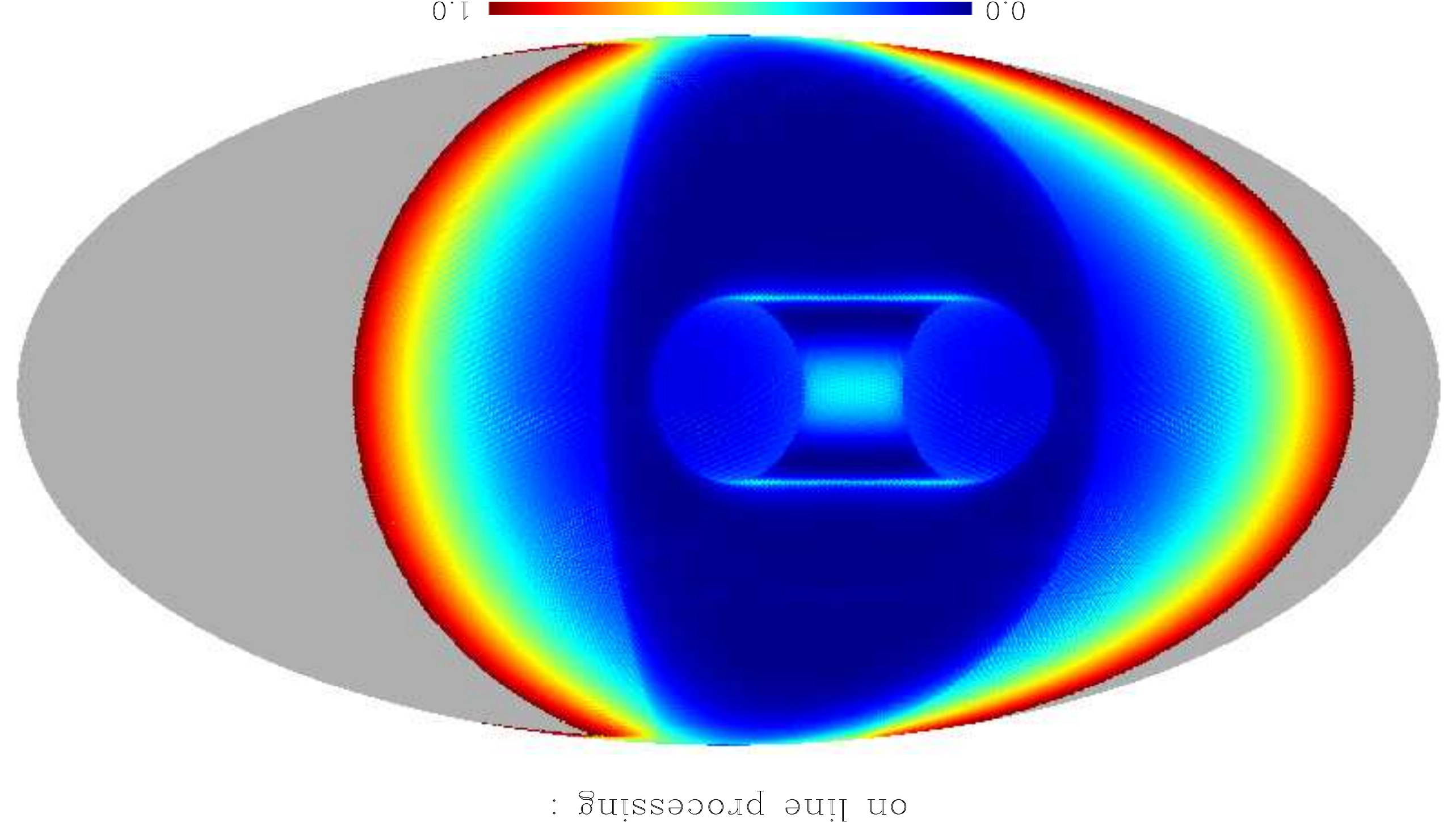}  &
\includegraphics[width=0.25\linewidth, trim=0cm 1cm 0cm 0cm, clip=true, angle=180]{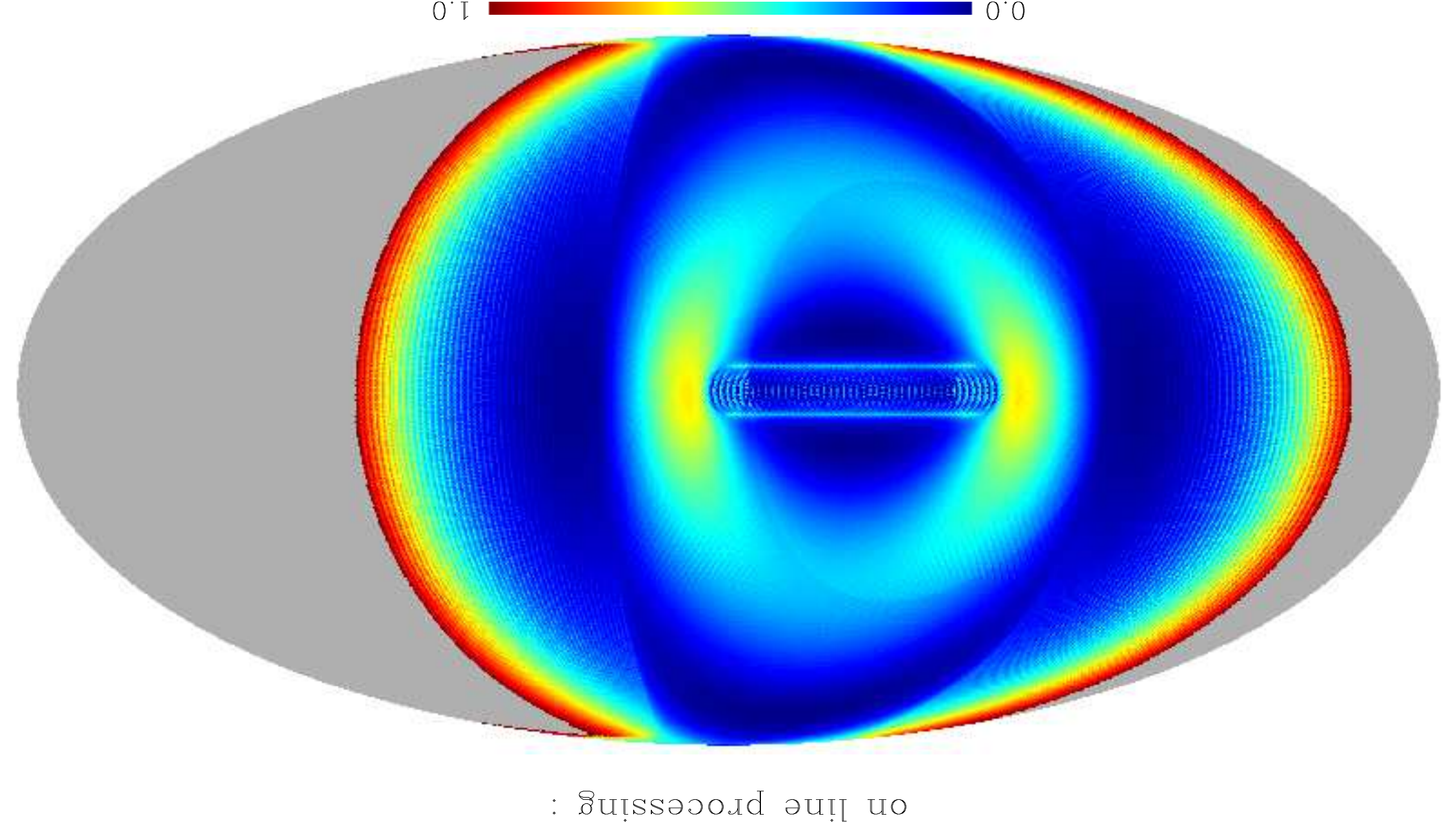}&
\includegraphics[width=0.25\linewidth, trim=0cm 1cm 0cm 0cm, clip=true, angle=180]{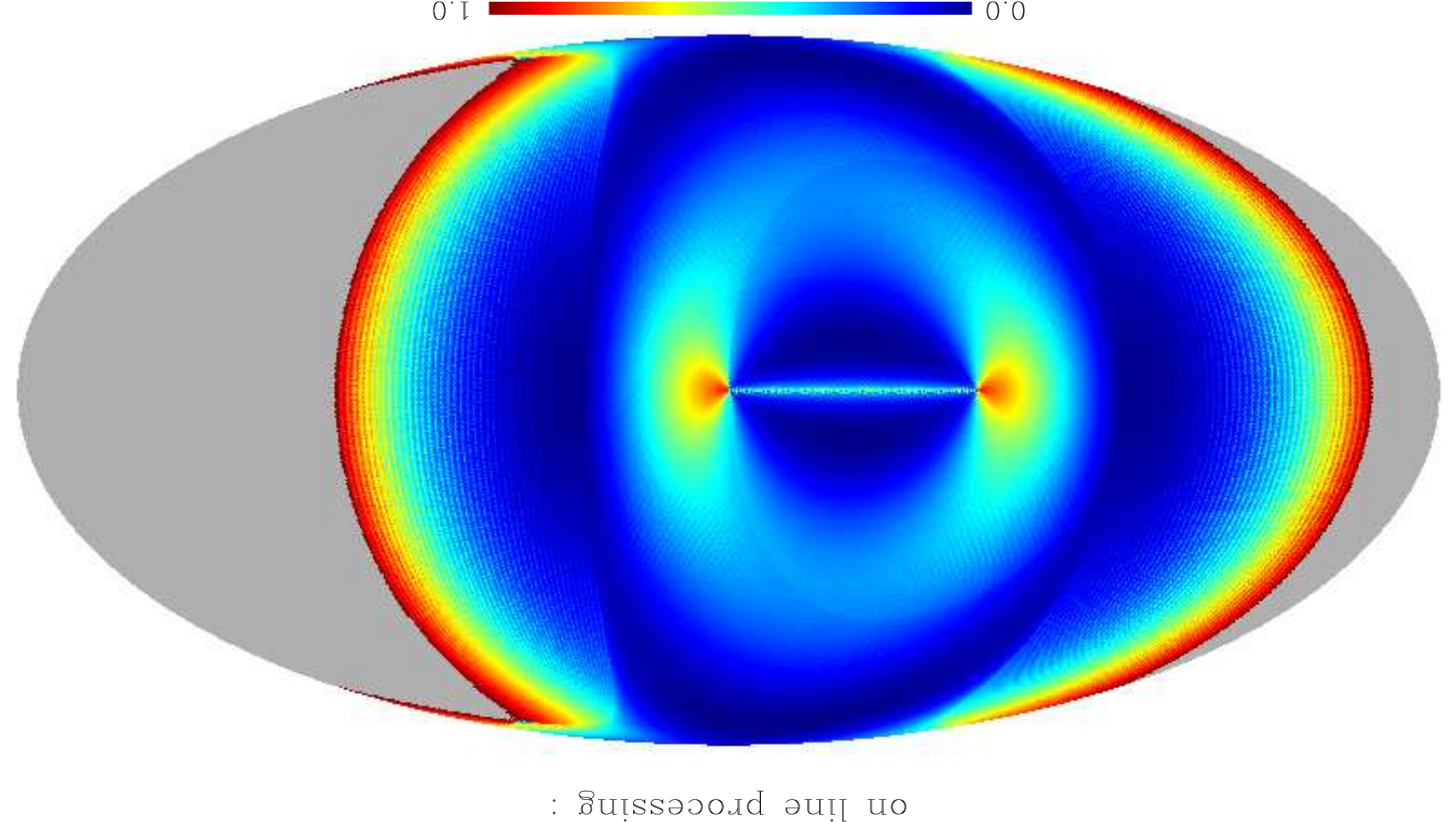}&
\includegraphics[width=0.25\linewidth, trim=0cm 1cm 0cm 0cm, clip=true, angle=180]{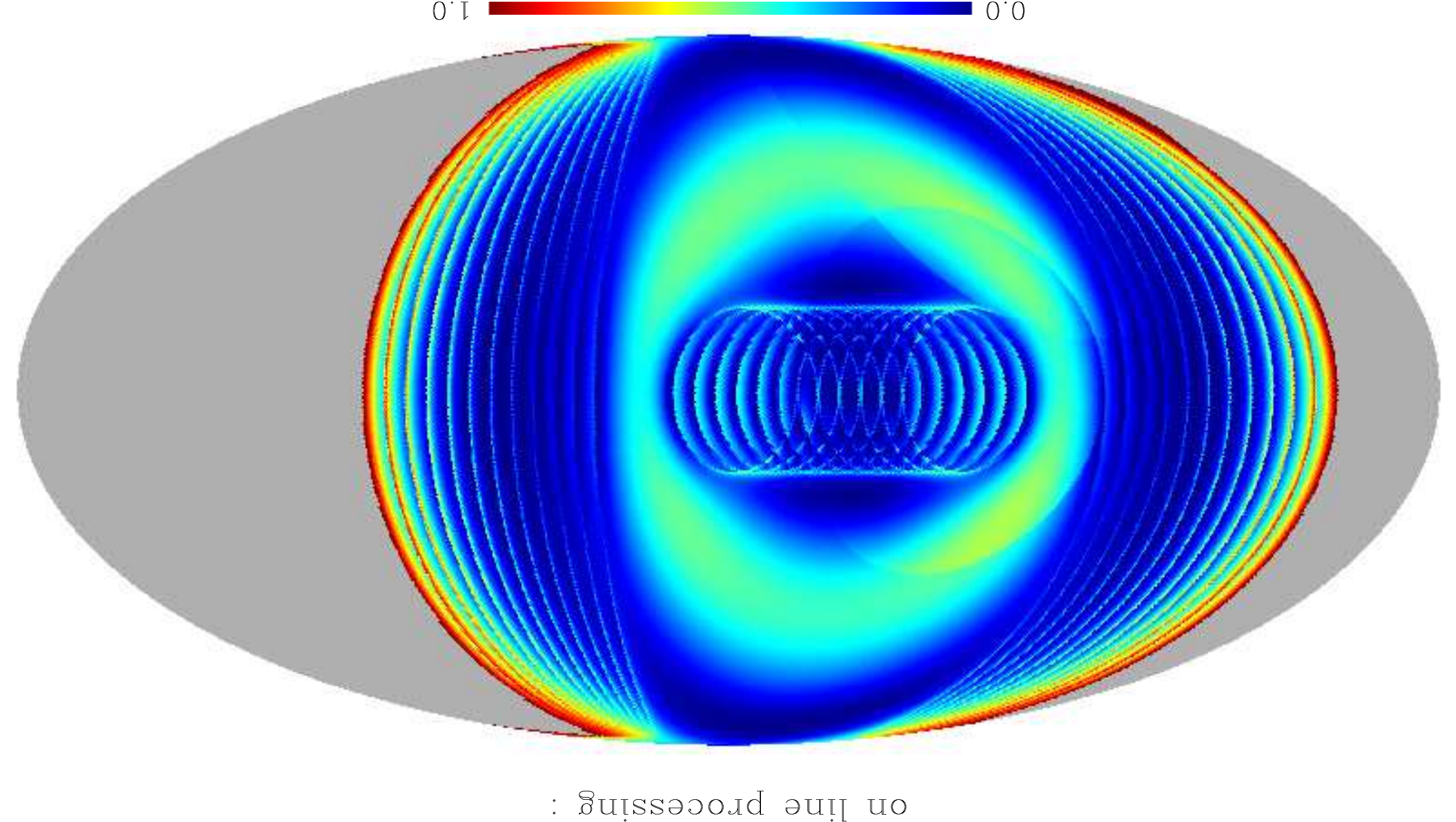}\\

\includegraphics[width=0.25\linewidth, trim=0cm 1cm 0cm 0cm, clip=true, angle=180]{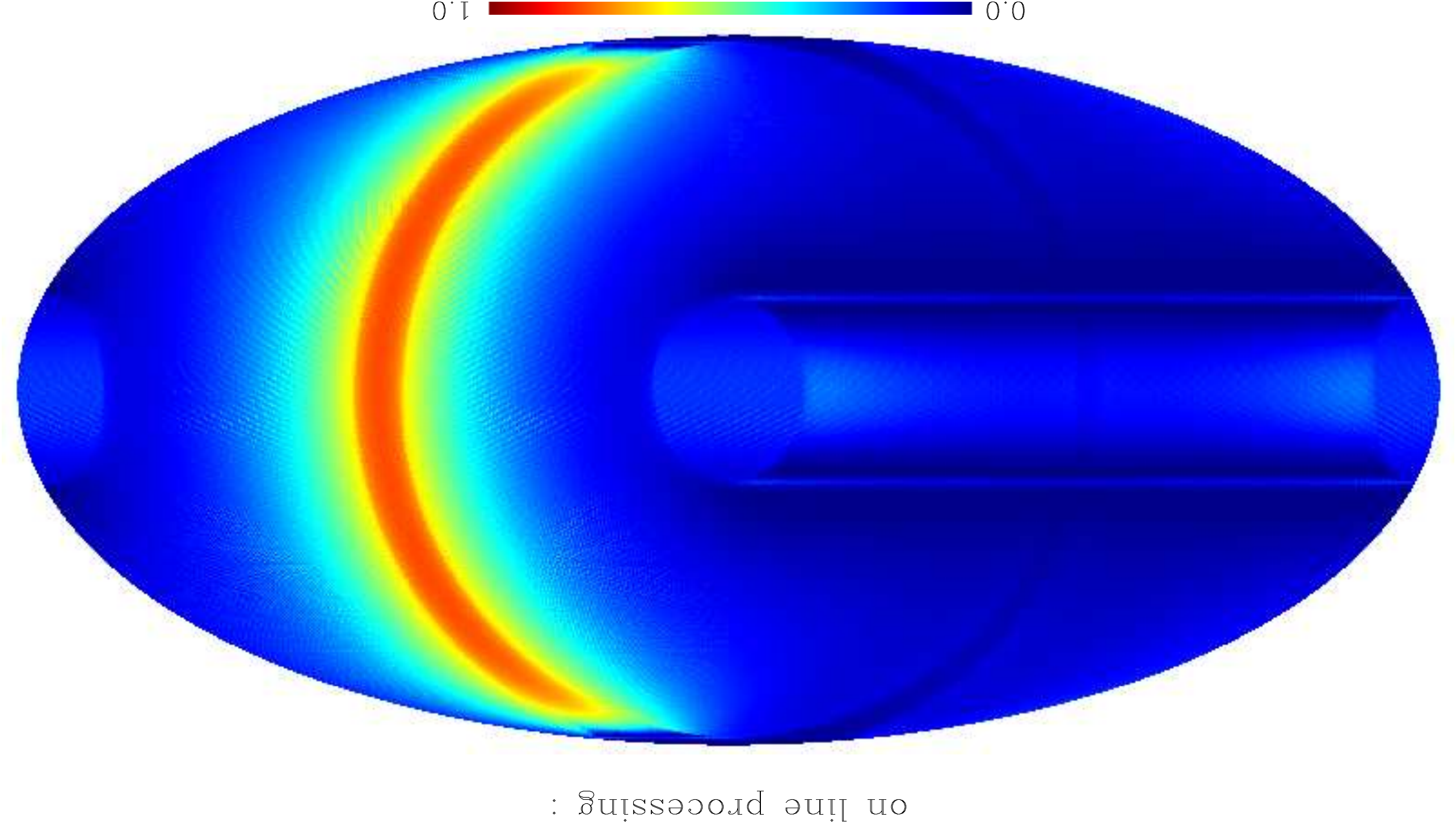}  &
\includegraphics[width=0.25\linewidth, trim=0cm 1cm 0cm 0cm, clip=true, angle=180]{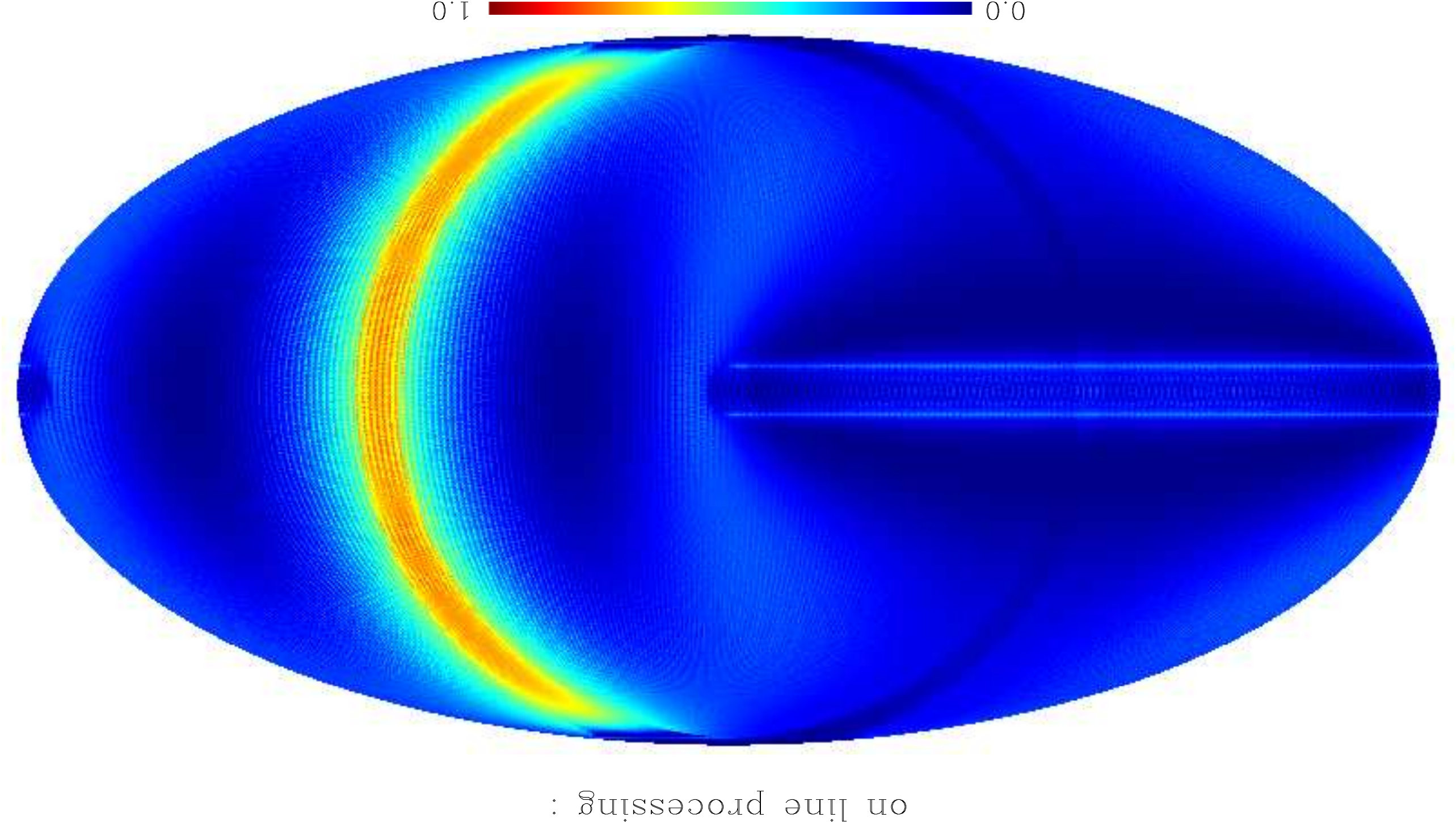}&
\includegraphics[width=0.25\linewidth, trim=0cm 1cm 0cm 0cm, clip=true, angle=180]{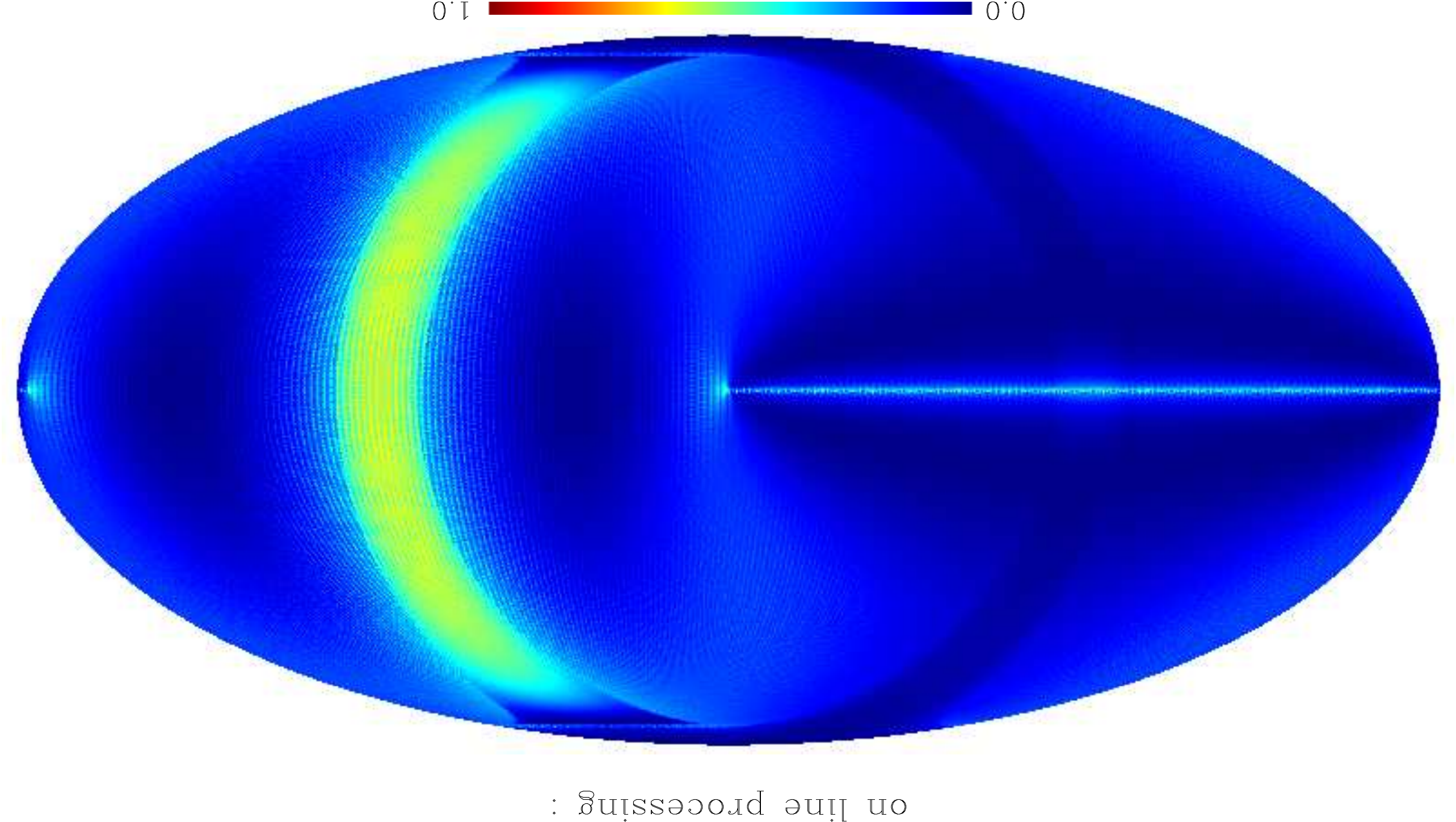}&
\includegraphics[width=0.25\linewidth, trim=0cm 1cm 0cm 0cm, clip=true, angle=180]{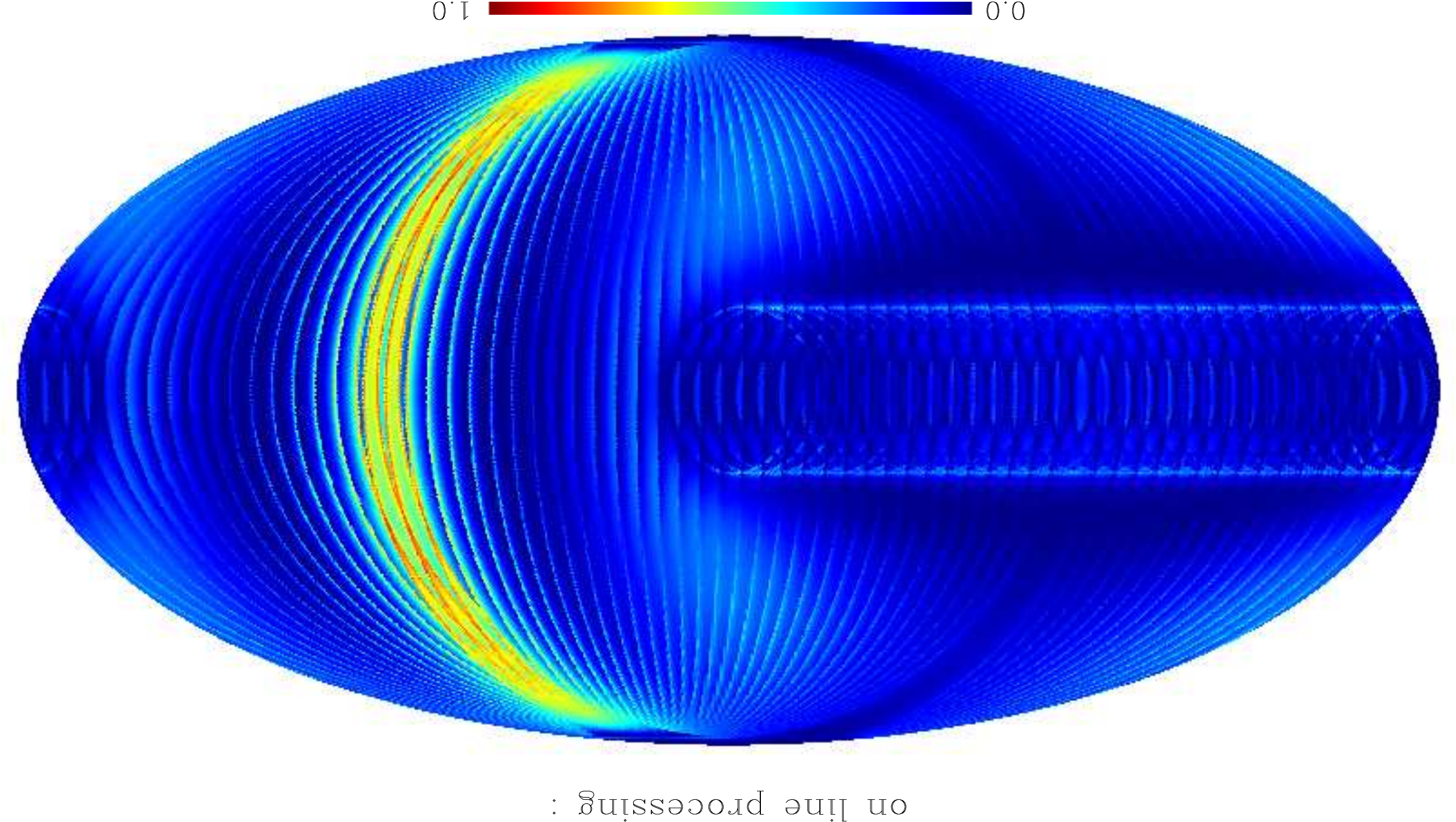}\\

\includegraphics[width=0.25\linewidth, trim=0cm 1cm 0cm 0cm, clip=true, angle=180]{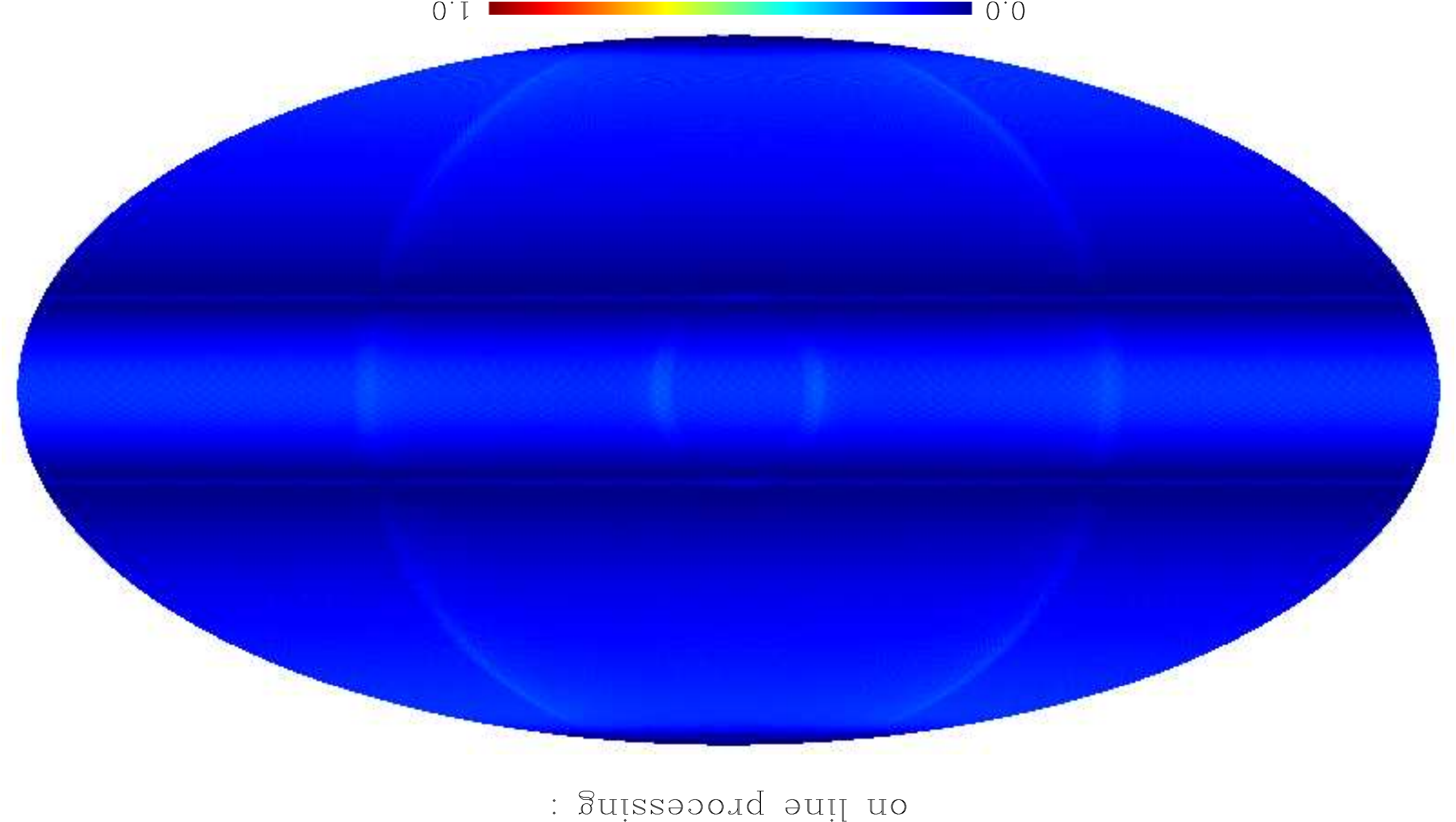}  &
\includegraphics[width=0.25\linewidth, trim=0cm 1cm 0cm 0cm, clip=true, angle=180]{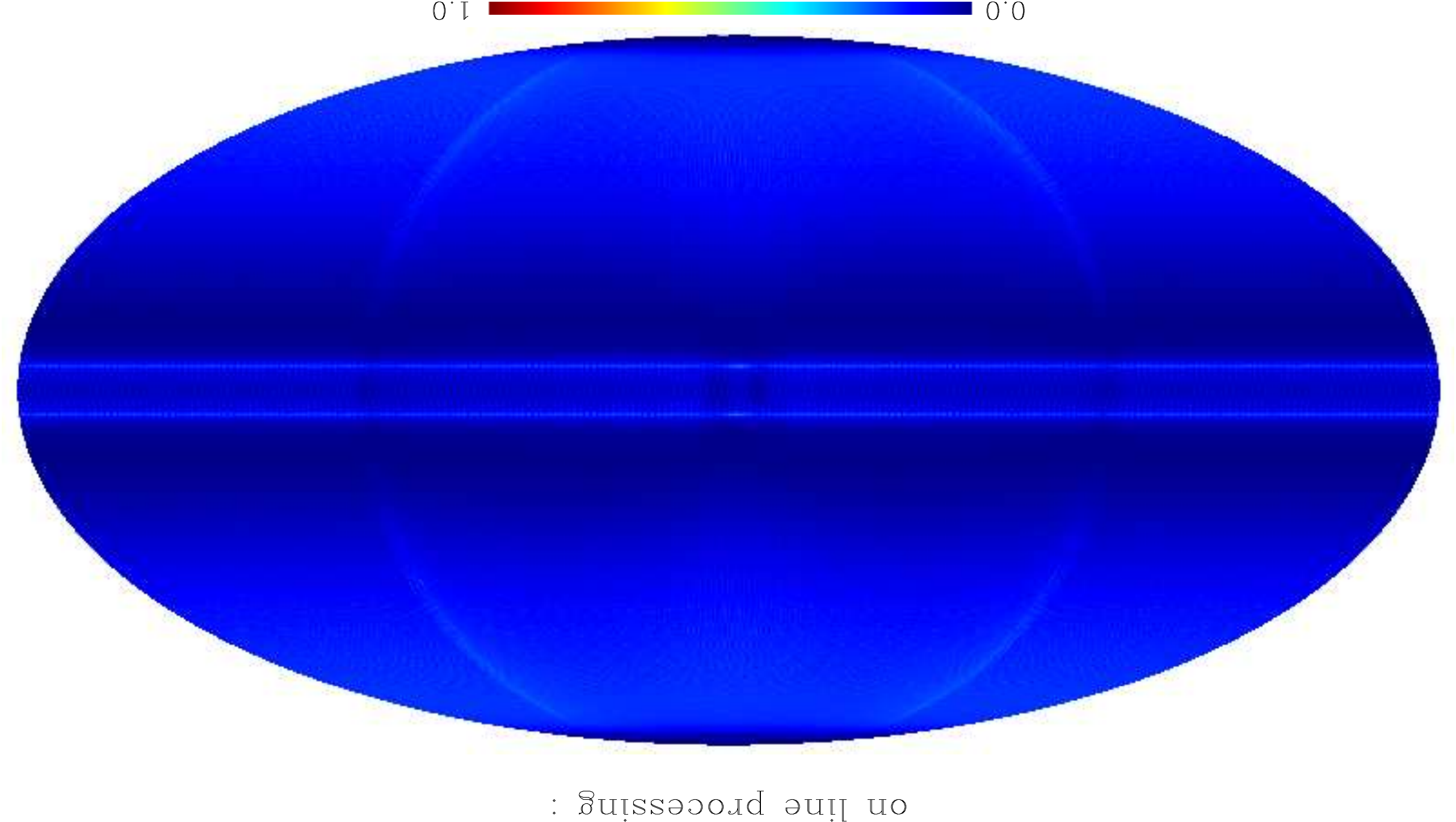}&
\includegraphics[width=0.25\linewidth, trim=0cm 1cm 0cm 0cm, clip=true, angle=180]{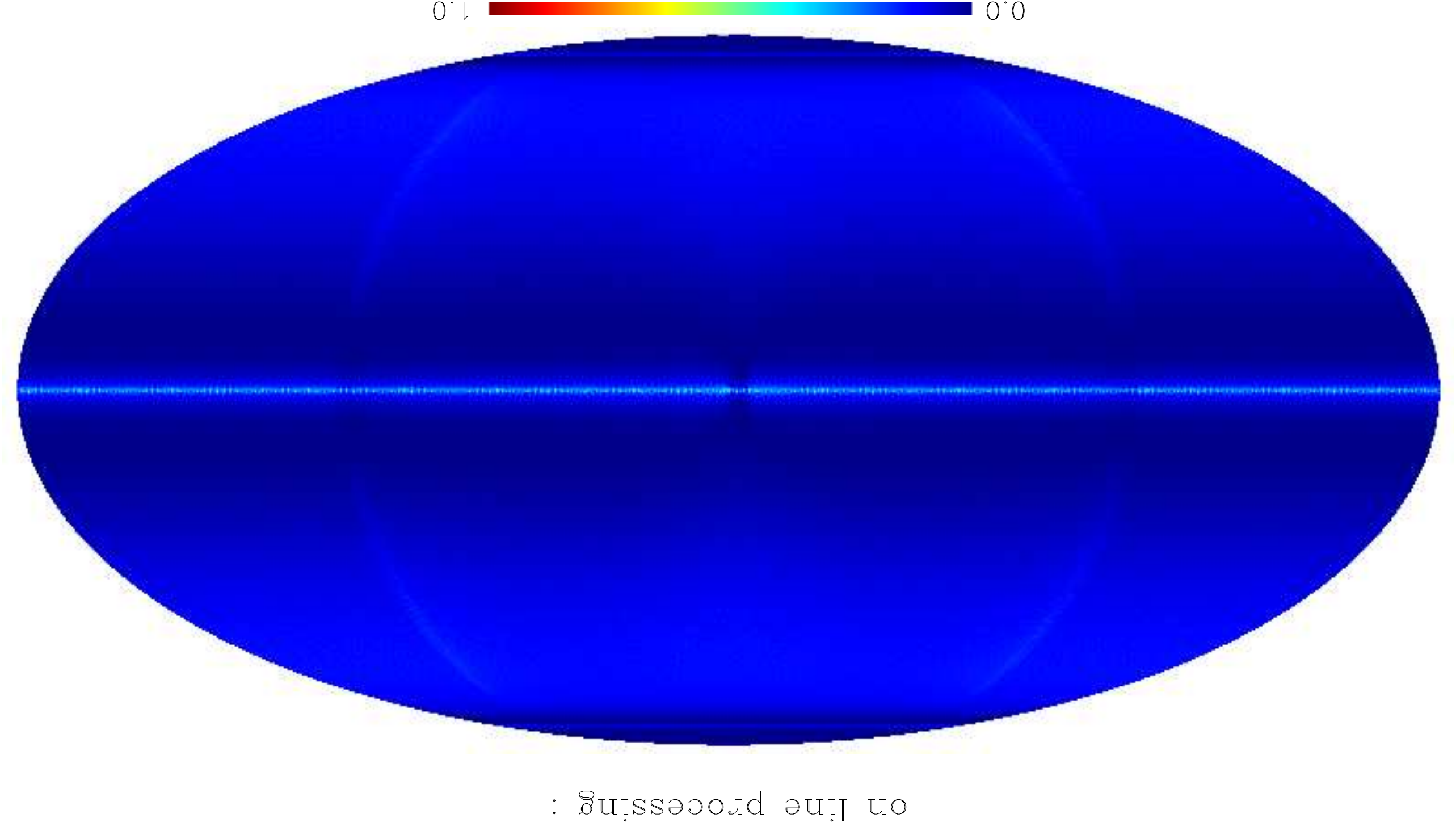}&
\includegraphics[width=0.25\linewidth, trim=0cm 1cm 0cm 0cm, clip=true, angle=180]{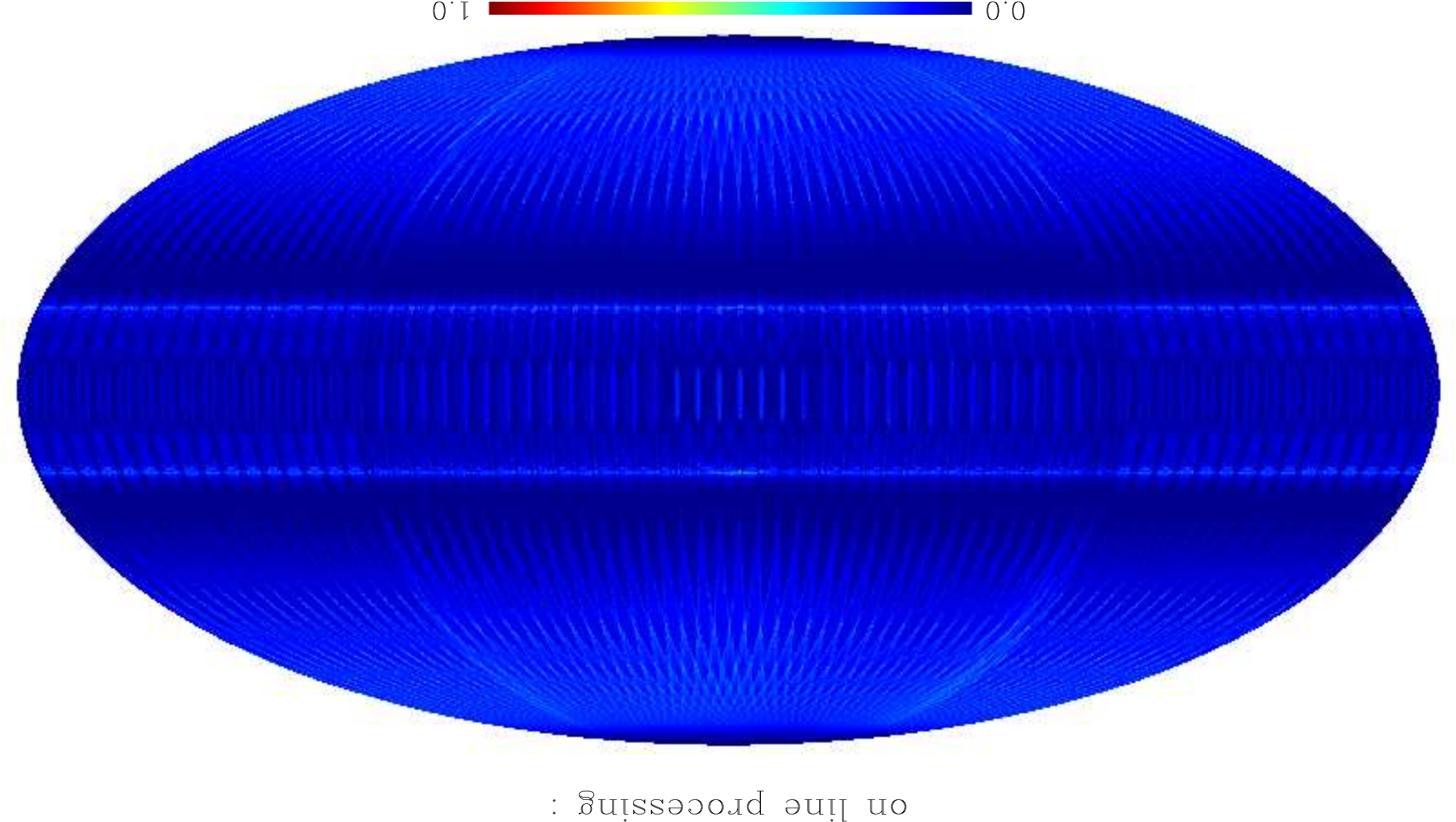}\\
\end{tabular}
\caption{The value of $\langle | h_{2}|^2\rangle$ of each scan listed
  in Table~\ref{tab:core_scan_params} on a map of $N_{\rm side} =$
  128. As in Figure~\ref{fig:core_scans_time_int} the rows correspond
  the scans running for 10 days, 1 month, 2 months, 6 months and 1
  year, from top to bottom respectively.}
\label{fig:core_scans_h_2}
\end{center}
\end{figure*}

%% file: sections/discussion.tex
% !TEX TS-program = compile

\section{Discussion and Conclusions} \label{sec:discussion}

The CMB $B$-mode power spectrum is approximately four orders of magnitude
fainter than the CMB temperature power spectrum. Any
instrumental imperfections that couple temperature fluctuations to
$B$-mode polarisation must therefore be carefully controlled and/or removed. Here
we have investigated the role that a scan strategy can have in mitigating
certain common systematics by averaging the error down with many
crossing angles. In Section~\ref{sec:ana}, we presented approximate
analytical forms for the error on the recovered $B$-mode power spectrum
that would result from differential gain, pointing and ellipticity if
two detector pairs are used in a polarisation experiment. By
minimising the $h$-values ($\tilde{h}_n$, see Table~\ref{tab:terms}),
of the scan strategy using multiple crossing angles, certain types of 
systematic effects can be averaged down. The different spins of the
systematic effects mean that different $h$-values are relevant for mitigating
different types of systematic effects.

By examining equations (\ref{eq:dg})--(\ref{eq:de}) we can see that
since differential gain is a spin-0 systematic effect, it is
suppressed by a factor $\langle |\tilde{h}_2|^2\rangle$, where the
average is over all the pixels in the scan. Differential pointing is
spin-1 and is therefore suppressed by factors involving $\langle
|\tilde{h}_1|^2\rangle$ and $\langle|\tilde{h}_3|^2\rangle
$. Differential ellipticity can couple temperature fluctuations to
$B$-mode polarisation in two ways: the first term in
equation~\eqref{eq:de} is independent of the scan strategy and
therefore if present, it will always result in spurious $B$-mode polarisation
regardless of the scan strategy. This has been shown a number of times
before~\citep{2007MNRAS.376.1767O,2008PhRvD..77h3003S,2014MNRAS.442.1963W}.
If the orientation of the differential ellipticity is parallel or
perpendicular to the polarisation sensitivity direction then the
coupling from temperature will be of a pure $E$-mode form. Any
misalignment and there will be also be coupling to $B$-mode polarisation. The
second term of equation~\eqref{eq:de} models the effect due to a difference in the
differential ellipticity between two pairs of detectors. This effect
couples temperature fluctuations to both $E$- and $B$-modes and can be
mitigated with an appropriate scan strategy through the $\langle
|\tilde{h}_4|^2\rangle$ term.

In Section~\ref{sec:sims}, we used simulations to calculate the error
on the $B$-mode power spectrum, shown in Fig.~\ref{fig:B_rec}, for
three example scan strategies. The larger precession angles and
smaller boresight angles of the WMAP and EPIC scans reduce the even
$h$-values and are therefore better at mitigating differential gain
and differential ellipticity as compared to the {\it Planck}
scan. {\color{black} The faster precession periods of the WMAP and EPIC scans reduce the 
odd $h$-values and are therefore better at mitigating differential pointing
as well.} In terms of future searches for
inflationary $B$-modes, our study suggests that differential gain is
potentially the most problematic effect as it affects lower $\ell$
ranges. 

Based on the validation by simulations in Section~\ref{sec:sims} as
well as the pseudo-$C_\ell$-based argument presented in
Appendix~\ref{sec:pseudo_cl_coupling}, we can be confident in using
the analytic predictions for the error on the $B$-mode polarisation
power spectrum of Section~\ref{sec:ana}. In Section
\ref{sec:scan_params} we combine the analytic analysis with a fast scan
strategy simulation code to search the scan strategy parameter space
for the optimal scan strategy, one which minimises the error on the cosmilogical parameters.
 A key result is Fig.~\ref{fig:h_num}
where we have presented the $\langle |\tilde{h}_n|^2\rangle$ values
for a range of satellite scan strategy parameters.

Our main conclusions of the general investigation in Section \ref{sec:scan_params} are as follows: (i) as long as $T_{\rm spin} \ll
T_{\rm prec} \ll 1~{\rm year}$ the exact values of the time scales are
unimportant for mitigating systematics by multiple crossing angles.
(ii) the main parameters of interest are the precession angle
($\alpha$) and the boresight angle ($\beta$). By lowering $\beta$ and increasing
$\alpha$ the scan strategy will make smaller circles on the
sky. These small circles are beneficial for creating a wide range of
orientation angles and are therefore effective in mitigating several
of the systematic effects that we have
considered. This is demonstrated in Fig.~\ref{fig:b_mode_grid} where we plot
the potential impact on the recovered value of the tensor-to-scalar ratio and
lensing parameter $A_{\rm lens}^{BB}$. For the particular levels of systematics that we have
assumed, we find little difference in our preferred scan strategy
regardless of whether we choose to target the inflationary $B$-mode
signal on large angular scales or the lensing $B$-mode signal on
smaller scales.

In Section \ref{sec:scan_investigation} we consider these general
observations and present the optimal scan strategies for three different
Lissajous orbits. The different orbits require different amounts of fuel
for injection, leaving different amounts to drive the scan. Also the different
orbits allow different precessions angles due to constraints from the data
transfer antenna (see Section~\ref{sec:prac_con}). We chose the scans
based on their ability to mitigate constant systematic errors in the
cosmological parameters (see Fig.~\ref{fig:b_mode_grid}). Further to this we require the scan
to allow us to make maps of half the sky on short time scales, in the order of tens of days.
Fig.~\ref{fig:core_scans_time_int} shows that the two ``medium orbit" scans
can make half sky maps short timescales while the other two require
the Earth to orbit around the Sun in order to observe half the sky.

During this optimisation we considered the practical constraints that
limit the possible scan strategies. It should be noted that the
constraint that limited our choice most is the constraint set by
equation \eqref{eq:aspect_angle}, where the precession angle is
limited to ensure the antenna can transmit data to the earth. This
forces us to limit the precession angle in the ``large orbit" to the
point where half maps cannot be made on short time scales.  The error
on the cosmological parameters were also much worse for the ``large
orbit" case than the other two because of the lower precession angle.
If the antenna was designed to allow a larger aspect angle with
respect to the Earth, then we could increase the precession angle. This
would improve both these issues. As the precession period on the
``large orbit" is much faster, due to the increased amount of fuel, it
would allow even faster half-sky maps, on the order of days instead of
tens of days.

Ultimately, the choice of scan strategy for any future CMB
polarization satellite mission will be a trade-off between these
various competing scientific requirements, the instrumental
capabilities, and the fuel and data rate resources required. The
results presented in this paper should prove very useful for rapidly
assessing the contribution of systematics mitigation in any such
trade-off exercise for a future CMB polarisation satellite.

%% file: sections/appendix.tex
% !TEX TS-program = compile

\clearpage
\appendix
\onecolumn
\section{pseudo-$C_\ell$ approach to calculating the temperature leakage} \label{sec:pseudo_cl_coupling}

In Section~\ref{sec:ana} we have made an assumption about the way in
which the scan strategy impacts on the leaked $B$-mode power
spectrum. Specifically, we assumed that the effect of mitigation by the scan strategy
can be approximated as a simple suppression of the leaked signal by the relevant
$\langle|\ft{h}_n|\rangle$ value (see Section~\ref{sec:ana}) and that
the direction of the leaked polarisation would mean half of the power is
of an  $E$-mode form and the other half is $B$-mode form. This
approximation allowed
us to derive a relatively simple set of equations to describe
the temperature-to-$B$-mode polarisation power spectrum leakage. This
simple set of equations was essential to quickly predict the
leaked $B$-mode power spectrum for any given scan strategy as we did
in Section~\ref{sec:scan_params}. Here we examine this approximation
further and derive an exact analytical form for the leaked $B$-mode power
spectrum for the case of differential gain. With this exact equation
we can examine the approximation further.

We start from equation~\eqref{eq:deltap_dg} which describes the
leakage in the polarisation map due to differential gain. We begin by
defining the spherical harmonic modes of the CMB temperature field and
of the differential gain as suppressed by the scan strategy:
\ba
\ft{a}^T_{\ell m} &=& \int d\Omega ~_{0}Y_{\ell m}^*(\Omega)~ T^B(\Omega),\\
H^{\pm2}_{\ell m} &=& \int d\Omega ~_{\pm2}Y_{ \ell m}^*(\Omega) ~\frac{1}{2}(\delta g_1 \pm i\delta g_2) \ft{h}_{\pm2}(\Omega).
\ea
We are interested in working out the error on the $B$-mode power
spectrum. We therefore calculate the decomposition of the leaked
polarisation in terms of spin weighted spherical harmonics,
\ba
_2\ft{a}_{\ell_1 m_1} &=& \int d\Omega ~  _{2}Y_{ \ell_1 m_1}^*(\Omega) ~\Delta P^{\rm g}(\Omega),\\
_{-2}\ft{a}_{ \ell_1 m_1} &=& \int d\Omega ~  _{-2}Y_{ \ell_1 m_1}^*(\Omega) ~\Delta P^{\rm g*}(\Omega).
\ea
Substituting these into our expression for $\Delta P^{\rm g}$, we find
\ba
_{\pm2}\ft{a}_{\ell_1 m_1} &=& \int d\Omega ~  _{\pm2}Y_{ \ell_1 m_1}^*(\Omega) ~\frac{1}{2}(\delta g_1 \pm i\delta g_2)\ft{h}_{\pm2}(\Omega) T^B(\Omega).
\ea
We now substitute the spherical harmonic decomposition of the smoothed
temperature field to find
\ba
_{\pm2}\ft{a}_{ \ell_1 m_1} &=& \int d\Omega ~  _{\pm2}Y_{ \ell_1 m_1}^*(\Omega) ~\frac{1}{2}(\delta g_1 \pm i\delta g_2)\ft{h}_{\pm2}(\Omega) \sum_{\ell_2 m_2} \ft{a}^T_{\ell_2 m_2}\,_{0}Y_{\ell_2 m_2}(\Omega)\\
&=& \sum_{\ell_2 m_2} K^{\pm~\ell_1 \ell_2}_{m_1 m_2} \ft{a}^T_{\ell_2 m_2},
\ea
where we have defined the coupling kernel,
\ba
K^{\pm~\ell_1 \ell_2}_{m_1 m_2} &=& \int d\Omega ~  _{\pm2}Y_{ \ell_1 m_1}^*(\Omega) ~\frac{1}{2}(\delta g_1 \pm i\delta g_2)\ft{h}_{\pm2}(\Omega) _{0}Y_{\ell_2 m_2}(\Omega).
\ea
We can now calculate the error on the recovered $B$-mode power
spectrum. We start with the error on the measured $B$-mode power spectrum,
\ba
\Delta \ft{C}^{BB}_{\ell_1} &=& \frac{1}{2\ell_1 + 1} \sum_{m_1} \ft{a}^{B}_{\ell_1 m_1}\ft{a}^{B*}_{\ell_1 m_1}\\
&=& \frac{1}{4(2\ell_1 + 1)} \sum_{m_1} (\ft{a}^{2}_{\ell_1 m_1} - \ft{a}^{-2}_{\ell_1 m_1}) (\ft{a}^{2}_{\ell_1 m_1} - \ft{a}^{-2}_{\ell_1 m_1})^*\\
&=& \frac{1}{4(2\ell_1 + 1)} \sum_{\substack {m_1{}\\ \ell_2 m_2{}\\ \ell'_2 m'_2}} [K^{+~\ell_1 \ell_2}_{m_1 m_2} \ft{a}^T_{\ell_2 m_2}K^{+~\ell_1 \ell'_2*}_{m_1 m'_2} \ft{a}^{T*}_{\ell'_2 m'_2} + K^{-~\ell_1 \ell_2}_{m_1 m_2} \ft{a}^T_{\ell_2 m_2}K^{-~\ell_1 \ell'_2*}_{m_1 m'_2} \ft{a}^{T*}_{\ell'_2 m'_2}\label{eq:big_couple}\\
 \nonumber  && \qquad\qquad\qquad\qquad+ \quad(K^{+~\ell_1 \ell_2}_{m_1 m_2} \ft{a}^T_{\ell_2 m_2}K^{-~\ell_1 \ell'_2*}_{m_1 m'_2} \ft{a}^{T*}_{\ell'_2 m'_2} + {\rm c.c.})],
\ea
where the brackets denote the term to which the ${\rm c.c.}$ applies to.
We can simplify this equation by requiring statistical isotropy of the
CMB temperature field. This allows us to write $\ft{a}^T_{\ell_1
m_1}\ft{a}^{T*}_{\ell_2 m_2} = \delta_{\ell_1 \ell_2}\delta_{m_1 m_2}
B^2_{\ell_1}\langle C_{\ell_1}^T\rangle$, where $B_{\ell}$
is the temperature beam window function and $\langle \rangle$
 denotes averaging over CMB realisations. Substituting this
result into equation \eqref{eq:big_couple} and evaluating the
Kronecker delta functions gives us
\ba
\langle \Delta \ft{C}^{BB}_{\ell_1}\rangle &=& \frac{1}{4(2\ell_1 + 1)} \sum_{\substack {m_1{}\\ \ell_2 m_2}} [K^{+~\ell_1 \ell_2}_{m_1 m_2} K^{+~\ell_1 \ell_2*}_{m_1 m_2} + K^{-~\ell_1 \ell_2}_{m_1 m_2} K^{-~\ell_1 \ell_2*}_{m_1 m_2} + (K^{+~\ell_1 \ell_2}_{m_1 m_2} K^{-~\ell_1 \ell_2*}_{m_1 m_2} + {\rm c.c.})] B^2_{\ell_2}\langle C^{TT}_{\ell_2} \rangle\\
&=& \sum_{\ell_2} M_{\ell_1 \ell_2} B^2_{\ell_2}\langle C^{TT}_{\ell_2} \rangle.
\ea 
To calculate the coupling operator we must first calculate a product
of two coupling kernels. We do this in
Appendix~\ref{sec:mult_kernals}. To calculate the error on the
recovered $B$-mode power spectrum we must then deconvolve for the
polarisation power spectrum. We assume that the temperature and
polarisation beam window functions are the same. This gives us,
\ba
\langle \Delta C^{BB}_{\ell_1}\rangle &=& \frac{1}{B^2_{\ell_1}}\sum_{\ell_2} M_{\ell_1 \ell_2} B^2_{\ell_2}\langle C^{TT}_{\ell_2} \rangle.\label{eq:full_dg}
\ea 
As with all pseudo-$C_\ell$ coupling operators this matrix can be
approximated by a diagonal matrix with values equal to the fraction of
sky covered in the experiment. This approximation works best when the CMB power 
spectrum is close to constant. Here the equivalent to the sky fraction
is simply the average of the modulus squared of the window function
$\frac{1}{2}(\delta g_1 + i\delta g_2)\ft{h}_2$. There is one
difference in that half of the spurious polarisation power will be in
a $E$-mode form and the other half in a $B$-mode form. Therefore we
have that
\ba
M_{\ell_1 \ell_2} \approx \frac{1}{8}|\delta g_1 + i\delta g_2|^2\langle|\ft{h}_2|^2\rangle\delta_{\ell_1 \ell_2},
\ea
With this approximation, equation~\eqref{eq:full_dg} reduces
to equation~\eqref{eq:dg}. In Fig.~\ref{fig:pscl_B_error} we show how well
this approximation holds for realistic scan strategies. We plot the
recovered $B$-mode power spectrum from simulations (as described in
Section~\ref{sec:sims}) assuming the {\it Planck}, WMAP
and EPIC scan strategies and including a differential gain systematic
error in each detector pair. We also plot the predictions for the
biased $B$-mode power,
\ba
C^{BB ~{\rm rec}}_{\ell} = C^{BB~{\rm true}}_{\ell} + \Delta C^{BB}_{\ell}.
\ea
We plot two predictions for the biased power: one where the predicted
error is of the simplified form of equation~\eqref{eq:dg} and a second
where the full coupling operator of equation~\eqref{eq:full_dg} is
used. The coupling operator based prediction is in excellent agreement
with the simulations. However, it is also clear that the simplified
formulae of Section~\ref{sec:ana} also provide a very good
approximation over the multipole range of interest. 
%In Section \ref{sec:scan_prams} we use a simplified
%equations to quickly predict the error on the recovered $B$-mode power
%spectrum for two $\ell$ ranges, 2 to 201 and 801 to 1000. This is be
%representative of a experiments ability to correctly recover
%inflationary $B$-modes and the lensing $B$-mode
%peak. Figure \ref{fig:pscl_B_error} shows a good agreement with the
%simplified equations and the full coupling operator, giving us
%confidence in the approximations made in Section \ref{sec:ana}.
\input{sections/fig/pseudo_cl}

\section{Product of two coupling kernels} \label{sec:mult_kernals}

In Appendix \ref{sec:pseudo_cl_coupling} we use the product of two
coupling kernels. Here we calculate this product. The
definition of the coupling kernel gives us,
\ba
K^{\pm~\ell_1 \ell_2}_{m_1 m_2} &=& \int d\Omega ~  _{\pm2}Y_{ \ell_1 m_1}^*(\Omega) ~\frac{1}{2}(\delta g_1 \pm i\delta g_2)\ft{h}_{\pm2}(\Omega) _{0}Y_{\ell_2 m_2}(\Omega),\\
&=& \sum_{\ell_3 m_3} H^{\pm 2}_{\ell_3 m_3}\int d\Omega ~  _{\pm2}Y_{ \ell_1 m_1}^*(\Omega) _{0}Y_{\ell_2 m_2}(\Omega)_{\pm2}Y_{ \ell_2 m_2}(\Omega),\\
&=& \sum_{\ell_3 m_3} (-1)^{m_1} H^{\pm 2}_{\ell_3 m_3} \sqrt{F_{\ell_1 \ell_2 \ell_3}} \threej{\ell_1}{\ell_2}{\ell_3}{-m_1}{m_2}{m_3}\threej{\ell_1}{\ell_2}{\ell_3}{\mp2}{0}{\pm2},
\ea
where the second equality comes from an identity found in \citet{varshalovich1988}, and we have defined,
\ba
F_{\ell_1 \ell_2 \ell_3} = \frac{(2\ell_1 + 1)(2\ell_2 + 1)(2\ell_3 + 1)}{4\pi}.
\ea
We are now in a position to calculate the product of two coupling kernels,
\ba
\sum_{m_1 m_2} K^{\pm~\ell_1 \ell_2}_{m_1 m_2} K^{\pm~\ell_1 \ell_2 *}_{m_1 m_2} &=& \sum_{\substack{m_1 m_2{}\\ \ell_3 m_3{}\\ \ell'_3 m'_3}}   H^{\pm 2}_{\ell_3 m_3} H^{\pm 2 *}_{\ell'_3 m'_3} \sqrt{F_{\ell_1 \ell_2 \ell_3}} \threej{\ell_1}{\ell_2}{\ell_3}{-m_1}{m_2}{m_3}\threej{\ell_1}{\ell_2}{\ell_3}{\mp2}{0}{\pm2} \\ \nonumber
&& \qquad \qquad \qquad \qquad \qquad \times \sqrt{F_{\ell_1 \ell_2 \ell'_3}} \threej{\ell_1}{\ell_2}{\ell'_3}{-m_1}{m_2}{m'_3}\threej{\ell_1}{\ell_2}{\ell'_3}{\mp2}{0}{\pm2}.
\ea
To simplify this result, we use the orthogonality relation,
\ba
\sum_{m_1 m_2}\threej{\ell_1}{\ell_2}{\ell_3}{m_1}{m_2}{m_3}\threej{\ell_1}{\ell_2}{\ell'_3}{m_1}{m_2}{m'_3} = \frac{1}{2\ell_3 +1} \delta_{\ell_3, \ell'_3} \delta_{m_3, m'_3}.
\ea
Evaluating the Kronecker delta function leads to 
\ba
\sum_{m_1 m_2} K^{\pm~\ell_1 \ell_2}_{m_1 m_2} K^{\pm~\ell_1 \ell_2 *}_{m_1 m_2} &=& \sum_{\ell_3}   F_{\ell_1 \ell_2 \ell_3}\mathcal{H}^{\pm~\pm}_{\ell_3}\threej{\ell_1}{\ell_2}{\ell_3}{\mp2}{0}{\pm2} \threej{\ell_1}{\ell_2}{\ell_3}{\mp2}{0}{\pm2},
\ea
where we have defined the spin power spectrum of the window function,
\ba
\mathcal{H}^{\pm~\pm}_{\ell_3} = \frac{1}{2\ell_3 + 1} \sum_{m_3} H^{\pm 2}_{\ell_3 m_3} H^{\pm 2 *}_{\ell_3 m_3}.
\ea

%% file: sections/fig/pseudo_cl.tex
% !TEX TS-program = compile

\begin{figure}
\begin{center}
\includegraphics[width=0.5\linewidth, trim=0cm 0cm 0cm 0cm, clip=true]{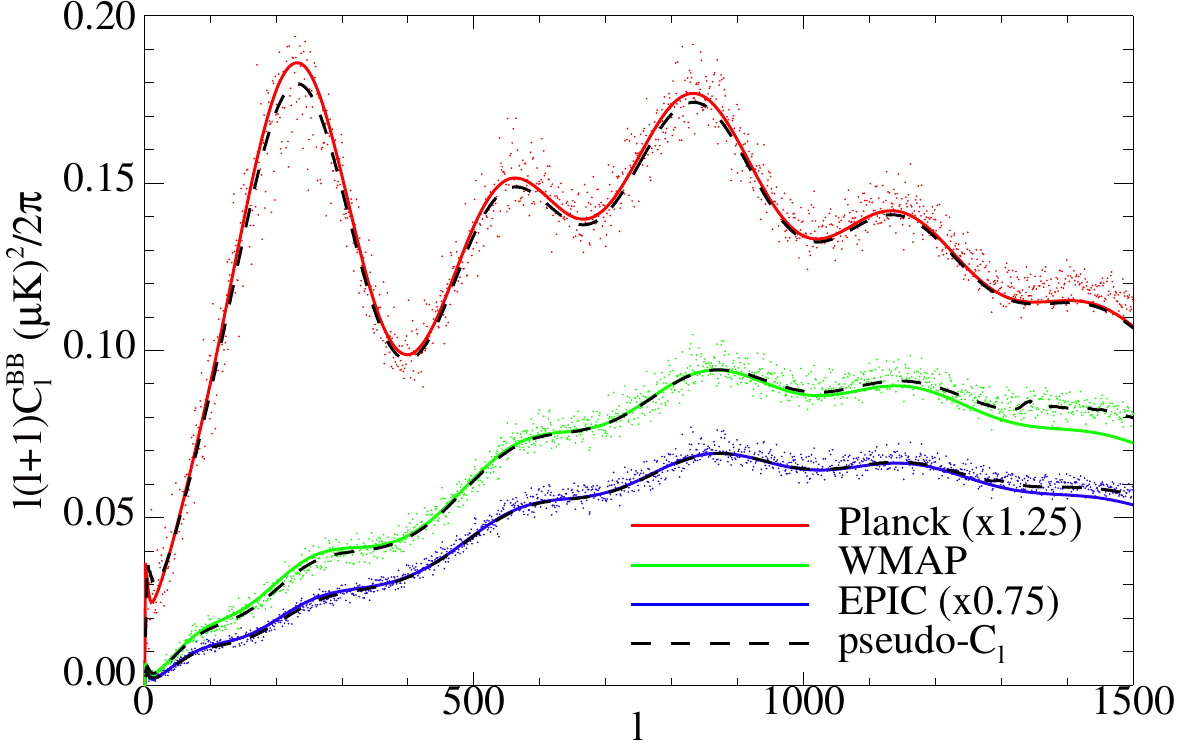}
\caption{The recovered $B$-mode power spectrum when a differential
  gain systematic error is present assuming the {\it
    Planck}, WMAP and EPIC scan strategies (shown in red, green and blue
  respectively). The points show the result for one simulation. The solid lines shows
  the predictions for the recovered power spectrum  using the approximate
  model of equation~\eqref{eq:dg}. The black dashed line shows the
  predictions using the full coupling operator of
  equation~\eqref{eq:full_dg}. There is good agreement between the
  two predictions for $\ell\lesssim1000$. This provides strong
  justification for using the simple equations of
  Section~\ref{sec:ana} to predict the error on the $B$-mode power
  spectrum in Section~\ref{sec:scan_params}. {\color{black}The small discrepancy
between the pseudo-$C_\ell$ prediction and the simulation result 
for {\it Planck} at high $\ell$ is a numerical artefact associated
with the simulation software.}}
\label{fig:pscl_B_error}
\end{center}
\end{figure}